\documentclass{myllncs-mixalis}
\usepackage{graphicx}
\usepackage{amsmath}
\usepackage{amssymb}
\usepackage[ruled, noline, algo2e, noend]{algorithm2e}
\usepackage[T1]{fontenc}
\usepackage{enumerate}
\usepackage[font=small, format=hang, labelfont=bf]{caption}
\usepackage{subfig}
\usepackage{color}
\usepackage{color}

\SetArgSty{}  \SetKw{KwOr}{or}
\SetKw{KwAnd}{and} \SetKw{KwInvariant}{invariant:}
\SetKw{KwDownTo}{down to} \SetKwInOut{Input}{input}
\SetKwInOut{Output}{output} \SetKwInOut{Require}{require}
\SetKwComment{Comment}{start}{end}
\SetKwIF{If}{ElseIf}{Else}{if}{then}{else if}{else}{endif}
\dontprintsemicolon

\newcommand{\eat}[1] {{}}

\newtheorem{prop}{Property}

\begin{document}
\parindent=0cm
\parskip=5pt

\title{Crossing-Optimal Acyclic HP-Completion for Outerplanar $st$-Digraphs}
\author{
Tamara  Mchedlidze
,  Antonios Symvonis 
}

\institute{%
    Dept. of Mathematics, National Technical University of Athens,
 Athens, Greece.\\
    \texttt{\{mchet,symvonis\}@math.ntua.gr}
}


 \maketitle

\vspace{-10pt}

\begin{abstract}

Given an embedded  planar acyclic digraph $G$, we define the problem
of \emph{acyclic hamiltonian path completion with crossing
minimization (Acyclic-HPCCM)} to be the problem of determining a
\emph{hamiltonian path completion set} of edges such that, when
these edges are embedded on $G$, they create the smallest possible
number of edge crossings and turn  $G$ to a hamiltonian acyclic
digraph. Our results include:
\begin{enumerate}
\item
We provide a characterization under which a planar $st$-digraph $G$
is hamiltonian.
\item For an outerplanar $st$-digraph $G$,
we define the \emph{$st$-polygon decomposition of $G$} and, based on
its properties, we develop a linear-time algorithm that solves the
Acyclic-HPCCM problem.
\item
For the class of planar $st$-digraphs, we establish an equivalence
between the Acyclic-HPCCM problem and the problem of determining an
upward 2-page topological book embedding with  minimum number of
spine crossings. We infer (based on this equivalence) for the class
of outerplanar $st$-digraphs an upward topological 2-page book
embedding with minimum number of spine crossings.
\end{enumerate}

To the best of our knowledge, it is the first time  that
edge-crossing minimization is studied in conjunction with the
acyclic hamiltonian completion problem and the first time that an
optimal algorithm with respect to spine crossing minimization is
presented for upward topological book embeddings.
\end{abstract}

\keywordname{~Hamiltonian path completion, planar graph, outerplanar
graph,  st-graph, crossing, topological book embedding, upward
drawing.}\\ \\




\section{Introduction}
In the \emph{hamiltonian path completion problem} (for short,
\emph{HP-completion}) we are given a graph $G$ (directed or
undirected) and we are asked to identify a set of edges (refereed to
as an \emph{HP-completion set}) such that, when these edges are
embedded on  $G$ they turn it to a hamiltonian graph, that is, a
graph containing a hamiltonian path\footnote{In the literature, a
\emph{hamiltonian graph} is traditionally referred to as a graph
 containing a hamiltonian cycle. In this paper, we refer to a
hamiltonian graph as a graph containing  a hamiltonian path.}. The
resulting hamiltonian graph $G^\prime$ is referred to as the
\emph{HP-completed graph} of $G$.
 When we treat the HP-completion problem as an
optimization problem, we are interested in an HP-completion set of
minimum size.

When the input graph $G$ is a planar embedded digraph, an
HP-completion set for $G$ must be naturally extended to include an
embedding of its edges on the plane, yielding to an embedded
HP-completed digraph $G^\prime$. In general, $G^\prime$ is not
planar, and thus, it is natural to attempt to minimize the number of
edge crossings of the embedding of the HP-completed digraph
$G^\prime$ instead of the size of the HP-completion set. We refer to
this problem as the \emph{HP-completion with crossing minimization
problem} (for short, \emph{HPCCM}).

 \eat{The HPCCM problem can be
further refined  by placing restrictions on the maximum number of
permitted crossings per edge of  $G$.}

When the input digraph $G$ is acyclic, we can insist on
HP-completion sets which leave the HP-completed digraph $G^\prime$
also acyclic. We refer to this version of the problem as the
\emph{acyclic HP-completion problem}.

A \emph{k-page book} is a structure consisting of a line, referred
to as \emph{spine}, and of $k$ half-planes, referred to as
\emph{pages}, that have the spine as their common boundary. A
\emph{book embedding} of a  graph $G$ is a drawing of $G$ on a book
such that the vertices are aligned along the spine, each edge is
entirely drawn on a single page, and edges do not cross each other.
If we are interested only in two-dimensional structures we have to
concentrate on  2-page book embeddings and to allow spine crossings.
These embeddings are  also referred to as  2-page\emph{ topological
}book embeddings.

For acyclic digraphs, an upward book embedding can be considered to
be a book embedding in which the spine is vertical and all edges are
drawn monotonically increasing in the upward direction. As a
consequence, in an upward book embedding of an acyclic digraph the
vertices appear along the spine in topological order.

The results on topological book embeddings that appear in the
literature focus on the number of spine crossings per edge required
to book-embed a graph on a 2-page book. However, approaching the
topological book embedding problem as an optimization problem, it
makes sense to also try to minimize the total number of spine
crossings.

In this paper, we introduce the problem of \emph{acyclic hamiltonian
path completion with crossing minimization} (for short,
\emph{Acyclic-HPCCM}) for planar embedded acyclic digraphs. To the
best of our knowledge, this is the first time that edge-crossing
minimization is studied in conjunction with the acyclic
HP-completion problem. Then, we provide a characterization under
which a  planar $st$-digraph is hamiltonian. For an outerplanar
 $st$-digraph $G$, we define the
\emph{$st$-polygon decomposition of $G$} and, based on the
decomposition's properties, we develop a linear-time algorithm that
solves the Acyclic-HPCCM problem.

In addition, for the class of planar $st$-digraphs, we establish an
equivalence between the acyclic-HPCCM problem and the problem of
determining an upward 2-page topological book embeddig with a
minimal number of spine crossings. Based on this equivalence, we can
infer  for the class of outerplanar  $st$-digraphs an upward
topological 2-page book embedding with minimum number of spine
crossings. Again, to the best of our knowledge, this is the first
time that an optimal algorithm with respect to spine crossing
minimization is presented for upward topological book embeddings.

\subsection{Problem Definition}
\label{sec:problemDefinition}
 Let $G=(V,E)$ be a  graph. Throughout
the paper, we use the term \emph{``graph''} we refer to both
directed and undirected graphs. We use the term \emph{``digraph''}
when we want to restict our attention to directed graphs. We assume
familiarity with basic graph theory~\cite{Harary72,Diestel05}. A
\emph{hamiltonian path} of $G$ is a  path that visits every vertex
of $G$ exactly once. Determining whether a graph has a hamiltonian
path or circuit is NP-complete~\cite{GareyJS74}. The problem remains
NP-complete for cubic planar graphs~\cite{GareyJS74}, for maximal
planar graphs~\cite{Wigderson82}  and for planar
digraphs~\cite{GareyJS74}. It can be trivially solved in polynomial
time for planar acyclic digraphs.

Given a graph $G=(V,E)$, directed or undirected, a non-negative
integer $k \leq |V|$ and two vertices $s,~t \in V$, \emph{the
hamiltonian path completion (HPC)} problem asks whether there exists
a superset $E^\prime$ containing $E$ such that $|E^\prime- E| \leq
k$  and the graph $G^\prime = (V, E^\prime)$ has a hamiltonian path
from vertex $s$ to vertex $t$. We refer to  $G^\prime$ and to the
set of edges $|E^\prime- E|$ as the \emph{HP-completed graph} and
the \emph{HP-completion set} of graph $G$, respectively. We assume
that all edges of a HP-completion set are part of the Hamiltonian
path of $G^\prime$, otherwise they can be removed. When $G$ is a
directed acyclic graph, we can insist on HP-completion sets which
leave the HP-completed digraph also acyclic. We refer to this
version of the problem as the \emph{acyclic HP-completion problem}.
The hamiltonian path completion problem is
NP-complete~\cite{GareyJ79}. For
acyclic digraphs 
the  HPC problem is solved in polynomial time~\cite{KarejanM80}.

A \emph{drawing} $\Gamma$  of graph $G$ maps every vertex $v$ of $G$
to a distinct point $p(v)$ on the plane and each edge $e=(u,v)$ of
$G$ to a simple Jordan curve joining $p(u)$ with $p(v)$. A drawing
in which every edge $(u,v)$ is a a simple Jordan curve monotonically
increasing in the vertical direction is an \emph{upward  drawing}. A
drawing $\Gamma$ of graph $G$ is \emph{planar} if no two distinct
edges intersect except at their end-vertices. Graph $G$ is  called
\emph{planar} if it admits a planar drawing $\Gamma$.

An embedding of a planar graph $G$ is the equivalence class of
planar drawings of $G$ that define the same set of faces or,
equivalently, of face boundaries. A planar graph together with the
description of a set of faces $F$ is called an \emph{embedded planar
graph}.

Let $G=(V,E)$ be an embedded planar graph, $E^\prime$ be a superset
of   edges containing $E$, and $\Gamma(G^\prime)$ be a drawing of~
$G^\prime=(V, E^\prime)$. When the deletion from $\Gamma(G^\prime)$
of the edges in $E^\prime -E$ induces the embedded planar graph $G$,
we say that $\Gamma(G^\prime)$ \emph{preserves the embedded planar
graph $G$}.

\begin{definition}
\label{def:HPCCM} Given an embedded planar  graph   $G=(V,E)$,
directed or undirected, a non-negative integer $c$, and two vertices
$s,~t \in V$, the \emph{hamiltonian path completion with edge
crossing minimization (HPCCM) problem}  asks whether there exists a
superset $E^\prime$ containing $E$ and a drawing $\Gamma(G^\prime)$
of graph $G^\prime = (V, E^\prime)$ such that (i) $G^\prime$ has a
hamiltonian path from vertex $s$ to vertex $t$, (ii)
$\Gamma(G^\prime)$ has at most $c$ edge crossings, and (iii)
$\Gamma(G^\prime)$ preserves the embedded planar graph $G$.
\end{definition}

We refer to the version of the HPCCM problem where the input is an
 acyclic digraph and we
are interested in HP-completion sets which leave the HP-completed
digraph also acyclic as the \emph{Acyclic-HPCCM} problem.

Over the set of all HP-completion sets for a graph $G$, and over all
of their different drawings that respect $G$, the one with a minimum
number of edge-crossings is called a \emph{crossing-optimal
HP-completion set}.

\begin{figure}[htb]
    \begin{minipage}{\textwidth}
    \centering
    \includegraphics[width=1\textwidth]{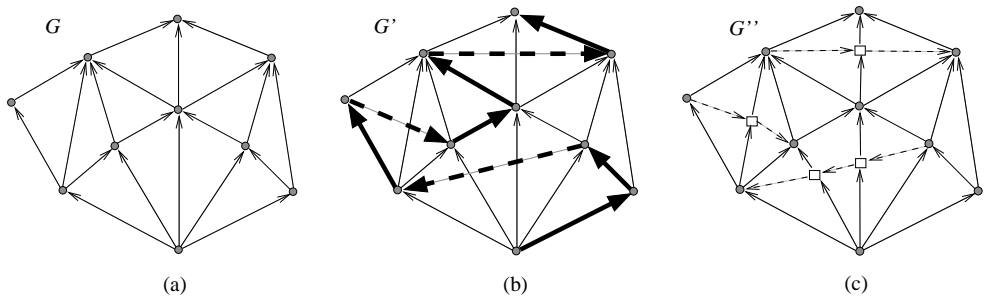}
    \caption{(a) A planar embedded digraph $G$.
    (b)  A drawing $\Gamma(G^\prime)$ of an HP-completed digraph $G^\prime$ of
    $G$. The edges of the hamiltonian path of $G^\prime$ appear
    bold, with the edges of the the HP-completion set  shown dashed.
    (c) The HP-extended digraph $G^{\prime\prime}$ of $G$ w.r.t. $\Gamma(G^\prime)$.
    The newly inserted vertices appear as squares.}
    \label{fig:HPextended}
  \end{minipage}
\end{figure}

Let $G=(V,E)$ be an embedded  planar graph, let $E_c$ be an
HP-completion set of $G$ and let $\Gamma(G^\prime)$ of $G^\prime =
(V, E\cup E_c)$ be a drawing with $c$ crossings that preserves  $G$.
The graph $G_c$ induced from drawing $\Gamma(G^\prime)$ by inserting
a new vertex at each edge crossing and by splitting the edges
involved in the edge-crossing is referred to as the
\emph{HP-extended graph of $G$ w.r.t. $\Gamma(G^\prime)$}. (See
Figure~\ref{fig:HPextended})

In this paper, we present a linear time algorithm for solving the
Acyclic-HPCCM problem for outerplanar  $st$-digrpahs. A planar graph
$G$ is \emph{outerplanar} if there exist a drawing of $G$ such that
all of $G$'s vertices appear on the boundary of the same face (which
is usually drawn as the external face). Let $G=(V,E)$ be a  digraph.
A vertex of $G$ with in-degree equal to zero (0) is called a
\emph{source}, while, a vertex of $G$ with out-degree equal to zero
is called a  \emph{sink}. An \emph{$st$-digraph} is an acyclic
digraph with exactly one source  and exactly one sink.
Traditionally, the source and the sink of an $st$-digraph are
denoted by $s$ and $t$, respectively. An $st$-digraph which is
planar (resp. outerplanar) and, in addition, it is embedded on the
plane so that both of its source and sink appear on the boundary of
its external face, is referred to as a \emph{planar $st$-digraph}
(resp. an \emph{outerplanar $st$-digraph}). \eat{An
\emph{triangulated outerplanar} graph is an outerplanar graph with
triangulated interior, i.e., all interior faces consist of 3
vertices and 3 edges.} It is known that a planar $st$-digraph admits
a planar upward drawing \cite{Kelly87,DiBattistaT88}. In the rest of
the paper, all $st$-digraphs will be drawn upward.

\subsection{Related Work}

For acyclic digraphs, the  Acyclic-HPC problem has been studied in
the literature in the context of partially ordered sets (posets)
under the terms  \emph{Linear extensions} and \emph{Jump Number}.
Each acyclic-digraph $G$ can be treated as a poset $P$. A linear
extension of $P$ is a total ordering $L =\{x_1 \ldots x_n\}$ of the
elements of $P$ such that $x_i < x_j$ in $L$ whenever $x_i < x_j$ in
$P$. We denote by $L(P)$ the set of all linear extensions of $P$. A
 pair $(x_i, x_{i+1})$ of consecutive elements of $L$ is called a \emph{\emph{jump} in
$L$} if $x_i$ is not comparable to $x_{i+1}$ in $P$. Denote the
number of jumps of $L$  by $s(P, L)$. Then, the \emph{jump number}
of $P$, $s(P)$, is defined as $s(P) = \min \{s(P, L): L \in L(P)\}$.
Call  a linear extension $L$ in $L(P)$ optimal if $s(P, L) = s(P)$.
The \emph{jump number problem} is to find $s(P)$ and to construct an
optimal linear extension of $P$.

From the above definitions, it follows that an optimal linear
extension of a poset $P$ (or its corresponding  acyclic digraph
$G$), is identical to an acyclic HP-completion set $E_c$ of minimum
size for $G$, and its jump number is equal to the size of $E_c$.
This problem has been widely studied, in part due to its
applications to scheduling. It has been shown to be NP-hard even for
bipartite ordered sets~\cite{Pulleyblank81} and the class of
interval orders~\cite{Mitas91}. Up to our knowledge, its
computational classification is still open for lattices.
Nevertheless, polynomial time algorithms are known for several
classes of ordered sets. For instance, efficient algorithms are
known for series-parallel orders~\cite{CogisH79}, N-free
orders~\cite{Rival83}, cycle-free orders~\cite{DuffusRW82}, orders
of width two~\cite{CheinH84}, orders of bounded
width~\cite{ColbournP85}, bipartite orders of dimension
two~\cite{SteinerS87} and K-free orders~\cite{ShararyZ91}.
Brightwell and  Winkler~\cite{BrightwellW91} showed that counting
the number of linear extensions is $\boldmath{\sharp}$P-complete. An
algorithm that generates all of the linear extensions of a poset in
a constant amortized time, that is in time $\mathcal{O}(|L(P)|)$,
was presented by Pruesse and Ruskey~\cite{PruesseR94}. Later,  Ono
and Nakano~\cite{OnoN05} presented an algorithm which generates each
linear extension in worst case constant time.

With respect to related work on book embeddings,
Yannakakis\cite{Yannakakis89} has shown that planar graphs have a
book embedding on a 4-page book and that there exist planar graphs
that require 4 pages for their book embedding. Thus, book embedding
for planar graphs are, in general, three-dimensional structures. If
we are interested only on two-dimensional structures we have to
concentrate on  2-page book embeddings and to allow spine crossings.
In the literature, the book embeddings where spine crossings are
allowed are referred to as \emph{topological book
embeddings}~\cite{EnomotoMO99}. It is known that every planar graph
admits a 2-page topological book embedding with only one spine
crossing per edge~\cite{DiGiacomoDLW05}.

For acyclic digraphs and posets, \emph{upward book embeddings} have
been also studied in the
literature~\cite{AlzohairiR96,HeathP97,HeathP99,HeathPT99,NowakowskiP89}.
An upward book embedding can be considered to be a book embedding in
which the spine is vertical and all edges are drawn monotonically
increasing in the upward direction.  The minimum number of pages
required by an upward book embedding of a planar acyclic digraph is
unbounded~\cite{HeathP97}, while, the minimum number of pages
required by an upward planar digraph is not
known~\cite{AlzohairiR96,HeathP97,NowakowskiP89}. Giordano et
al.~\cite{GiordanoLMS07} studied \emph{upward topological book
embeddings} of embedded upward planar digraphs, i.e., topological
2-page book embedding where all edges are drawn monotonically
increasing in the upward direction. They have showed how to
construct in linear time an upward topological book embedding for an
embedded triangulated planar $st$-digraph with at most one spine
crossing per edge. Given that (i) upward planar digraphs are exactly
the subgraphs of planar $st$-digraphs~\cite{DiBattistaT88,Kelly87}
and (ii) embedded upward planar digraphs  can be augmented to become
triangulated planar $st$-digraphs in linear
time~\cite{GiordanoLMS07}, it follows that any embedded upward
planar digraph has a topological book embedding with one spine
crossing per edge.

We emphasize that the presented bibliography is in no way
exhaustive. The topics of \emph{hamiltonian paths}, \emph{linear
orderings} and \emph{book embeddings} have been studied for a long
time and an extensive body of literature has been accumulated.

\subsection{Our Results}
\label{sec:ourResults} 

\begin{figure}[htb]
    \centering
    \includegraphics[width=0.99\textwidth]{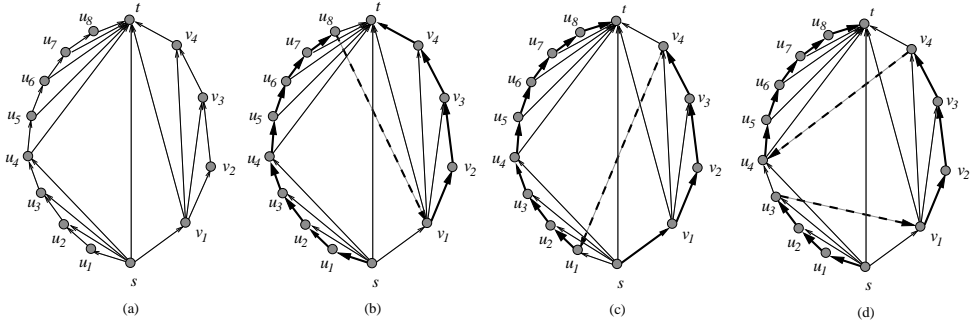}
    \caption{Two crossing per edge is needed to minimize the total number of crossing.
    The  edges of the HP-completion sets appear dashed. The resulting hamiltonian path are shown in bold. }
    \label{fig:counterexample_no_of_crossings}
\end{figure}

In our previous work on Acyclic-HPCCM
problem~\cite{MchedlidzeS09,MchetSymArxiv08} we reported a linear
time algorithm that solves this problem for the class of outerplanar
\textit{ triangulated} $st$-digraph provided  that \textit{each edge
of the initial graph can be crossed at most once} by the edge of the
crossing-optimal HP-completion set.
Figure~\ref{fig:counterexample_no_of_crossings}.a  gives an example
of an outerplanar triangulated $st$-digraph for which an
HP-completion set with smaller number of crossings can be found if
there is no restriction on the number of crossings per edge. In
particular, the $st$-digraph  becomes hamiltonian by adding one of
the following completion sets: $A=\{(u_8, v_1)\}$, $B=\{(v_4,
u_1)\}$ or $C=\{(u_3,v_1), (v_4,u_4)\}$ (see
Figures~\ref{fig:counterexample_no_of_crossings}.b-d). Sets $A$ and
$B$ creates 5 crossings with one crossing per edge of $G$ while, set
$C$ creates 4 crossings with at most 2 crossings per edge of $G$.

In addition to relaxing the restriction of {at most one crossing per
edge of the $st$-digraph}, the algorithm presented in this paper
does not require its input outerplanar $st$-digraph to be
triangulated, extending in this way the class of graphs for which we
are able to compute a crossing-optimal HP-completion set.

\begin{figure}[htb]
    \centering
    \includegraphics[width=0.8\textwidth]{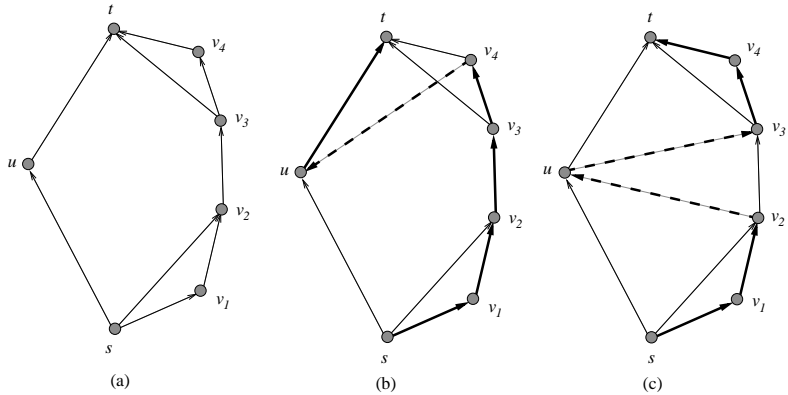}
    \caption{A non-triangulated $st$-polygon that has a crossing optimal HP-completion set of size 2
    that creates no crossings.
    Any  HP-completion set of size 1 creates 1 crossing.}
    \label{fig:counterexample_triangulated}
\end{figure}

For the non-triangulated outerplanar $st$-digraph of
Figure~\ref{fig:counterexample_triangulated}.a, every acyclic
HP-completion set of size 1 creates 1 edge crossing (see
Figure~\ref{fig:counterexample_triangulated}.b) while, it is
possible to obtain an acyclic HP-completion set of size 2 without
any crossings (see Figure~\ref{fig:counterexample_triangulated}.c).

In this work we show that (i) for any $st$-polygon (i.e., an
outerplanar $st$-digraph with no edge connecting its two opposite
sides) there is always a  crossing-optimal acyclic HP-completion set
of size at most 2 (Section~\ref{sec:st-polygons},
Theorem~\ref{thm:st-polygon-size-2}),  and,  (ii)   any crossing
optimal acyclic HP-completion set for an outerplanar $st$-digraph
$G$ creates at most 2 crossings per edge of $G$
(Section~\ref{sec:HP-completion-set-properties},
Theorem~\ref{thm:allproperties}). Based on these properties and
introduced $st$-polygon decomposition of an outerplanar $st$-digraph
(Section~\ref{sec:st-polygon-decomp}, we derive a linear time
algorithm that solves the Acyclic-HPCCM problem for outerplanar
$st$-digraphs.

In~\cite{MchedlidzeS09} we established an equivalence between the
acyclic-HPCCM problem and the problem of determining an upward
2-page topological book embedding with a minimal number of spine
crossings. Based on this equivalence and the algorithm in this
paper, we can infer for the class of outerplanar triangulated
$st$-digraphs an upward topological 2-page book embedding with
minimum number of spine crossings. To the best of our knowledge,
this is the first time that an optimal algorithm with respect to
spine crossing minimization is presented for upward topological book
embeddings without restrictions the number of crossings per edge.

\section{Hamiltonian Planar $st$-Digraphs}

In this section, we develop the necessary and sufficient condition
for a planar $st$-digraph to be hamiltonian. The provided
characterization will be later used in the development of
crossing-optimal HP-completion sets for outerplanar $st$-digraphs.

It is well known\cite{TamassiaT86} that for every vertex $v$ of a
planar $st$-digraph, its incoming (outgoing) incident edges appear
consecutively around $v$. For any  vertex $v$,  we denote by $Left(v)$ (resp. $Right(v)$)
the face to the left (resp. right) of the leftmost (resp. rightmost)
incoming and outgoing edges incident to $v$.
For any  edge $e=(u,v)$,  we denote by $Left(e)$ (resp. $Right(e)$)
the face to the left (resp. right) of edge $e$ as we move from $u$ to $v$.
The \emph{dual} of an $st$-digraph $G$, denoted by  $G^{*}$, is a digraph
such that: (i) there is a vertex in $G^*$ for each face of G; (ii)
for every edge $e \neq (s, t)$ of $G$,  there is an edge $e^* = (f,g)$ in
$G^*$, where $f = Left(e)$ and $g = Right(e)$; (iii) egde $(s^*, t^*)$ is in $G^*$.
The following lemma is
a direct consequence from Lemma~7 by Tamassia and
Preparata~\cite{TamassiaP90}.

\begin{lemma}
\label{lem:TamassiaPreparata} Let $u$ and $v$ be two vertices of a
planar $st$-digraph such that there is no directed path between them
in either direction. Then, in the dual $G^*$ of $G$ there is either
a path from $Right(u)$ to $Left(v)$ or a path from $Right(v)$ to
$Left(u)$. \qed
\end{lemma}

The following lemma demonstrates a property of planar $st$-digraphs.

\begin{lemma}
\label{lemma:mutualDisconnected} Let $G$ be a planar $st$-digraph
that does not have a hamiltonian path. Then, there exist two
vertices in $G$ that are not connected by a directed path in either
direction.
\end{lemma}

\begin{proof}

Let $P$ be a  longest path from $s$ to $t$ and let $a$ be a vertex
that does not belong in $P$. Since $G$ does not have a hamiltonian
path, such a vertex always exists.  Let $s^\prime$ be the last
vertex in $P$ such that there exists a path
$P_{s^\prime\rightsquigarrow a}$  from $s^\prime$ to $a$ with no
vertices in $P$. Similarly, define $t^\prime$ to be the first vertex
in $P$ such that there exists a path $P_{ a\rightsquigarrow
t^\prime}$ from $a$ to $t^\prime$ with no vertices in $P$. Since $G$
is acyclic, $s^\prime$ appears before $t^\prime$ in $P$ (see
Figure~\ref{fig:mutualDisconnected}). Note that $s^\prime$ (resp.
$t^\prime$) might be vertex $s$ (resp. $t$). From the construction
of $s^\prime$ and $t^\prime$ it follows that any vertex $b$,
distinct from $s^\prime$ and $t^\prime$, that is located on path $P$
between vertices  $s^\prime$ and $t^\prime$, is not connected with
vertex $a$ in either direction. Thus, vertices $a$ and $b$ satisfy
the property of the lemma.

\begin{figure}[htb]
    \centering
    \includegraphics[width=0.2\textwidth]{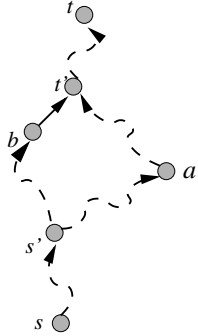}
    \caption{Subgraph used in the proof of
    Lemma~\ref{lemma:mutualDisconnected}. Vertices $a$ and $b$ are
    not connected by a path in either direction.}
    \label{fig:mutualDisconnected}
\end{figure}

Note that such a vertex $b$ always exists. If this was not the case,
then  path $P$ would contain edge $(s^\prime,t^\prime)$.  Then, path
$P$ could be extended by replacing  $(s^\prime, t^\prime)$ by path
$P_{s^\prime\rightsquigarrow a}$ followed by path
$P_{s^\prime\rightsquigarrow a}$. This would lead to new path
$P^\prime$ from $s$ to $t$ that is longer than $P$, a contradiction
since $P$ was assumed to be of maximum length. \qed
\end{proof}

\begin{figure}[htb]
    \begin{minipage}{0.47\textwidth}
    \centering
    \includegraphics[width=0.75\textwidth]{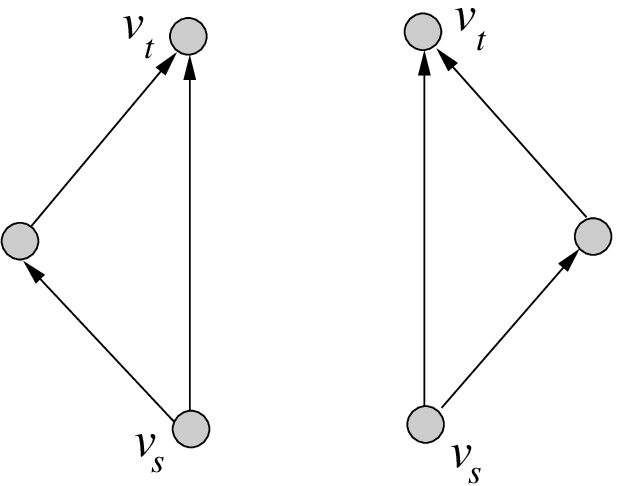}
    \caption{Left and  right-sided embedded triangles.}
    \label{fig:triangle}
  \end{minipage}
\hfill
    \begin{minipage}{0.47\textwidth}
    \centering
    \includegraphics[width=.8\textwidth]{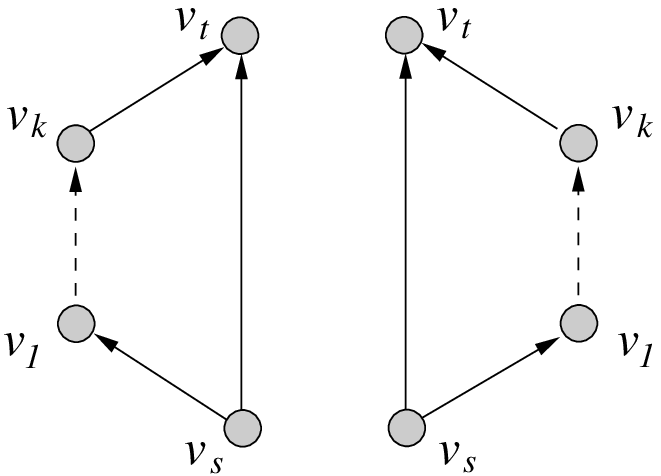}
    \caption{Left and  right-sided embedded  generalized triangles.}
    \label{fig:generalized_triangle}
  \end{minipage}
\end{figure}

Every face of a planar $st$-digraph consists of two sides, each of
them directed from its source to its sink. When  one side of the
face is a single edge and the other side (the longest) contains
exactly one vertex, the face is referred to as \emph{triangle} (see
Figure~\ref{fig:triangle}). In the case where the longest edge
contains more than one vertex, the face is referred to as a
\emph{generalized triangle} (see
Figure~\ref{fig:generalized_triangle}). We call both a triangle and
a generalized triangle \emph{left-sided} (rest. \emph{right-sided})
if its left (resp. right) side is its longest side, i.e., it
contains at least one vertex.

\begin{figure}[htb]
    \begin{minipage}{0.47\textwidth}
    \centering
    \includegraphics[width=0.6\textwidth]{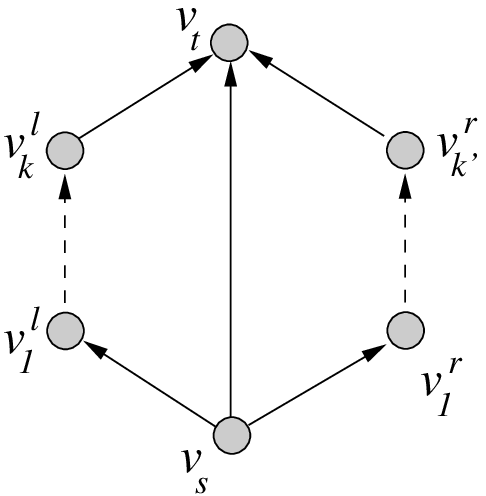}
    \caption{A strong rhombus.}
    \label{fig:rhombus_strong}
  \end{minipage}
\hfill
    \begin{minipage}{0.47\textwidth}
    \centering
    \includegraphics[width=.6\textwidth]{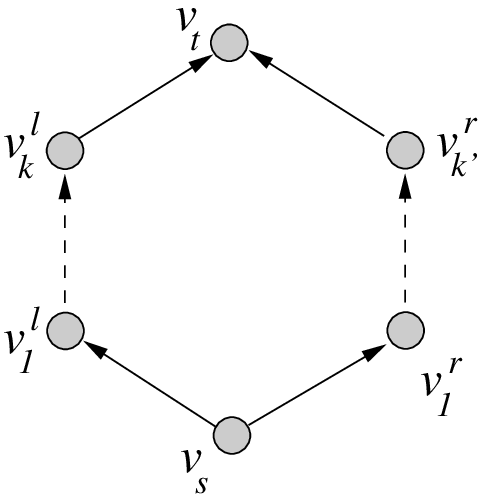}
    \caption{A weak rhombus.}
    \label{fig:rhombus_weak}
  \end{minipage}

\end{figure}

 The outerplanar $st$-digraph of Figure~\ref{fig:rhombus_strong} is called a \emph{strong rhombus}.
 It consists of two generalized triangles (one left-sided and one right-sided) which have
 their $(s,t)$ edge in common.  The edge $(s,t)$ of a rhombus is referred to as
 its \emph{median} and is always drawn in the interior of
its drawing. The outerplanar $st$-digraph resulting from the
deletion of the median of a strong rhorbus is referred to as a
\emph{weak rhombus}. Thus, a  weak rhombus  is  an outerplanar
$st$-digraph consisting of a single face that has at least one
vertex at each its side (see Figure~\ref{fig:rhombus_weak}). We use
the term {\em rhombus} to refer to either a strong or a weak
rhombus.

\eat{
\begin{figure}[htb]
    \begin{minipage}{0.47\textwidth}
    \centering
    \includegraphics[width=0.6\textwidth]{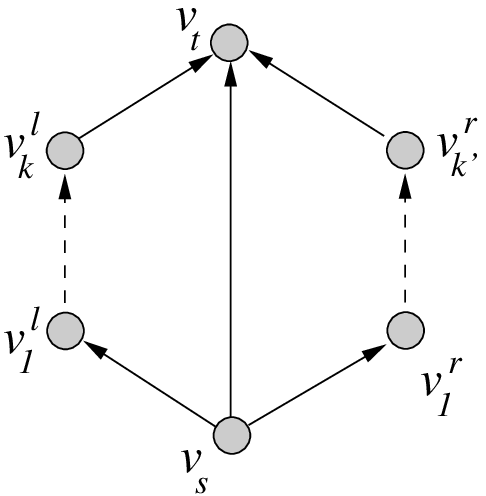}
    \caption{The rhombus embedded digraph.}
    \label{fig:rhombus_strong}
  \end{minipage}
\hfill
    \begin{minipage}{0.47\textwidth}
    \centering
    \includegraphics[width=.65\textwidth]{images/st-polygon-3.eps}
    \caption{An $st$-polygon.}
    \label{fig:st-polygon}
  \end{minipage}

\end{figure}

 An \emph{$st$-polygon with median} is an outerplanar
 $st$-digraph  that always contains edge $(s,t)$
connecting its source $s$ to its sink $t$. Edge $(s,t)$ is referred
to as the \emph{median of the $st$-polygon} and it always lies in
the interior of the its drawing. As a consequence, an $st$-polygon
with median must have at least 4 vertices.
Figure~\ref{fig:st-polygon} shows an $st$-polygon with median. A
\emph{generalized $st$-polygon} is an $st$-polygon with median from
which the edge $(s,t)$(median) was removed (see
Figure~\ref{fig:st-polygon_gen}). As a result a generalized
$st$-polygon do not contain any edge connecting its left and its
right sides. When we refer to an $st$-polygon we will mean both
general and with median $st$-polygons. An $st$-polygon (that is a
subgraph of some embedded planar digraph) which cannot be extended
by the addition of more vertices to its external boundary is called
a \emph{maximal $st$-polygon}.

\begin{figure}[htb]
    \begin{minipage}{0.47\textwidth}
    \centering
    \includegraphics[width=0.6\textwidth]{images/rhomb_gen.eps}
    \caption{The rhombus embedded digraph.}
    \label{fig:rhombus_gen}
  \end{minipage}
\hfill
    \begin{minipage}{0.47\textwidth}
    \centering
    \includegraphics[width=.65\textwidth]{images/st-polygon_gen.eps}
    \caption{An $st$-polygon.}
    \label{fig:st-polygon_gen}
  \end{minipage}

\end{figure}
}

\eat{ In the following we study only triangulated $st$-digraphs, so
any rhombus will be as in Figure~\ref{fig:trinagulated_rhombus}, and
any $st$-polygon will be triangulated. So for simplicity, in the
following when we say rhombus we mean the graph of
Figure~\ref{fig:trinagulated_rhombus}, and when saying $st$-polygon
we mean a triangulated $st$-polygon, so let the triangularity be the
part of definition of this graphs.

} 

\eat{
\begin{lemma}
\label{lemma:oneRombus} An $st$-polygon contains exactly one
rhombus.
\end{lemma}
\begin{proof}
 By definition, an $st$-polygon is a triangulated outerplanar
graph with $s$ and $t$ as its source and sink, respectively. It
follows, that the median of the $st$-polygon is also a median of a
rhombus formed by the two faces to the left and the right of edge
$(s,t)$. Assume now that there exists another rhombus. Then, due to
the fact that the $st$-polygon is outerplanar, all of its vertices
would lie on one side of the $(s,t)$-polygon. However, this is not
possible without violating the acyclicity of the $st$-polygon.
 \qed
\end{proof}
}eat

 The following theorem provides a characterization of
 $st$-digraphs that have a hamiltonian path.

\begin{theorem}
\label{thm:rhombus} Let $G$ be a planar $st$-digraph. $G$ has a
hamiltonian path if and only if $G$ does not contain any rhombus
(strong or weak) as a subgraph.
\end{theorem}

\begin{proof}

\begin{figure}[htb]
\begin{minipage}{\textwidth}
    \centering
    \includegraphics[width=0.25\textwidth]{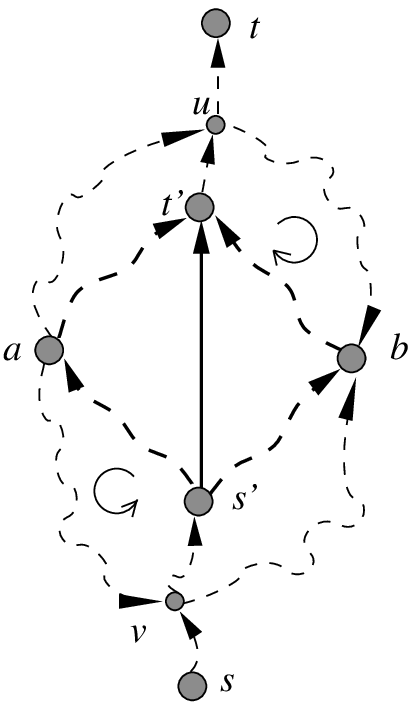}
    \caption{The subgraph containing a rhombus which is used in the proof
of Theorem~\ref{thm:rhombus}. In the case of a weak rhombus, edge
$(s^\prime,t^\prime)$ is not present.}
  \label{fig:genrhombusProof}
  \end{minipage}
\end{figure}

$(\Rightarrow)$~~~  We assume that $G$ has a hamiltonian path and we
 show  that it contains no  rhombus (strong or weak) as an
embedded subgraph. For the sake of contradiction,  assume first that
$G$ contains a strong rhombus  characterized by  vertices $s^\prime$
(its source), $t^\prime$ (its sink),   $a$ (on its left side) and
$b$ (on its right side) (see Figure~\ref{fig:genrhombusProof}).
Then, vertices $a$ and $b$ of the strong rhombus are not connected
by a directed path in either direction. To see this, assume wlog
that there was a path connecting $a$ to $b$. Then, this path has to
lie outside the rhombus and intersect either the path from
$t^\prime$ to $t$ at a vertex $u$ or the path from $s$ to $s^\prime$
at a vertex $v$. In either case, there must exist a cycle in $G$,
contradicting the fact that $G$ is acyclic.

Assume now, for the sake of contradiction again, that $G$ contains a
weak rhombus characterized by   vertices $s^\prime, ~t^\prime, ~a,~
\mbox{and}~ b$.  Then, by using the same argument as above, we
conclude that vertices $a$ and $b$ of the weak rhombus are not
connected by a directed path that lies outside the rhombus in either
direction. Note also that the vertices $a$ and $b$ can not be
connected by a path that lies in the internal of the weak rhombus
since the weak rhombus consists, by definition, of a single face.

So, we have shown that vertices $a$ and $b$ of the  rhombus (strong
or weak) are not connected  by a directed path in either direction,
and thus, there cannot exist any hamiltonian path in $G$, a clear
contradiction.

$(\Leftarrow)$ We assume that $G$ contains neither a strong nor a
weak  rhombus as an embedded subgraph and we  prove that $G$ has a
hamiltonian path. For the sake of contradiction, assume that $G$
does not have a hamiltonian path. Then, from
Lemma~\ref{lemma:mutualDisconnected}, if follows that there exist
two vertices $u$ and $v$ of $G$ that are not connected by a directed
path in either direction. From Lemma~\ref{lem:TamassiaPreparata}, it
then follows that there exists in the dual $G^*$ of $G$ a directed
path from either $Right(u)$ to $Left(v)$, or from $Right(v)$ to
$Left(u)$. Wlog, assume that the path in the dual $G^*$ is from
$Right(u)$ to $Left(v)$  (see Figure~\ref{fig:rhombusProofBack}.a)
and let $f_0, ~f_1, ~\ldots, ~f_k$ be the faces the path passes
through, where $f_0=Right(u)$ and $f_k=Left(v)$. We denote the path
from $Right(u)$ to $Left(v)$ by $P_{u,v}$. Note that each face of
the digraph $G$ and therefore of the path $P_{u,v}$ is a generalized
triangle, because as we supposed $G$ do not contain any weak
rhombus.

\begin{figure}[tb]
    \begin{minipage}{\textwidth}
    \centering
    \includegraphics[width=1\textwidth]{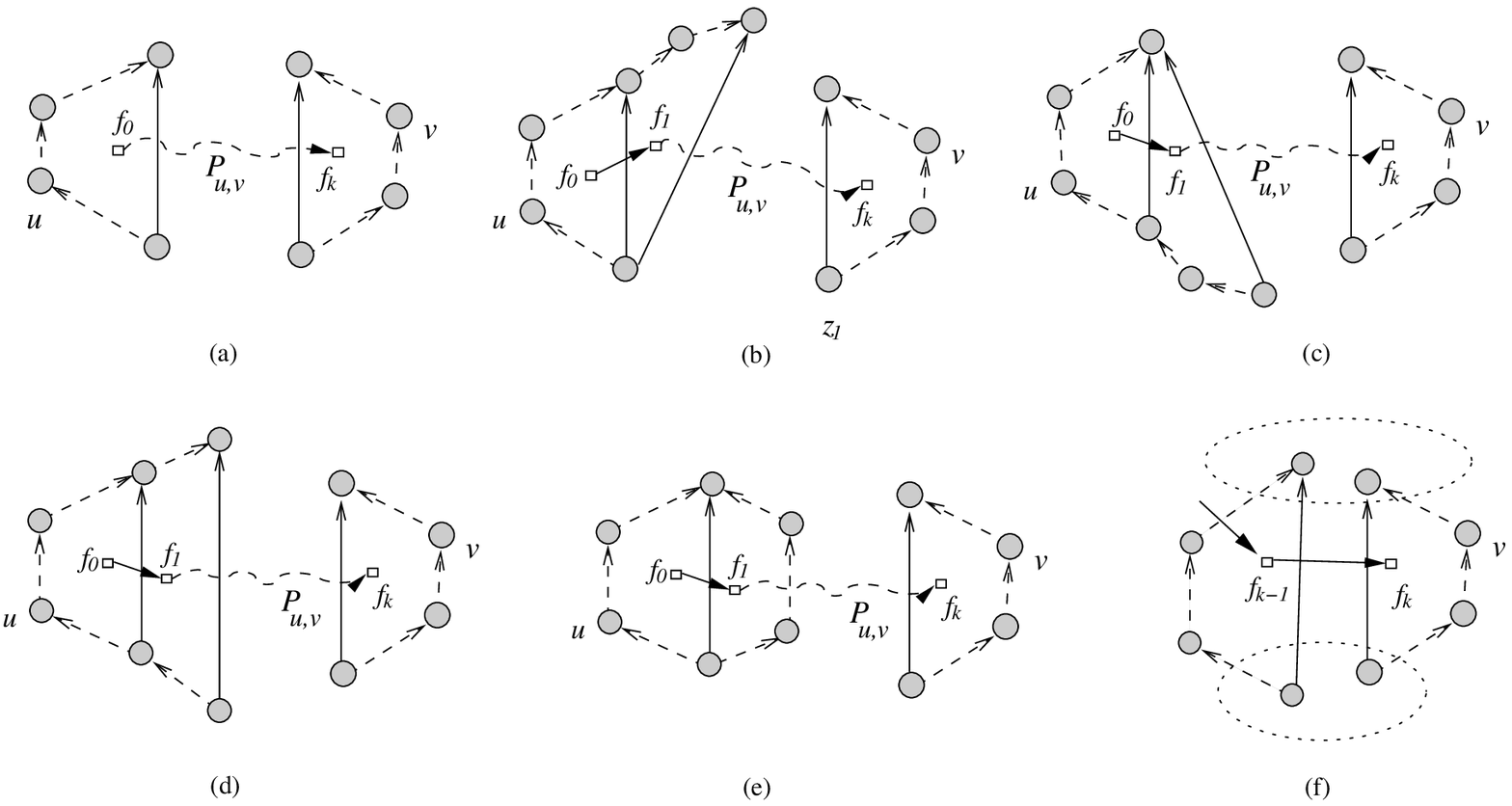}
    \caption{The different cases occurring in  the construction of path $P_{u,v}$ as described in
    the proof of Theorem~\ref{thm:rhombus}.}
    \label{fig:rhombusProofBack}
  \end{minipage}
\end{figure}

Note that path $P_{u,v}$ can exit face $f_0$ only through the solid
edge (see Figure~\ref{fig:rhombusProofBack}.a). The path then enters
a new face and, in the rest of the proof, we  construct the sequence
of faces it goes through.

The next face $f_1$ of the path,  consists of the solid edge of face
$f_0$ and some other edges. There are 2 possible cases to consider
for the face $f_1$:
\begin{description}
 \item[\textit{Case~1}:] Face $f_1$ is left-sided.  Then,
path $P_{u,v}$ enters $f_1$ through one of the edges on its left
side (see
Figure~\ref{fig:rhombusProofBack}.b,~\ref{fig:rhombusProofBack}.c,
~\ref{fig:rhombusProofBack}.d for possible configurations).\\
Observe that, since $f_1$ is left-sided, $f_1$ has only one outgoing
edge in $G^*$. Thus, in all of these cases, the only edge through
which  path $P_{u,v}$ can leave $f_1$ is the single edge on the
right side of the generalized triangle $f_1$.

\item[\textit{Case~2}:] The face $f_1$ is right-sided. Then the only
edge through which the path $P_{u,v}$ can enter $f_1$ is the the
only edge of the left side (see
Figure~\ref{fig:rhombusProofBack}.e). Note that in this case, $f_0$
and $f_1$ form a strong rhombus. Thus, this case cannot occur, since
we assumed that $G$ has no strong rhombus as an embedded subgraph.
\end{description}

A characteristic of the first case that allow to further continue
the identification of the faces path  $P_{u,v}$ goes through, is
that there is \emph{a single} edge that  exits face $f_1$. Thus, we
can continue identifying the faces path $P_{u,v}$ passes through,
building in such a way  a unique sequence $f_0, ~f_1, ~\ldots,
~f_{k-1}$. Note that all of these faces are left-sided otherwise $G$
contains a rhombus with median.

At the end, path $P_{u,v}$ has to leave the left-sided face
$f_{k-1}$  and  enter the right-sided face $f_k$. As the only way to
enter a right-sided face is to cross the single edge on its left
side, we have that the single edge on the right side of $f_{k-1}$
and the single edge on the left side of $f_k$ coincide forming a
strong rhombus (see Figure~\ref{fig:rhombusProofBack}.f).  This is a
clear contradiction since we assumed that $G$ has no strong rhombus
as an embedded subgraph.\qed
\end{proof}

\section{Optimal Acyclic Hamiltonian Path Completion for Outerplanar Triangulated st-digraphs}

In this section we present an algorithm that computes a
crossing-optimal acyclic HP-completion set  for an outerplanar
$st$-digraph. Let $G=(V^l \cup V^r \cup \{s,t\}, E)~$ be an
outerplanar
 $st$-digraph, where $s$ is its source, $t$ is its sink
and  the vertices in $V_l$ (resp. $V_r$) are located on the left
(resp. right) part of the boundary of the external face. Let $V^l =
\{ v^l_1,~\ldots, v^l_k\}$ and $V^r = \{  v^r_1,~\ldots, v^r_m\}$,
where the subscripts indicate the order in which the vertices appear
on the left (right) part of the external boundary. By convention,
source and the sink are considered to lie on both the left and the
right sides of the external boundary. Observe that each face of $G$
is also an outerplanar $st$-digraph.  We refer to an edge that has
both of its end-vertices on the same side of $G$ as an
\emph{one-sided} edge. All remaining edges are referred to as
\emph{two-sided} edges. The edges exiting the source and the edges
entering the sink are treated as one-sided edges.

The following lemma presents an essential property of an acyclic
HP-completion set of an outerplanar $st$-digraph $G$.

\begin{lemma}
\label{lem:sideInOrder} The HP-completion set of an outerplanar
$st$-digraph $G=(V^l \cup V^r \cup \{s,t\}, E)~$ induces a
hamiltonian path that visits  the vertices of $V_l$ (resp. $V_r$) in
the order they appear on the left side (resp. right side) of $G$.
\end{lemma}
\begin{proof}
Let $E_c$ be an acyclic HP-completion set for $G$ and let $G_c$ be
the induced HP-completed acyclic digraph. Consider two vertices
$v_1$ and $v_2$ that appear on that order on the same side (left or
rigth) of $G$. Then, in $G$ there is a path $P_{v_1,v_2}$ from $v_1$
to $v_2$ since each side of an outerplanar $st$-digraph is a
directed path from its source to its sink. For the sake of
contradiction, assume that $v_2$ appears before $v_1$ in the
hamiltonian path induced by the acyclic HP-completion set of $G$.
Then,  the hamiltonian path contains  a sub-path $P_{v_2,v_1}$ from
$v_2$ to $v_1$. Thus, paths $P_{v_1,v_2}$ and $P_{v_2,v_1}$ form a
cycle in $G_c$, a clear contradiction since $G_c$ is acyclic. \qed
\end{proof}

\subsection{st-polygons}
\label{sec:st-polygons}

\begin{figure}[htb]
    \begin{minipage}{0.47\textwidth}
    \centering
    \includegraphics[width=.70\textwidth]{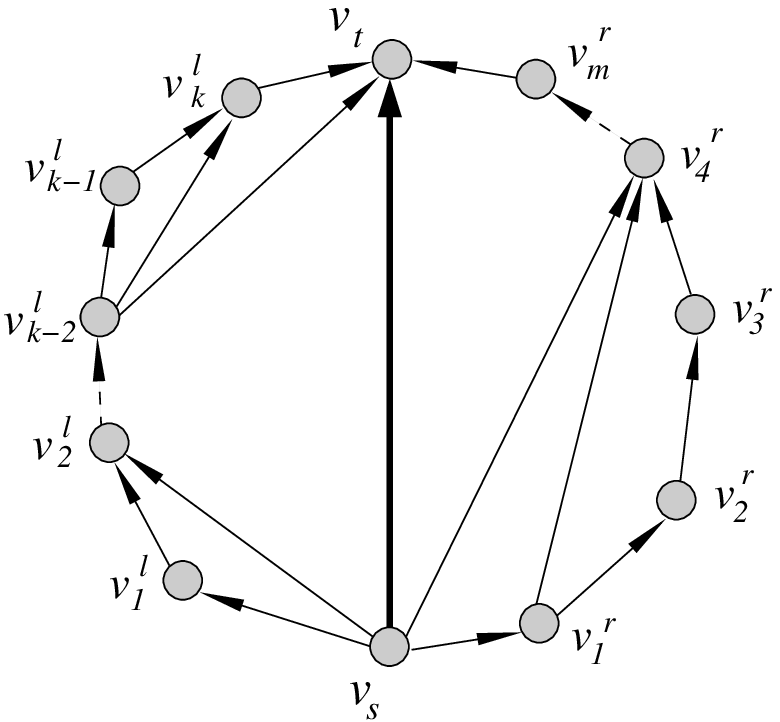}
    \caption{A strong $st$-polygon.}
    \label{fig:strong-st-polygon}
  \end{minipage}
\hfill
    \begin{minipage}{0.47\textwidth}
    \centering
    \includegraphics[width=.70\textwidth]{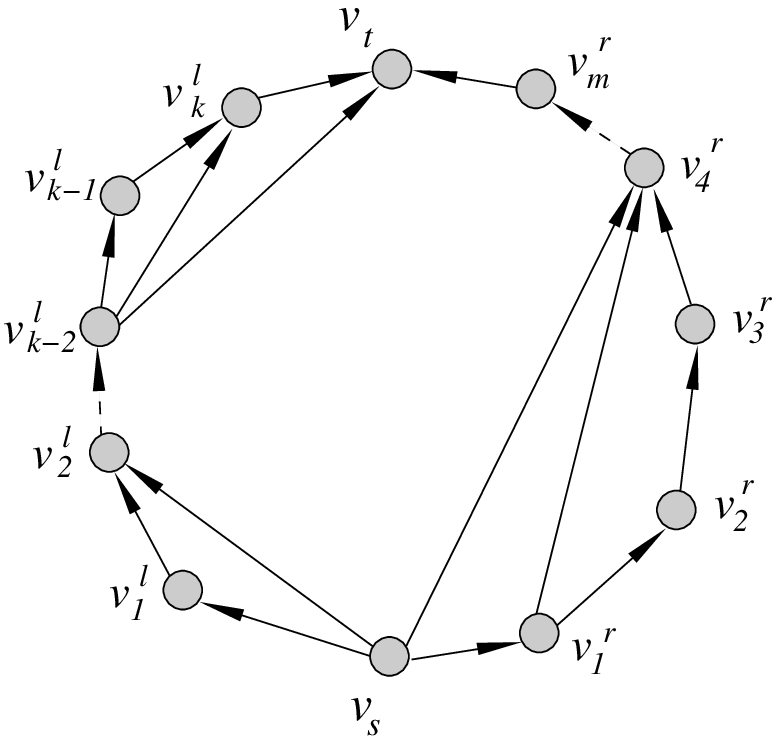}
    \caption{A weak  $st$-polygon.}
    \label{fig:weak-st-polygon}
  \end{minipage}
\end{figure}

 A  \emph{strong $st$-polygon} is an outerplanar
 $st$-digraph  that always contains edge $(v_s, v_t)$
connecting its source $v_s$ to its sink $v_t$ )(see
Figure~\ref{fig:strong-st-polygon}). Edge $(v_s,v_t)$ is referred to
as its \emph{median} and it always lies in the interior of  its
drawing. As a consequence , in a strong $st$-polygon no edge
connects a vertex on its left side to a vertex on its right side.
The outerplanar $st$-digraph that results from the deletion of the
median of a strong $st$-polygon is referred to as a \emph{weak
$st$-polygon} (see Figure~\ref{fig:weak-st-polygon}). We use the
term \emph{$st$-polygon} to refer to both a strong and a weak
$st$-polygon. Observe that each $st$-polygon has at least 4
vertices.

\begin{figure}[htb]
    \centering
    \includegraphics[width=.25\textwidth]{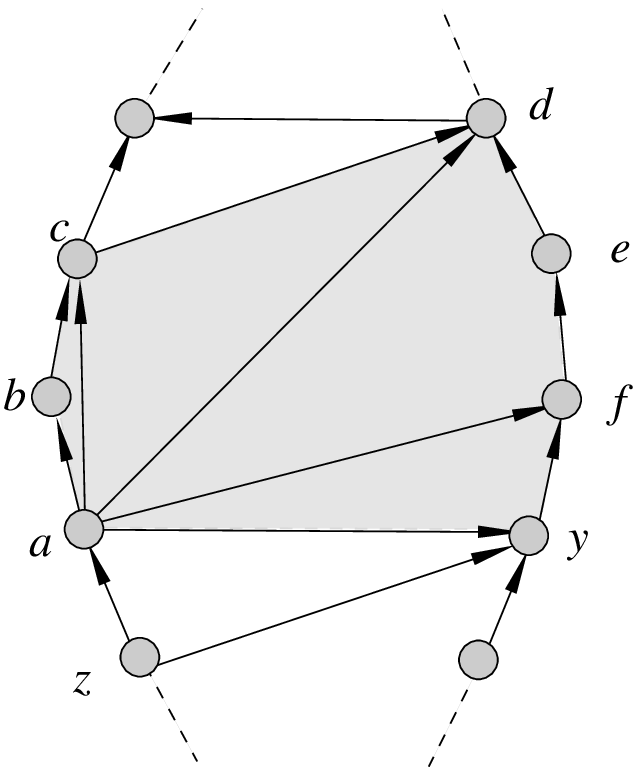}
    \caption{The $st$-polygon  with vertices $a ~\mbox{(source)},~b, ~c, ~d
~\mbox{(sink)},~ e, ~f, ~\mbox{and}~y$ on its boundary is maximal.}
    \label{fig:maximal-st-polygon}
\end{figure}

Consider an outerplanar $st$-digraph $G$ and  one of its embedded
subgraphs $G_p$ that is an $st$-polygon (strong or weak). $G_p$ is
called a \emph{maximal $st$-polygon} if it cannot  be extended (and
still remain an $st$-polygon) by the addition of more vertices to
its external boundary. In Figure~\ref{fig:maximal-st-polygon}, the
$st$-polygon $G_{a,d}$ with vertices $a ~\mbox{(source)},~b, ~c, ~d
~\mbox{(sink)},~ e,~\mbox{and}~f$ on its boundary is not maximal
since the subgraph $G^\prime_{a,d}$ obtained by adding vertex $y$ to
it is still an $st$-polygon. However, the $st$-polygon
$G^\prime_{a,d}$ is maximal since the addition of either vertex $x$
or $y$ to it does not yield another $st$-polygon.

Observe that  an  $st$-polygon that is a subgraph of an outerplanar
$st$-digraph $G$ fully occupies a \emph{``strip''} of it that is
limited by two edges (one adjacent to its source and one to its
sink), each having its endpoints at different sides of $G$. We refer
to these two edges as the \emph{limiting edges} of the $st$-polygon.
Note that the limiting edges of an $st$-polygon that is an embeded
subgraph of an outerplanar graph are sufficient to define it. In
Figure~\ref{fig:maximal-st-polygon}, the maximal $st$-polygon with
vertex $a$ as its source and vertex $d$ as its sink in limited by
edges $(a,y)$ and $(c,d)$.

\begin{lemma}
\label{lemma:oneRombus} An $st$-polygon contains exactly one
rhombus.
\end{lemma}
\begin{proof}
 Suppose a weak $st$-polygon $G_p$. By definition it contains a weak
 rhombus. Suppose that this is not the only weak rhombus contained
 in $G_p$ and let $R$ be a second one. As $G_p$ is an outerplanar graph
 and does not contain edges connecting its two opposite sides, we have that all
 the vertices of $R$ must lie on the same side of $G_p$, say its left side. But
 then we have that the sink of $R$ is another sink in $G_p$ or that the source of $R$
 is another source of $G_p$(see
 Figure~\ref{fig:maximalstpolygon}). This contradicts the fact that $G_p$ is an $st$-polygon.
 Suppose now that $R$ is a strong rhombus. This case  also leads to
 a contradiction, as $R$ can be converted to a weak rhombus by deleting its median.

\begin{figure}[htb]
    \centering
    \includegraphics[width=0.5\textwidth]{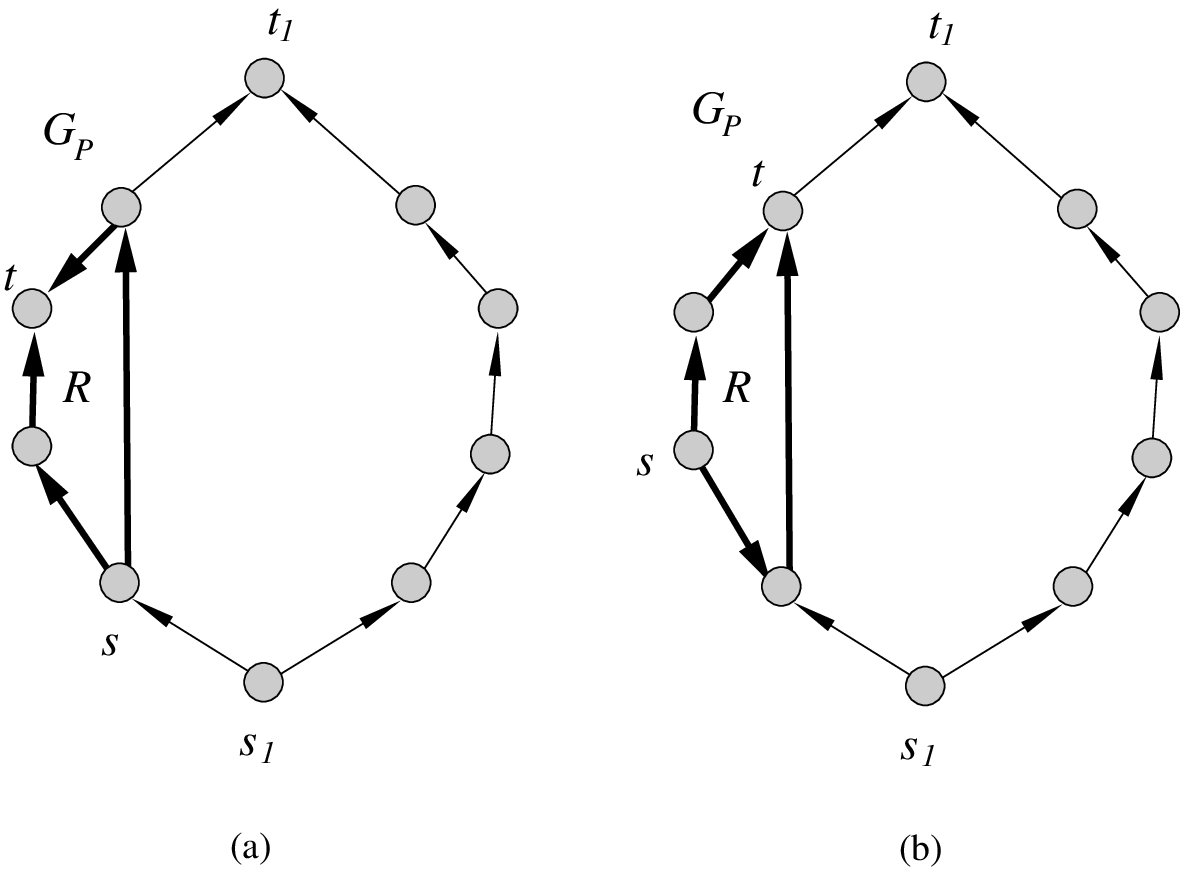}
    \caption{Two possible ways for the embedding of a second rhombus into an $st$-polygon.
    Both lead to a configuration that
    contradicts the definition of an $st$-polygon.}
    \label{fig:maximalstpolygon}
\end{figure}

If $G_p$ is a strong $st$-polygon, then by the same argument we
 show that $G_p$ can not contain a second rhombus (strong or weak).
 \qed
\end{proof}

The following lemmata concern a crossing-optimal acyclic
HP-completion set for a single $st$-polygon. They state that there exist crossing optimal acyclic
HP-completion sets containing at most two edges.

\begin{lemma}\label{lemma:3to1}
Let  $R=(V^l \cup V^r \cup \{s,t\}, E)~$ be an $st$-polygon. Let $P$
be an acyclic HP-completion set for $R$ such that $|P|=2\mu+1,
~\mu\geq 1$. Then, there exists another acyclic HP-completion set
$P^\prime$ for $R$ such that $|P^\prime|=1$ and the edges of
$P^\prime$ create at most as many crossings with the edges of $R$ as
the edges of $P$ do. In addition, the hamiltonian paths induced by
$P$ and $P^\prime$ have in common their first and last edges.
\end{lemma}

\begin{figure}[htb]
    \centering
    \includegraphics[width=0.6\textwidth]{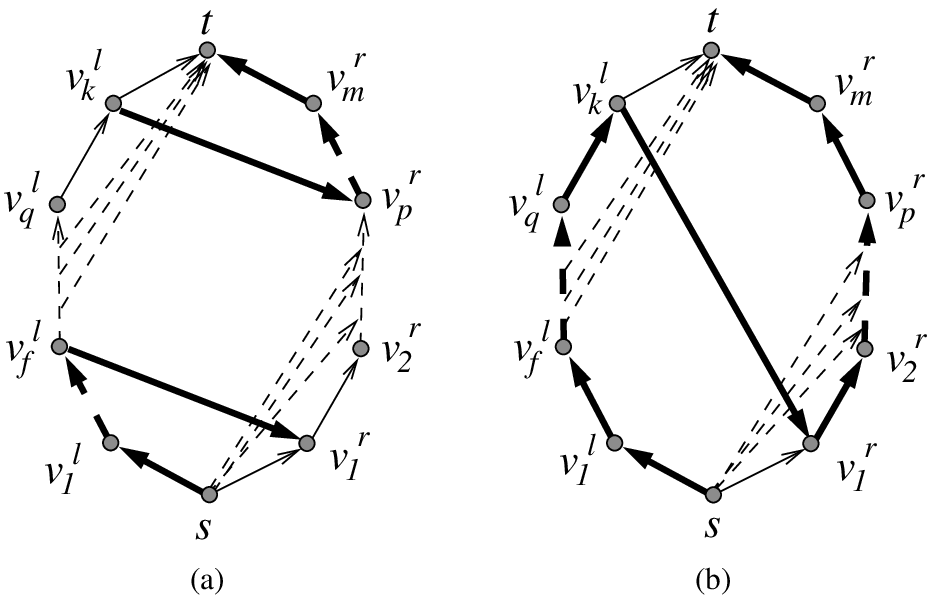}
    \caption{An acyclic HP-completion set of odd size for an $st$-polygon and an equivalent acyclic HP-completion set
    of size 1. }
    \label{fig:3to1}
\end{figure}

\begin{proof}
First observe that, as a consequence of Lemma~\ref{lem:sideInOrder},
any acyclic HP-completion set for $R$ does not contain any one-sided
edge.  Thus, all $2\mu +1$ edges of $P$ are two-sided edges.
Moreover, since  $P$ contains an odd number of edges, both  the
first and the last edge of $P$ have the same direction. Without loss
of generality, let the lowermost edge of $P$   be directed from left
to right (see Figure~\ref{fig:3to1}(a)). By
Lemma~\ref{lem:sideInOrder}, it follows that the destination of the
lowermost edge of $P$ is the lowermost vertex on the right side of
$R$ (i.e., vertex $v_1^r$) while the origin of the topmost edge of
$P$ is the topmost vertex of the left side of $R$ (i.e., vertex
$v_k^l$).

Observe   that $P^\prime = \left\{ (v_k^l,v_1^r)  \right\}$ is an
acyclic HP-completion set for $R$. The induced hamiltonian path is
$(s \dashrightarrow v_k^l \rightarrow v_1^r \dashrightarrow
t)$\footnote{A dashed-arrow "$\dashrightarrow$" indicates a path
that is on the left or the right side of an $st$-polygon (or
outerplanar graph) and might contain intermediate vertices.}.

In order to complete the proof, we show that edge $(v_k^l,v_1^r)$
does not cross more edges of $R$ than the edges of $P$ do. To see
that, observe that edge $(v_k^l,v_1^r)$ crosses all edges in set
$\{(s,v): v \in V^r \setminus \{v_1^r\} \}$ as well as  all edges in
set $\{(v,t): v \in V^l \setminus \{v_m^l\} \}$, provided they exist
(see Figure ~\ref{fig:3to1}(b)). However, the edges in these two
sets are also crossed by the lowermost and the topmost edges of $P$,
respectively. Thus, edge $(v_k^l,v_1^r)$ creates at most as many
crossings with the edges of $R$ as the edges of $P$ do. Observe also
that the hamiltonian paths induced by $P$ and $P^\prime$ have in
common their first and last edges. \qed
\end{proof}

\begin{lemma}\label{lemma:4to2}
Let  $R=(V^l \cup V^r \cup \{s,t\}, E)~$ be an $st$-polygon. Let $P$
be an acyclic HP-completion set for $R$ such that $|P|=2\mu, ~\mu
\geq 1$. Then, there exists another acyclic HP-completion set
$P^\prime$ for $R$ such that $|P^\prime|=2$ and the edges of
$P^\prime$ create at most as many crossings with the edges of $R$ as
the edges of $P$ do. In addition, the hamiltonian paths induced by
$P$ and $P^\prime$ have in common their first and last edges.
\end{lemma}

\begin{figure}[htb]
    \centering
    \includegraphics[width=0.7\textwidth]{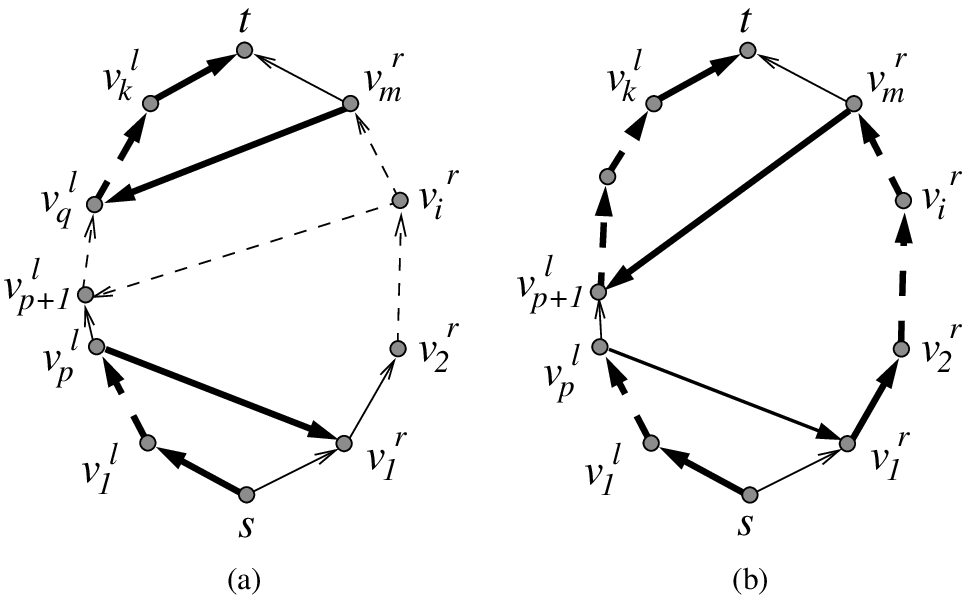}
    \caption{An acyclic HP-completion set of even size for an $st$-polygon and an equivalent acyclic HP-completion set
    of size 2. }
    \label{fig:4to2}
\end{figure}

\begin{proof}
As in the case of an HP-completion set of odd size
(Lemma~\ref{lemma:3to1}), the  $2\mu$ edges of $P$ are two-sided
edges. Moreover, since  $P$ contains an even number of edges, the
first and the last edge of $P$ have opposite direction. Without loss
of generality, let the lowermost edge of $P$ be directed from left
to right (see Figure~\ref{fig:4to2}(a)). By
Lemma~\ref{lem:sideInOrder}, it follows that the destination of the
lowermost edge of $P$ is the lowermost vertex on the right side of
$R$ (i.e., vertex $v_1^r$) while the origin of the topmost edge of
$P$ is the topmost vertex of the right side of $R$ (i.e., vertex
$v_m^r$). Let the lowermost edge of $P$ be $(v_p^l, v_1^r)$. Then,
from Lemma~\ref{lemma:3to1} it follows that the HP-completion set
$P$ also contains  edge $(v_i^r, v_{p+1}^l)$ for some $1<i\leq m$.
If $i=m$, then $P$ contains exactly 2 edges and lemma is trivially
true. So, we consider the case where $i < m$.

Observe   that, for the case where $|P|>3$, the set of edges
$P^\prime = \left\{ (v_p^l,v_1^r),~ (v_m^r, v_{p+1}^l) \right\}$ is
an acyclic HP-completion set for $R$. The induced hamiltonian path
is $(s \dashrightarrow v_p^l \rightarrow v_1^r \dashrightarrow v_m^r
\rightarrow v_{p+1}^l \dashrightarrow t)$.

In order to complete the proof, we show that edges $(v_p^l,v_1^r)$
and $(v_m^r, v_{p+1}^l)$ does not cross more edges of $R$ than the
edges of $P$ do. The edges of $E$ that are crossed by the two edges
of $P\prime$ can be classified in the following disjoint groups.
\vspace*{-0.3cm}
\begin{enumerate}[a)]
\item
\textit{Edges having their origin below edge $(v_p^l,v_1^r)$ and
their destination above edge $(v_m^r, v_{p+1}^l)$.} All of these
edges are crossed by both edges in $P^\prime$. But, they are also
crossed by at least  edges $(v_p^l,v_1^r)$ and $(v_i^r, v_{p+1}^l)$
of $P$.

\item
\textit{Edges having their origin below edge $(v_p^l,v_1^r)$ and
their destination between  edges  $(v_p^l,v_1^r)$ and $(v_m^r,
v_{p+1}^l)$.} All of these edges are crossed by only edge
$(v_p^l,v_1^r)$ in $P^\prime$. But,  $(v_p^l,v_1^r)$ also belongs in
$P$.
\item
\textit{Edges having their origin between  edges $(v_p^l,v_1^r)$ and
$(v_m^r, v_{p+1}^l)$ and  their destination above edge $(v_m^r,
v_{p+1}^l)$.} All of these edges are crossed by  only edge $(v_m^r,
v_{p+1}^l)$ of $P^\prime$. But, they are also crossed by at least
the topmost edge $(v_m^r, v_{q}^l)$ of $P$.

\end{enumerate}
Thus, the edges in $P^\prime$  create at most as many crossings with
the edges of $R$ as the edges of $P$ do. Observe also that the
hamiltonian paths induced by $P$ and $P^\prime$ have in common their
first and last edges. \qed
\end{proof}

The following theorem follows directly from Lemma~\ref{lemma:3to1}
and Lemma~\ref{lemma:4to2}.

\begin{theorem}
\label{thm:st-polygon-size-2} Any $st$-polygon has a crossing
optimal acyclic HP-completion set of size at most 2. \qed
\end{theorem}

\subsection{$st$-polygon decomposition of an outerplanar $st$-digraph}
\label{sec:st-polygon-decomp}

\begin{lemma}
\label{lem:medianComputation} Assume an outerplanar $st$-digraph
$G=(V^l \cup V^r \cup \{s,t\}, E)~$ and an arbitrary edge
$e=(s^\prime, t^\prime) \in E$. If $O(V)$ time is available for the
preprocessing of $G$,  we can decide in $O(1)$ time whether $e$ is a
median edge of some strong $st$-polygon. Moreover, the two vertices
(in addition to $s^\prime$ and $t^\prime$) that define
 a maximal strong $st$-polygon that has edge $e$ as its median can be also computed in $O(1)$ time.
\end{lemma}

\begin{proof}
We can preprocess graph $G$ in linear time so that for each of its
vertices  we know the first and last (in clock-wise order) in-coming
and out-going edges.

\begin{figure}[htb]
    \centering
    \includegraphics[width=0.7\textwidth]{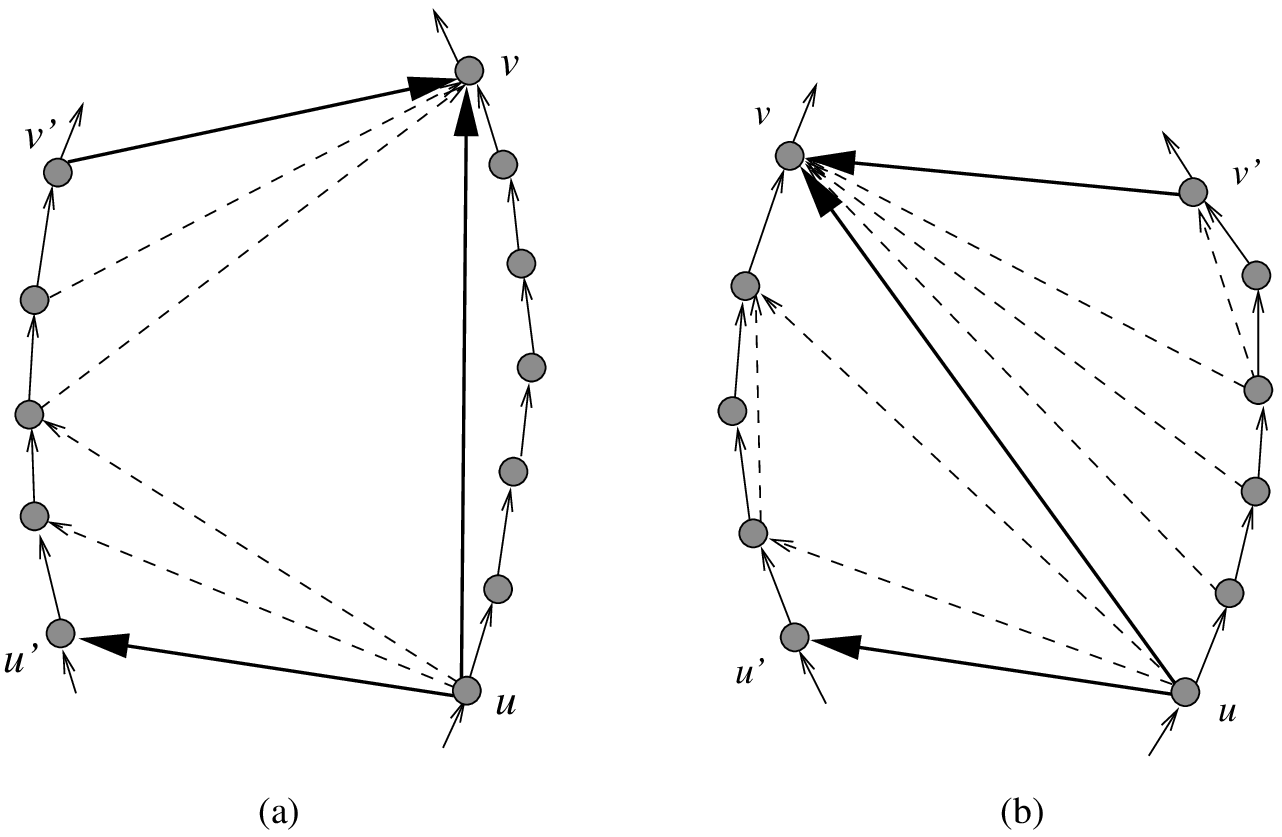}
    \caption{The  two edges that bound the $st$-polygon and its median are shown in bold.}
    \label{fig:detect-st-polygon}
\end{figure}

Observe that an one-sided edge $(u,v)$ is a median of an strong
$st$-polygon if the following hold (see
Figure~\ref{fig:detect-st-polygon}.a): \vspace*{-0.2cm}
\begin{quote}
\begin{enumerate}[a)]
\item $u$ and $v$ are not successive vertices of the side of $G$.
\item $u$ has a two-sided outgoing edge.
\item $v$ has a two-sided incoming edge.
\end{enumerate}
\end{quote}

Similarly, observe that a two-sided edge $(u,v)$ with $u \in V^R$
(resp.  $u \in V^L$) is a median of a strong $st$-polygon if the
following hold (see Figure~\ref{fig:detect-st-polygon}.b):
\vspace*{-0.2cm}
\begin{quote}
\begin{enumerate}[a)]
\item $u$ has a two-sided outgoing edge that is clock-wise before (resp. after)
$(u,v)$.
\item $v$ has a two-sided incoming edge that is clock-wise before (resp. after) $(u,v)$.
\end{enumerate}
\end{quote}

All of the above conditions can be trivially tested in $O(1)$ time.
Then, the two remaining vertices that define the maximal strong
$st$-polygon having $(u,v)$ as its median can be found in $O(1)$
time and, moreover, the strong $st$-polygon can be reported in time
proportional to its size. \qed
\end{proof}

\begin{lemma}
\label{lem:rhombusComputation} Assume an outerplanar $st$-digraph
$G=(V^l \cup V^r \cup \{s,t\}, E)~$ and a face $f$ with source $u$
and sink $v$. If $O(V)$ time is available for the preprocessing of
$G$, we can decide in $O(1)$ time whether $f$ is a weak rhombus.
Moreover, the two vertices (in addition to $u$ and $v$) that define
a maximal weak $st$-polygon that contains $f$ can be also computed
in $O(1)$ time.
\end{lemma}

\begin{proof}

By definition, a weak rhombus is a face that has at least one vertex
on each of its sides. Thus, we can test whether face $f$ is a weak
rhombus in $O(1)$ time, if for each face we have available the lists
of vertices on its left and right sides.

As it was noted in the pervious proof, we can preprocess graph $G$
in linear time so that for each of its vertices  we know its first
and last (in clock-wise order) in-coming and out-going edges. Then,
the two remaining vertices that define the maximal weak $st$-polygon
having $f$ as a subgraph can be found in $O(1)$ time and it can be
reported in time proportional to its size.  For example,  in
Figure~\ref{fig:detect-weak-rhombus} where vertices $u$ and $v$ are
both on the right side, the limiting edges of the maximal weak
$st$-polygon are the first outgoing edge from $u$ and the last
incoming edge to $v$.\qed
\begin{figure}[htb]
    \centering
    \includegraphics[width=0.3\textwidth]{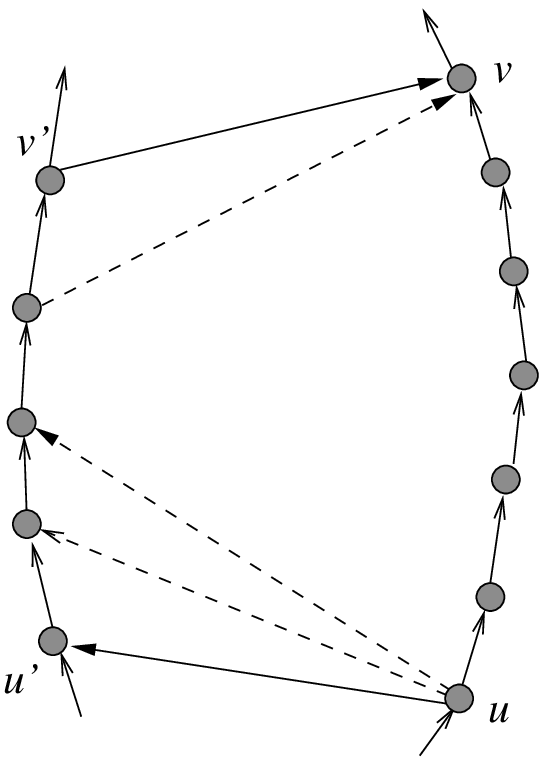}
    \caption{The weak rhombus with $u$ and $v$ as its source and sink, respectively,
    and the maximal $st$-polygon containing it.}
    \label{fig:detect-weak-rhombus}
\end{figure}

\end{proof}

Observe also that,  as we extend a weak (resp. strong)  rhombus to
finally obtain the maximal weak (resp. strong) $st$-polygon that
contains it, we include all edges that are outgoing from $u$ and
incoming to $v$. During this procedure,  all faces  attached to the
rhombus are generalized triangles.

\begin{lemma}
\label{lem:areaDisjointPolygons}
 The maximal $st$-polygons contained in an
outerplanar $st$-digraph $G$ are mutually area-disjoint.
\end{lemma}

\begin{proof}

We first observe that a maximal $st$-polygon can not fully contain
another. If it does, then we would have a maximal $st$-polygon
containing two rhombuses, which is impossible due to
Lemma~\ref{lemma:oneRombus}.

For the shake of contradiction, assume two  $st$-polygons $P_1$ and
$P_2$ that have a partial overlap. We denote by $(s_1,u_1^l,\dots,
u_k^l, u_1^r, \dots, u_m^r,t_1)$ and $(s_2,v_1^l,\dots, v_k^l,
v_1^r, \dots, v_m^r,t_2)$ the vertices of $P_1$ and $P_2$
respectively. Throughout   the proof we refer to
Figure~\ref{fig:areadisjoint}.

Due to the assumed partial overlap of  $P_1$ and $P_2$,  an edge of
one of them, say $P_1$, must be contained within the other (say
$P_2$). Below we show that none of the two possible upper limiting
edges $(u_k^l,s_1)$ and $(u_m^r,s_1)$ of $P_1$ can  be contained in
$P_2$.

\begin{figure}[htb]
    \centering
    \includegraphics[width=0.4\textwidth]{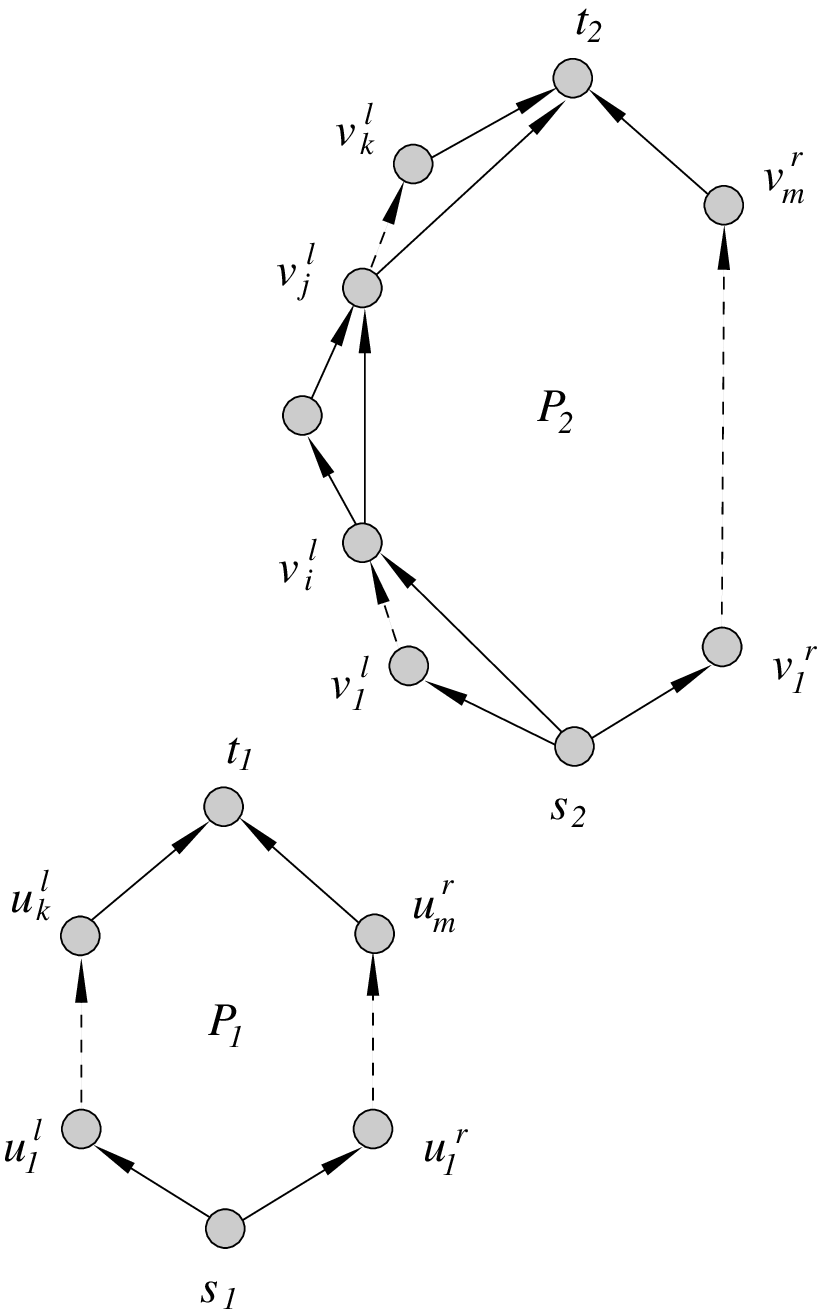}
    \caption{Two $st$-polygons from the proof of Lemma~\ref{lem:areaDisjointPolygons}}
    \label{fig:areadisjoint}
\end{figure}

We have to consider three cases.
\begin{description}
\item[\textit{Case 1:}] \textit{One of the edges $(u_k^l,t_1)$ and $(u_m^r,t_1)$ of
$P_1$ coincide with the internal edge of $P_2$ connecting $s_2$ with
a vertex $v^l_i$ on its left side (the case where it is on its right
side is  symmetrical).}  Edge $(u_k^l,t_1)$ can not coincide with
$(s_2,v_i^l)$, since then,  edge $(u_m^r,t_1)$ has to be inside
$P_2$ and therefore to connect the left side of $P_2$ with its right
side. This is  a   contradiction since $P_2$ is an $st$-polygon and
it can not contain any such edge.

Now assume that  edge $(u_m^r,t_1)$ of $P_1$ coincides with  edge
$(s_2,v_i^l)$ of $P_2$, Then, edge $(s_2,v_1^l)$ is inside $P_1$ and
joins its right  with its left side, which is again impossible as we
supposed $P_1$ to be an  $st$-polygon.

\item[\textit{Case 2:}] \textit{ One of edges $(u_k^l,t_1)$ and $(u_m^r,t_1)$ of
$P_1$ coincides with the internal edge of $P_2$ connecting two
vertices on its same side.} Let it again be the left side and denote
the edge by $(v_i^l,v_j^l)$. Assume first that $(u_m^r,t_1)$
coincides with $(v_i^l,v_j^l)$. As  graph $P_2$ is outerplanar,  we
have that all the remaining vertices of $P_1$ have to be placed
above  vertex $v_i^l$ and below  vertex $v_j^l$  on the left side of
$P_2$. Therefore $P_1$ is fully contained in $P_2$, which is
impossible.

Assume now that $(u_k^l,t_1)$ coincides with  edge $(v_i^l,v_j^l)$.
Then,  edge $(u_m^r,t_1)$ of $P_1$ coincides with  edge
$(v_{i^\prime}^l,v_j^l)$ of $P_2$, $i^\prime<i$. This is impossible
as it was covered in the above paragraph. Note also that
$v_{i^\prime}^l$ cannot be $s_2$ since this configuration was shown
to be impossible in Case~1.

\item[\textit{Case 3:}] \textit{One of edges $(u_k^l,t_1)$ and $(u_m^r,t_1)$ of
$P_1$ coincide with the internal edge of $P_2$ connecting the vertex
on its side (suppose again on its left side) with the sink $t_2$.}
Let this edge be denoted by $(v_j^l,t_2)$. Suppose first that
$(u_k^l,t_1)$ coincides with $(v_j^l,t_2)$. If vertex $u_m^r$ is on
the right side of $P_2$ then $P_1$ is not maximal as $P_1$ can be
extended (and still remain an $st$-polygon) by including vertices
$v^l_j$ to $v_k^l$. So, assume that $u_m^r$ is on the left side of
$P_2$. Then, as covered in  Case~2, $P_1$ must be fully contained in
$P_2$ which leads to a contradiction.

Assume now that edge $(u_m^r,t_1)$ coinsides with edge
$(v_j^l,t_2)$. Due to outer-planarity of $P_2$, we have again that
all the vertices of $P_1$ have to be placed above $v_j^l$ and below
$t_2$ on the left side of $P_2$. So $P_1$ is again fully contained
in $P_2$, leading again to a contradiction.
\end{description}

We have managed to show that none of  edges $(u_k^l,s_1)$ and
$(u_m^r,s_1)$ is contained in $P_2$. Therefore,  there can be no
partial overlap between $P_1$ and $P_2$.\qed
\end{proof}

\begin{figure}[htb]
    \centering
    \includegraphics[width=1\textwidth]{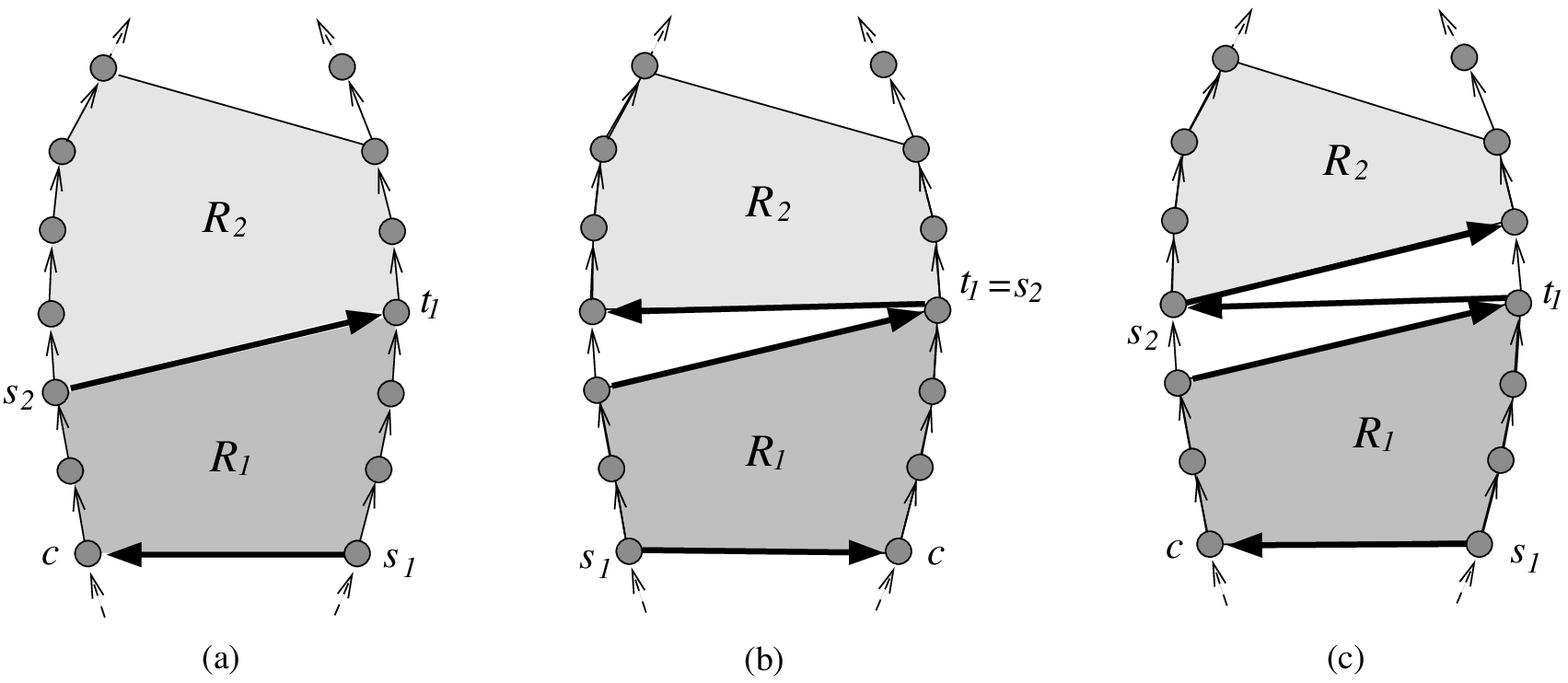}
    \caption{Configurations of adjacent $st$-polygons of an outerplanar $st$-digraph.}
    \label{fig:succesivePolygons}
\end{figure}

Denote by $\mathcal{R}(G)$ the set of all maximal $st$-polygons  of
an outerplanar $st$-digraph $G$. Observe than not every vertex of
$G$ belongs to one of its maximal $st$-polygons. We refer to  the
vertices of $G$ that are not part of any maximal $st$-polygon as
\emph{free vertices} and we denote them by $\mathcal{F}(G)$.
 Also observe that
an ordering can be imposed on the maximal $st$-polygons of an
outerplanar $st$-digraph $G$ based on the ordering of the area
disjoint strips occupied by each $st$-polygon. The vertices which do
not belong to some $st$-polygon are located in the area between the
strips occupied by consecutive $st$-polygons which. \eat{, as we
will see shortly, is triangulated (see
Figure~\ref{fig:succesivePolygons}.a).}

\begin{lemma}
\label{lem:freevertices} Let $R_1$ and $R_2$ be two consecutive
maximal $st$-polygons of an outerplanar $st$-digraph $G$ which do
not share an egde and let $V_f \subseteq \mathcal{F}(G)$ be the set
of free vertices lying between $R_1$ and $R_2$. Denote by $(u,t_1)$
and $(s_2,v)$ the upper limiting edge and the lower limiting edge of
$R_1$  and $R_2$, respectively. For the embedded subgraph  $G_f$ of
$G$ induced by the vertices of $V_f \cup \{u,t_1,s_2,v\}$ it holds:
\vspace*{-.3cm}
\begin{quote}
\begin{enumerate}[a)]
\item
$G_f$ is an outerplanar $st$-digraph having   vertices $u$ and $v$
as its source and sink, respectively.
\item  $G_f$ is hamiltonian.
\end{enumerate}
\end{quote}
\end{lemma}

\begin{proof}
We first show that statement (a) is true, that is,  $G_f$ is an
outerplanar $st$-digraph having vertices $u$ and $v$ as its source
and sink, respectively. Without loss of generality, assume that the
limiting edge $(s_2,v)$ of the upper maximal $st$-polygon $R_2$ is
directed towards the right side of the outerplanar $st$-digraph $G$.
We consider cases based on whether $R_1$ and $R_2$ share a  common
vertex.
\begin{figure}[htb]
    \centering
    \includegraphics[width=1\textwidth]{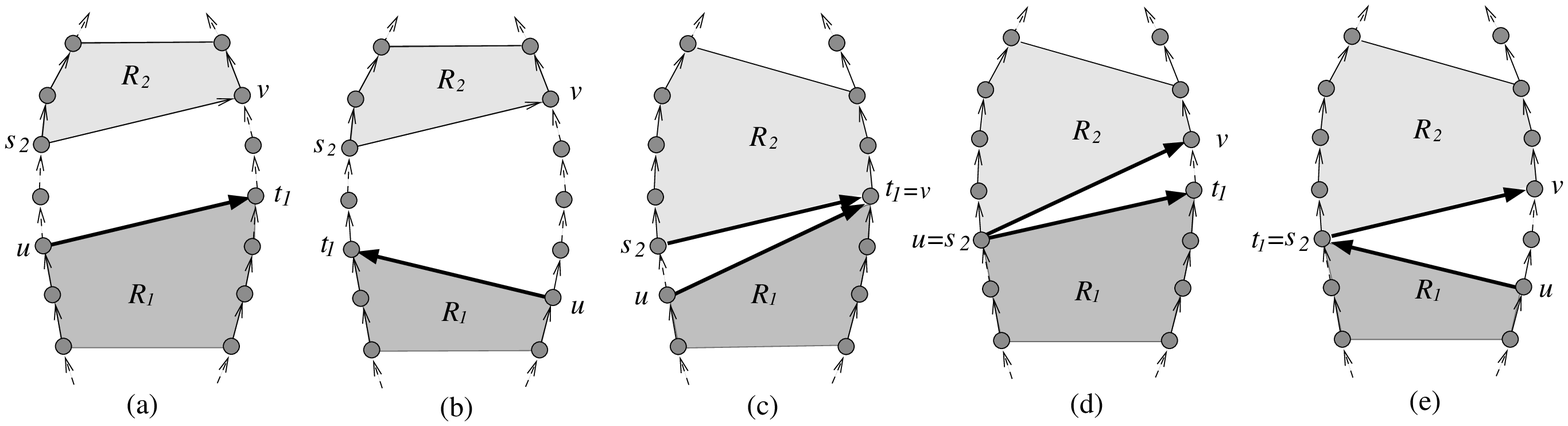}
    \caption{The $st$-polygon  with vertices $a ~\mbox{(source)},~b, ~c, ~d
~\mbox{(sink)},~ e, ~f, ~\mbox{and}~y$ on its boundary is maximal.}
    \label{fig:between-st-polygons}
\end{figure}

\begin{description}
\item[\textit{Case 1: $R_1$ and $R_2$ share no  common
vertex.}] Based on the direction of the limiting edge $(u,t_1)$ we
can further distinguish the following two cases:

\begin{description}
\item[\textit{Case 1a: $(u,t_1)$ is  directed towards the rigth side of $G$.}]
See Figure~\ref{fig:between-st-polygons}.a.  Observe that there are
two path from $u$ to $v$, one going through all vertices of $V_f$ on
the left side of $G$ to $s_2$ and then to $v$, and another to $t_1$
and then to $v$ passing through all vertices of $V_f$ on the right
side of $G$. Thus, for every vertex $w$ in $V_f$ (either on the left
or the rigth side of $G$) there is a path from $u$ to $w$ and a path
from $w$ to $v$, and thus, the embedded subgraph $G_f$ of $G$
induced by the vertices of $V_f \cup \{u,t_1,s_2,v\}$ is an
outerplanar $st$-digraph having vertices $u$ and $v$ as its source
and sink, respectively. Note that this holds even for the case where
one (or both) sides of $G$ contribute no vertices to $V_f$.
\item[\textit{Case 1b: $(u,t_1)$ is  directed towards the left side of $G$.}]
See Figure~\ref{fig:between-st-polygons}.b. Just observe that there
are again two path from $u$ to $v$, one going through all vertices
of $V_f$ on the right side of $G$, and another to $t_1$, then
passing through all vertices of $V_f$ on the left side of $G$ to
$s_2$ and then to $v$. The rest of the proof is identical to that of
Case~1a.
\end{description}

\item[\textit{Case 2: $R_1$ and $R_2$ share one  common
vertex.}] First observe that the limiting edge $(u,t_1)$ of $R_1$ is
directed towards the left side of $G$. To see that, assume for the
shake of contradiction that edge $(u,t_1)$ is directed towards the
right side of $G$. If $v$ coincides with $t_1$ (see
Figure~\ref{fig:between-st-polygons}.c) then the $st$-polygon $R_1$
could be extended (and still remain an $st$-polygon) by adding to it
the area between the two polygons $R_1$ and $R_2$, contradicting the
fact that $R_1$ is maximal. If $u$ coincides with $s_2$ (see
Figure~\ref{fig:between-st-polygons}.d) then the $st$-polygon $R_2$
could be extended (and still remain an $st$-polygon) by adding to it
the area between the two polygons $R_1$ and $R_2$, contradicting the
fact that $R_2$ is maximal.

Thus, the limiting edge $(u,t_1)$ of $R_1$ is directed towards the
left side of $G$ and $s_2$ coinsides with $t_1$ (see
Figure~\ref{fig:between-st-polygons}.e). The rest of the proof is a
special case of Case~1b (where no vertices of $V_f$ exist on the
left side of $G$).
\end{description}

So $G_f$ is an outerplanar $st$-digraph with the source $u$ and the
sink $v$.  Note also that $G_f$ does not contain a rhombus. If it
does, then it would be an $st$-polygon, contradicting the fact that
$R_1$ and $R_2$ are consecutive maximal $st$-polygons. Then, from
Theorem~\ref{thm:rhombus} it follows that $G_f$ is hamiltonian.\qed
\end{proof}

\begin{lemma}
\label{lem:commonedge} Let $R_1$ and $R_2$ be two consecutive
maximal $st$-polygons of an outerplanar $st$-digraph $G$ that share
a common edge. Let $t_1$ be the sink of $R_1$ and $s_2$ be the
source of $R_2$. Then,  edge $(s_{2},t_1)$ is their common edge.
\end{lemma}

\begin{proof}
Let the upper limiting edge of $R_1$ be edge $(u,t_1)$ and the lower
limiting edge of $R_2$ be edge $(s2,v)$. Since these are the only
two edges that can coincide, we conclude that $v$ coincides with
$t_1$ and $u$ coincides with $s_2$. Thus,  edge $(s_{2},t_1)$ is the
 edge shared by  $R_1$ and $R_2$.\qed
\end{proof}

\begin{lemma}
\label{lem:pahtProperties}
 Assume an outerplanar $st$-digraph $G$. Let
$R_1$ and $R_2$ be two of $G$'s consecutive maximal $st$-polygons
and let $V_f \subset \mathcal{F}(G)$ be the set of free vertices
lying between $R_1$ and $R_2$. Then, the following statements are
satisfied: \vspace*{-.3cm}
\begin{quote}
\begin{enumerate}[a)]
\item For any pair of vertices $u,~v\in V_f$ there is either a path
from $u$ to $v$ or from $v$ to $u$.
\item For any vertex $v\in V_f$ there is a path from the sink of $R_1$
to $v$ and from $v$ to the source of $R_2$.
\item If $V_f=\emptyset$, then there is a path from source of $R_1$
to the source of $R_2$.
\end{enumerate}
\end{quote}
\end{lemma}

\begin{proof}
$~$\\\vspace*{-0.8cm}

\begin{enumerate}[a)]
\item

From Lemma~\ref{lem:freevertices} we have that the subgraph $G_f$ of
$G$ is hamiltonian, so we have that all their vertices are connected
by a directed path.

\item

Follows directly from Lemma~\ref{lem:freevertices}.

\item Note that there are 3 configuration  in which no free vertex exists between two consecutive
$st$-polygons (see Figures~\ref{fig:succesivePolygons}.b-d). Denote
by $s_1$ and $s_2$ the sources of $R_1$ and $R_2$, respectively. If
$s_1$ and $s_2$ lie on the same side of $G$ then the claim is
obviously true since $G$ is an OT-$st$-digraph. If they belong to
opposite sides of $G$, observe that the lower limiting edge
$(s_1,c)$ of $R_1$ leads to the side of $G$ which contains $s_2$.
Since there is a path from $c$ to $s_2$, it follows that there is a
path from $s_1$ to $s_2$.\qed
\end{enumerate}
\end{proof}

We refer to the source vertex $s_i$ of each maximal $st$-polygon
$R_i \in \mathcal{R}(G),~ 1 \leq i \leq |\mathcal{R}(G)|$ as the
\emph{representative} of $R_i$ and we denote it by $r(R_i)$. We also
define the representative of a free vertex $v\in \mathcal{F}(G)$ to
be $v$ itself, i.e. $r(v)=v$. For any two distinct elements $x, ~y~
\in \mathcal{R}(G) \cup \mathcal{F}(G)$, we define the  relation
$\angle_p$ as follows: \emph{$x \angle_p y$ iff there exists a path
from}  $r(x)$ \emph{to} $r(y)$.

\begin{lemma}
\label{lem:totalOrder}
 Let $G$ be an $n$ node outerplanar $st$-digraph. Then,
relation $\angle_p$ defines a total order on the elements
$\mathcal{R}(G) \cup \mathcal{F}(G)$. Moreover, this total order can
be computed in $O(n)$ time.
\end{lemma}

\begin{proof}
The fact that $\angle_p$ is a total order on $\mathcal{R}(G) \cup
\mathcal{F}(G)$
 follows from Lemma~\ref{lem:pahtProperties}. The  order of
 the element of $\mathcal{R}(G) \cup
\mathcal{F}(G)$ can be easily derived by the
 numbers assigned to the representatives of the elements (i.e., to
 vertices of $G$) by a topological sorting of the vertices of $G$.
 To complete the proof, recall that an $n$ node acyclic planar graph can be
 topologically sorted in $O(n)$ time.
\qed
\end{proof}

\begin{definition}
Given an outerplanar $st$-digraph $G$,  the \emph{$st$-polygon
decomposition} $\mathcal{D}(G)$ of $G$ is defined to be the total
order of its maximal $st$-polygons and its free vertices induced by
relation $\angle_p$.
\end{definition}

The following theorem follows directly from
Lemma~\ref{lem:medianComputation},
Lemma~\ref{lem:rhombusComputation} and Lemma~\ref{lem:totalOrder}.

\begin{theorem}
\label{thm:STpolygonDecomposition} An $st$-polygon decomposition of
an $n$ node OT-$st$-digraph $G$ can be computed in $O(n)$ time.
\end{theorem}

\subsection{Properties of a crossing-optimal acyclic HP-completion set}
\label{sec:HP-completion-set-properties}

In this section, we present  some properties of a crossing optimal
acyclic HP-completion for an outerplanar $st$-digraph that will be
taken into account in our algorithm. Assume an outerplanar
$st$-digraph $G=(V^l \cup V^r \cup \{s,t\}, E)~$ and its
$st$-polygon decomposition $\mathcal{D}(G)= \{ o_1,~ \ldots,
o_\lambda \}$.  By  $G_i$  we denote the graph induced by the
vertices of elements $o_1, \ldots, o_i,~i\leq \lambda$.

\begin{prop}
\label{prop:3crossings} Let $G=(V^l \cup V^r \cup \{s,t\}, E)~$  be
an outerplanar $st$-digraph. Then, no edge of $E$ is crossed by more
than 2 edges of a crossing-optimal acyclic HP-completion set for
$G$.
\end{prop}

\begin{figure}[htb]
  \begin{minipage}{\textwidth}
    \centering
    \includegraphics[width=\textwidth]{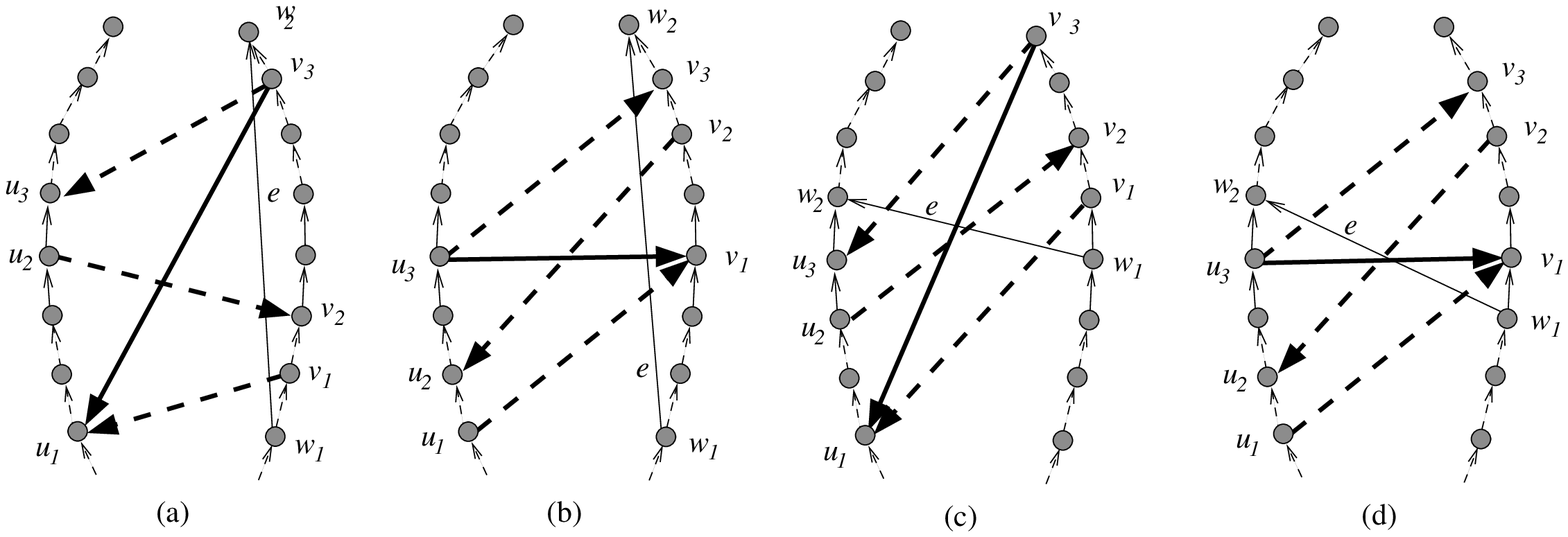}
    \caption{Configurations of crossing edges used in the proof of   Property~\ref{prop:3crossings}.}
    \label{fig:3crossings}
  \end{minipage}
\end{figure}

\begin{proof}
For the shake of contradiction, assume that  $P_{\mathrm{opt}}$ is a
crossing-optimal acyclic HP-completion set for  $G$, the edges of
which cross some  edge $e=(w_1,w_2)$ of $G$ three times. We will
show that we can obtain an acyclic HP-completion set for $G$ that
induces a smaller number of crossings that $P_{\mathrm{opt}}$, a
clear contradiction. We assume that all edges of $P_{\mathrm{opt}}$
participate  in the hamiltonian path of $G$; otherwise they can be
discarded.

We distinguish two cased based on whether edge $e$ is a one-sided or
a two-sided edge.

\begin{description}
 \item[\textit{Case 1:} \emph{The edge $e$ is one-sided.}]
Suppose without lost of generality that $e$ is on the right side. We
further distinguish two cases based on the orientation of   the edge,
say $e_1$, which appears first on the hamiltonian path of $G$ (out
of the 3 edges crossing edge $e$).

\begin{description}
\item[\textit{Case 1a: Edge $e_1$ is directed from  right to left.}]

Let $e_1$ be edge $(v_1,u_1)$ and let $(u_2,v_2)$ and $(v_3,u_3)$ be
the next two edges on the hamiltonian path which cross $e$ (see
Figure~\ref{fig:3crossings}.a). It is clear that these three edges
have alternating direction. Observe  that the path $P_{v_1, u_3}=
(v_1 \rightarrow u_1 \dashrightarrow u_2 \rightarrow v_2
\dashrightarrow v_3 \rightarrow u_3)$ is a sub-path of the
hamiltonian path of $G$. Also,  by Lemma~\ref{lem:sideInOrder},
vertex $u_2$ is immediately below vertex $u_3$ on the left side of
$G$  and vertex $v_2$ is immediately above vertex $v_1$ on the right
side of $G$.

Now, we show that the substitution of path $P_{v_1, u_3}$ of the
hamiltonian path of $G$ by  path $P^\prime_{v_1, u_3}= (v_1
\rightarrow v_2 \dashrightarrow v_3 \rightarrow u_1 \dashrightarrow
u_2 \rightarrow u_3)$  results to a reduction of the total number of
crossings by at least 2. Thus, there exists an HP-completion set
that crosses edge $e$ only once and causes 2 less crossings with
edges of $G$ compared to $P_{\mathrm{opt}}$, a clear contradiction.

Let us examine the edges of $G$ that are crossed by the new edge
$(v_3,u_1)$. These edge can be grouped as follows: (i) The one-sided
edges on the right side of $G$ that have their source below $v_3$
and their sink above $v_3$. Note that these edges are also crossed
by edge $(v_3,u_3)$. In addition, the edges that belong in this
group and have their origin below $w_1$ and their sink above $w_1$
are crossed by all three  edges $(v_1,u_1)$,  $(u_2,v_2)$ and
$(v_3,u_3)$ in the original HP-completion set. The edges on the
right side of $G$ that have their source below $v_3$ and their sink
above $v_3$. (ii)The two sided edges that have their source  below
$w_1$ on the right side of $G$  and their sink above $v_1$ on the
left side of $G$. These edges are also crossed by at least  edge
$(v_1,u_1)$ (and possibly by one or both of edges $(u_2,v_2)$ and
$(v_3,u_3)$). (iii)The one-sided edges on the left side of $G$ that
have their source below $u_1$ and their sink above $u_1$. Note that
these edges are also crossed by at least  edge $(v_1,u_1)$ (and
possibly by one or both of edges $(u_2,v_2)$ and $(v_3,u_3)$). (iv)
The two sided edges that have their source below $u_1$  on the left
side of $G$ and their sink above $w_2$ on the right side of $G$.
These edges are also crossed by all three  edges  $(v_1,u_1)$
$(u_2,v_2)$ and $(v_3,u_3)$). Thus, we have shown that edge
$(v_3,u_1)$ crosses at most as many edges of $G$ as the  three edges
$(v_1,u_1)$, $(u_2,v_2)$, $(v_3,u_3)$ taken together.

\item[\textit{Case 1b: Edge $e_1$ is directed from  left to right.}]
Let $e_1$ be edge $(u_1,v_1)$ and let $(v_2,u_2)$ and $(u_3,v_3)$ be
the next two edges on the hamiltonian path which cross $e$ (see
Figure~\ref{fig:3crossings}.b).  Observe  that the path $P_{u_1,
v_3}= (u_1 \rightarrow v_1 \dashrightarrow v_2 \rightarrow u_2
\dashrightarrow u_3 \rightarrow v_3)$ is a sub-path of the
hamiltonian path of $G$. Also,  by Lemma~\ref{lem:sideInOrder},
vertex $u_2$ is immediately above vertex $u_1$ on the left side of
$G$ and vertex $v_2$ is immediately below vertex $v_3$ on the right
side of $G$. By arguing in  a way similar to that of Case~1a, we can
show that the substitution of path $P_{u_1, v_3}$ of the hamiltonian
path of $G$ by  path $P^\prime_{u_1, v_3}= (u_1 \rightarrow u_2
\dashrightarrow u_3 \rightarrow v_1 \dashrightarrow v_2 \rightarrow
v_3)$  results to a reduction of the total number of crossings by at
least 2.

\eat{ The first edge $(u_1,v_1)$ is directed from the left to the
right. We take the two above consecutive edges $(v_2,u_2)$ and
$(u_3,v_3)$(see Figure~\ref{fig:3crossings}.b).  Now we will show
that these three edges can be substituted by the edge $(u_3,v_1)$
decreasing the number of crossings. Let us see the edges that can be
crossed by the new edge $(u_3,v_1)$. They are: $(a)$ The edges on
the right side that has a source below $v_1$ and sink above $v_1$,
but they are also crossed by the edge $(u_1,v_1)$. $(b)$ The edges
from $w_1$ and below to the left side - above $u_3$. These edges are
also crossed by all of three edges. $(c)$ The edges on the left side
that have a source below $u_3$ and sink above $u_3$. But these edges
are crossed at least by $(u_3,v_3)$. $(d)$ The edges with the source
below $u_3$ and sink in $w_2$ and above. But they are also crossed
by at least by $(u_3,v_3)$. So, summing the cases above, we have
that the edge $(u_3,v_1)$ has less number of crossings than
$(u_1,v_1)$, $(v_2,u_2)$, $(u_3, v_3)$ token together. }

\end{description}

\item[\textit{Case 2:} \emph{The edge $e$ is two-sided.}]
Suppose without lost of generality that $e$ is directed from  right
to left. We again distinguish two cases based on the orientation of
the edge, say $e_1$, which appears first on the hamiltonian path of
$G$ (out of the 3 edges crossing edge $e$).

\begin{description}
\item[\textit{Case 2a: Edge $e_1$ is directed from  right to left.}]
The proof is identical to that of Case~1a.

\eat{ We take the two above consecutive edges $(u_2,v_2)$ and
$(v_3,u_3)$(see Figure~\ref{fig:3crossings}.c). Now we will show
that these three edges can be substituted by the edge $(v_3,u_1)$
decreasing the number of crossings. Let us see the edges that can be
crossed by the new edge $(v_3,u_1)$. They are: $(a)$ The edges on
the right side with the source below $v_3$ and the sink above $v_3$.
But they are also crossed by $(v_3,u_3)$. $(b)$ Two sided edges that
has one its end vertex on the right side between the vertices $w_1$
and $v_3$ and the other end vertex on the left side above $w_2$.
These edges are also crossed by at least $(v_3,u_3)$. $(c)$ The
edges on the left side with the source below $u_1$ and the sink
above $u_1$. They are crossed by at least $(v_1,u_1)$.  $(d)$ The
edges that have one end point on the left side between $u_1$ and
$w_2$ and the other end point in $w_1$ and below it. These edges are
crossed at least by $(v_1,u_1)$. So we have that the edge
$(v_3,u_1)$ has less number of crossings than $(v_1,u_1)$,
$(u_2,v_2)$, $(v_3,u_3)$ token together.}

\item[\textit{Case 2b: Edge $e_1$ is directed from  left to right.}]
The proof is identical to that of Case~1b.

\eat{$(ii)$ The first edge $(u_1,v_1)$ is directed from the left to
the right. We take the two above consecutive edges $(v_2,u_2)$ and
$(u_3,v_3)$(see Figure~\ref{fig:3crossings}.d). Now we will show
that these three edges can be substituted by the edge $(u_3,v_1)$
decreasing the number of crossings. Let us see the edges that can be
crossed by the new edge $(u_3,v_1)$. They are: $(a)$ The edges on
the left side with the source below $u_3$ and the sink above $u_3$.
But they are also crossed by $(u_3,v_3)$. $(b)$ Two sided edges that
has one its end vertex on the right side between the vertices $w_1$
and $v_1$ and the other end vertex on the left side above $w_2$.
These edges are also crossed by at least $(u_1,v_1)$. $(c)$ The
edges on the right side with the source below $v_1$ and the sink
above $v_1$. They are crossed by at least $(u_1,v_1)$.  $(d)$ The
edges that have one end point on the left side between $u_3$ and
$w_2$ and the other end point in $w_1$ and below it. These edges are
crossed at least by $(u_3,v_3)$. So we have that the edge
$(u_3,v_1)$ has less number of crossings than $(u_1,v_1)$,
$(v_2,u_2)$, $(u_3,v_3)$ token together.}

\end{description}

\end{description}
\qed

\end{proof}

\begin{prop}
\label{prop:1edge-out} Let $G=(V^l \cup V^r \cup \{s,t\}, E)~$  be
an outerplanar $st$-digraph and let $\mathcal{D}(G)= \{ o_1,~
\ldots, o_\lambda \}$ be its $st$-polygon decomposition. Then, there
exists a crossing optimal acyclic HP-completion set for $G$ such
that, for every maximal $st$-polygon $o_i \in \mathcal{D}(G),~ i\leq
\lambda$, the HP-completion set does not contain  any  edge  that crosses the upper
limiting edge of $o_i$ and leaves $G_i$.
\end{prop}

\begin{proof}
Let $e=(x,t_i)$ be the upper limiting edge of $o_i$ and assume
without loss of generality that it is directed from right to left.
Also assume a crossing optimal acyclic HP-completion set
$P_{\mathrm{opt}}$ which violates the stated property, that is, it
contains an edge $\tilde{e}=(u,v), ~u \in G_i$, which crosses the
limiting edge $e$. Based on Lemma~\ref{lem:sideInOrder}, we conclude
that edge $\tilde{e}$ is a two sided edge, otherwise the vertices of
a single side appear out of order in the hamiltonian path induced by
$P_{\mathrm{opt}}$. We distinguish two cases based on direction of
the two-sided edge $\tilde{e}$.

\begin{figure}[htb]
  \begin{minipage}{\textwidth}
    \centering
    \includegraphics[width=0.5\textwidth]{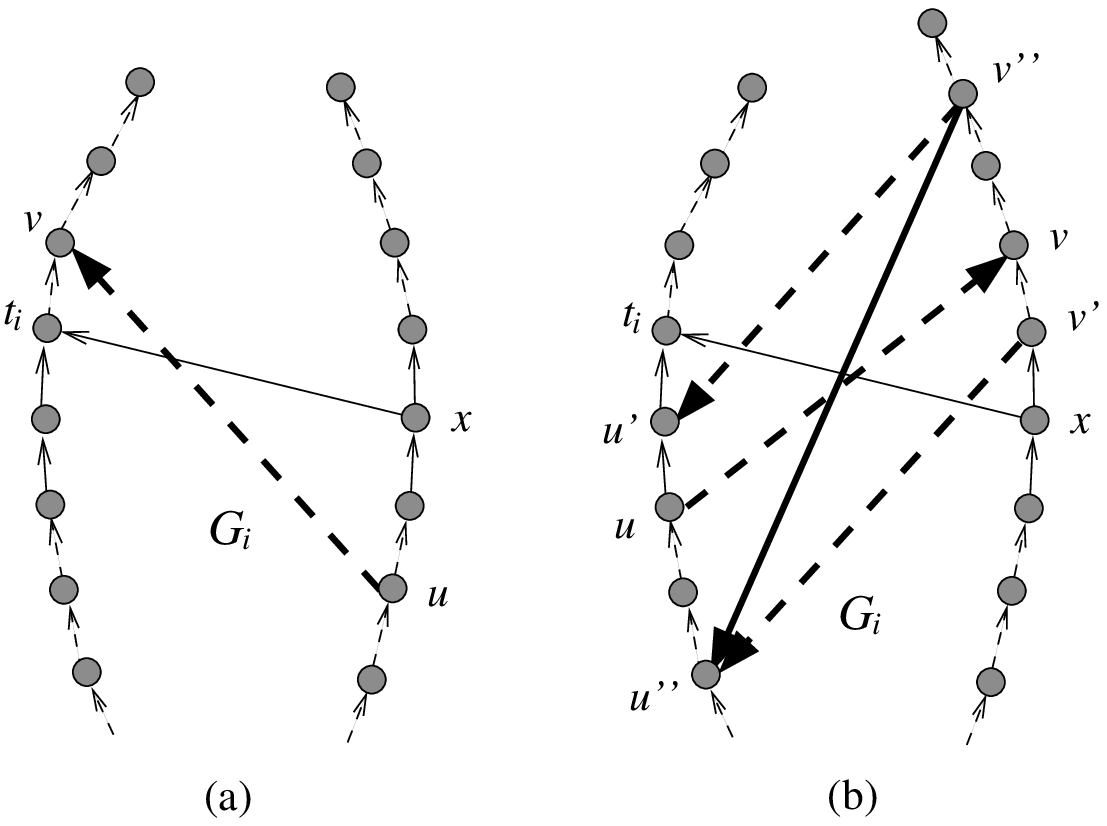}
    \caption{Configurations of crossing edges used in the proof of   Property~\ref{prop:1edge-out}.}
    \label{fig:no-outward-prop}
  \end{minipage}
\end{figure}

\begin{description}
\item[\textit{Case 1: Edge $\tilde{e}=(u,v)$ is directed from  right to left.}]
See Figure~\ref{fig:no-outward-prop}.a.

By Lemma~\ref{lem:sideInOrder},  in the hamiltonian path induced by
$P_{\mathrm{opt}}$ vertex $x$ is visited after vertex $u$ . So, in
the resulting HP-completed digraph, there must be a path from $v$ to
$x$ which, together with $(x,t_i)$ and the path $(t_i
\dashrightarrow v$) on the left side of $G$ form a cycle. This
contradicts the fact that $P_{\mathrm{opt}}$ is an acyclic
HP-completion set.

\item[\textit{Case 2: Edge $\tilde{e}=(u,v)$ is directed from  left to right.}]
See Figure~\ref{fig:no-outward-prop}.b. Denote by $v^\prime$ the
vertex positioned immediately below vertex $v$ (note that $v^\prime$
may coincide with $x$) and  by $u^\prime$ the vertex that is
immediately above $u$ (note that $u^\prime$ may coincide with
$t_i$).

Consider the hamiltonian path induced by $P_{\mathrm{opt}}$. By
Lemma~\ref{lem:sideInOrder} it follows that before crossing to the
right side of $G$ using edge  $(u,v)$  it had visited all vertices
on the right side which are placed below $v$, and thus, there is an
edge $(v^\prime,u^{\prime \prime}) \in P_{\mathrm{opt}}$, where
$u^{\prime \prime}$ is some vertex below $u$ on the left side of
$G$. Now note that, by Lemma~\ref{lem:sideInOrder},  vertex
$u^\prime$ has to appear in the hamiltonian path after vertex $u$, and thus,
there exists an edge  $(v^{\prime \prime}, u ^\prime) \in
P_{\mathrm{opt}}$ where $v^{\prime \prime}$ is a vertex above $v$ on
the right side of $G$.

By arguing in  a way similar to that of
Property~\ref{prop:3crossings}, we can show that the substitution of
path $P_{v^\prime, u^{\prime}}= ( v^\prime \rightarrow
u^{\prime\prime} \dashrightarrow u \rightarrow v \dashrightarrow
v^{\prime\prime} \rightarrow u^\prime )$ of the hamiltonian path of
$G$ by path $P^\prime_{v^\prime, u^{\prime}}= ( v^\prime
\dashrightarrow  v^{\prime\prime} \rightarrow u^{\prime\prime}
\dashrightarrow u^\prime)$ does not result in an increase of the
number of edge crossings. More specifically, when $v^\prime$ does
not coincide with $x$ and/or $u^\prime$ does not coincide with
$t_i$, the resulting new path causes at least one less crossing,
contradiction the optimality of $P_{\mathrm{opt}}$. In the case
where $v^\prime$  coincides with $x$ and $u^\prime$ coincides with
$t_i$ and the two hamiltonian paths cause the same number of
crossings, the new HP-completion set  has the desired property, that is, none of its edges crosses the limiting
edge $(x,t_i)$ and leaves $G_i$.\qed
\end{description}

\end{proof}

\begin{prop}
\label{prop:at-most-1edge}  Let $G=(V^l \cup V^r \cup \{s,t\}, E)~$
be an outerplanar $st$-digraph and let $\mathcal{D}(G)= \{ o_1,~
\ldots, o_\lambda \}$ be its $st$-polygon decomposition. Then, in
every crossing optimal acyclic HP-completion set for $G$ and for
every maximal $st$-polygon $o_i \in \mathcal{D}(G),~ i\leq \lambda$,
at most one edge crosses the upper limiting edge of $o_i$.
\end{prop}

\begin{proof}
Let edge $e=(x,t_i)$ be the upper limited edge of $o_i$. Without
loss of generality assume that it is directed from the right to the
left side of $G$,  and let $v$ be the vertex immediately above $x$
on the right side of $G$ and $u$ be the vertex immediately below
$t_i$ on the left side of $G$. By Property~\ref{prop:3crossings}, we
have that the edges of a crossing optimal acyclic HP-completion set
for $G$ do not cross $e$ three or more times.

For the shake of contradiction assume that there is  a crossing
optimal acyclic HP-completion  set $P_{\mathrm{opt}}$ for $G$ that
crosses edge  $e$ twice. Let $e_1, e_2~\in P_{\mathrm{opt}}$ be the
edges which cross $e$. Clearly, these two edges cross $e$ in the
opposite direction and do not cross each other. Let $e_1$ be the
edge that crosses $e$ and leaves $G_i$. Observe that $e_1$ has
opposite direction to that of $e$, otherwise a cycle is created.
Then, since $e_1$ does not cross $e_2$, edge $e_1$ does not coincide
with $(u,v)$. However, for the case where $e_1 \neq (u,v)$, we
established in the proof of Property~\ref{prop:1edge-out} (Case~2)
that we are always able to build an acyclic HP-completion set that
induced less crossings than $P_{\mathrm{opt}}$, a clear
contradiction\footnote{The proof is identical and for this reason it
is not repeated}.\qed
\end{proof}

\eat{  
\begin{prop}
\label{prop:1edgedirection} \marginpar{DO WE NEED IT?}

Let $G=(V^l \cup V^r \cup \{s,t\}, E)~$ be an outerplanar
$st$-digraph and let $\mathcal{D}(G)= \{ o_1,~ \ldots, o_\lambda \}$
be its $st$-polygon decomposition. Moreover, let $e_i$ be the upper
limiting edge of any maximal $st$-polygon $o_i \in \mathcal{D}(G),~
i\leq \lambda$. Then, in every crossing optimal acyclic
HP-completion set for $G$ and for every maximal $st$-polygon $o_i
\in \mathcal{D}(G),~ i\leq \lambda$, any edge that crosses $e_i$ and
enters $G_i$ has the same direction as $e_i$.
\end{prop}

\begin{proof}
Without loss of generality assume that $e_i=(x, t_i)$ is directed
from right to left. For the shake of contradiction, assume that a
crossing optimal acyclic HP-completion set for $G$ contains an edge
$\tilde{e}=(u,v)$ entering $G_i$ that is directed in the opposite
direction of $e_i$, that is, $u$ is above $t_i$ on the left side of
$G$ and $v$ is below $x$ on the right side of $G$. Then, cycle $(t_i
\dashrightarrow u \rightarrow v \dashrightarrow x \rightarrow t_i)$
is formed in the acyclic HP-completed graph, a clear contradiction.
\qed
\end{proof}

}

The following theorem states that there always exists a crossing optimal acyclic HP-completion set
for  outerplanar $st$-digraphs that has certain properties. The algorithm which we present in the next section, focusses
only on HP-completion set satisfying these properties.

\begin{theorem}
\label{thm:allproperties} Let $G=(V^l \cup V^r \cup \{s,t\}, E)~$  be
an outerplanar $st$-digraph and let $\mathcal{D}(G)= \{ o_1,~
\ldots, o_\lambda \}$ be its $st$-polygon decomposition. Then, there
exists a crossing optimal acyclic HP-completion set $P_{\mathrm{opt}}$ for $G$ such
that it satisfies the following properties: \vspace*{-.3cm}
\begin{quote}
\begin{enumerate}[a)]
\item
Each edge of $E$ is crossed by at most two edges of $P_{\mathrm{opt}}$.
\item
Each upper limiting edge $e_i$ of any maximal $st$-polygon $o_i,~ i\leq \lambda$, is crossed by at  most one
 edge of $P_{\mathrm{opt}}$. Moreover, the edge crossing $e_i$, if any, enters $G_i$.
\end{enumerate}
\end{quote}
\end{theorem}
\begin{proof}
Follows directly from  Properties~\ref{prop:3crossings},~\ref{prop:1edge-out} and~\ref{prop:at-most-1edge}.  \qed
\end{proof}

\subsection{The Algorithm}

The algorithm for obtaining a crossing-optimal acyclic HP-completion
set for an outerplanar $st$-digraph $G$ is a dynamic programming
algorithm based on the $st$-polygon decomposition $\mathcal{D}(G)=
\{ o_1,~ \ldots, o_\lambda \}$ of $G$. The following lemmata allow
us to compute a crossing-optimal acyclic HP-completion set for an
$st$-polygon and  to obtain   a crossing-optimal acyclic
HP-completion set for  $G_{i+1}$ by combining an optimal solution
for $G_{i}$ with an optimal solution for $o_{i+1}$.

\eat{

\begin{lemma}
\label{lem:NoMixedSidesBook} Assume an $st$-polygon $R=(V^l \cup V^r
\cup \{s,t\}, E)~$. In a crossing-optimal acyclic HP-completion set
of $R$ with at most 1 edge crossing per initial edge,  the vertices
of
  $V^l$ are visited before the vertices of $V^r$, or, vice
  versa.
\end{lemma}

\begin{proof}
In an $st$-polygon, due to the existence of the median $(s,t)$, no
edges exists between vertices of $V^l$ and $V^r$. Thus, the edges of
a Hamiltonian path that connect vertices at different sides all
belong in the HP-completion set and they all cross the median. Since
we insist on having at most one crossing per edge of $R$, the
HP-completion set must consist of a single edge. This implies that
all vertices of $V^l$ are visited before the vertices of $V^r$, or,
vice versa.\qed
\end{proof}

}

Assume an outerplanar $st$-digraph $G$. We denote by $S(G)$ the
hamiltonian path on the HP-extended digraph of $G$ that results when
a crossing-optimal HP-completion set is added to $G$. Note that if
we are only given $S(G)$ we can infer the size of the HP-completion
set and the number of edge crossings. Denote by $c(G)$ the number of
edge crossings caused by the HP-completion set inferred by $S(G)$.
If we are restricted to Hamiltonian paths that enter the sink of G
from a vertex on the left (resp. right) side of $G$, then we denote
the corresponding size of HP-completion set as $c(G,L)$ (resp.
$c(G,R)$). Obviously, $c(G)= \min \{ c(G,L),~c(G,R) \}$. Moreover,
the notation can be extended so that,  if the size of the
HP-completion set is $s$, then we denote by $c^i(G,L)~(c^i(G,R))$
the corresponding number of crossings for HP-completion sets that
contain exactly $i$ edges, $i \leq s$. By
Theorem~\ref{thm:st-polygon-size-2}, we know that the size of a
crossing-optimal acyclic HP-completion set for an $st$-polygon is at
most 2. This notation that restricts the size of the HP-completion
set will be used only for $st$-polygons and thus, only the terms
$c^1(G,L), ~c^1(G,R),~ c^2(G,L)$ and $c^2(G,R)$ will be utilized.

We use the operator $\oplus$ to indicate the concatenation of two
paths. By convention, the hamiltonian path of a single vertex is the
vertex itself.

\begin{lemma}
\label{lem:spineCrossingSTpolygon} Assume an $n$ vertex
$st$-polygon $o=(V^l \cup V^r \cup \{s,t\}, E)~$. A crossing-optimal
acyclic HP-completion set for $o$ and the corresponding number of
crossings can be computed in $O(n)$ time.
\end{lemma}

\begin{proof}
\begin{figure}[htb]
    \centering
    \includegraphics[width=.5\textwidth]{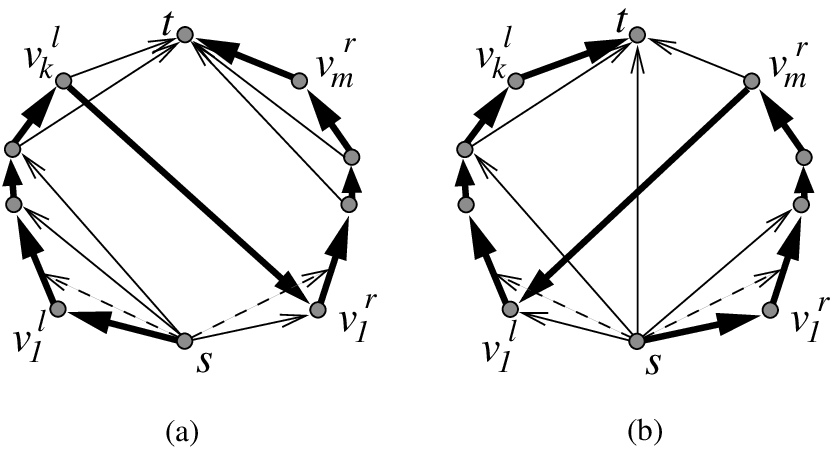}
    \caption{HP-completion set of an $st$-polygon with one edge.}
    \label{fig:HPcompletionPolygon}
\end{figure}

From Lemma~\ref{lemma:3to1} and Lemma~\ref{lemma:4to2} it follows
that it is sufficient to look through all HP-completion sets with
one or two edges in order to find a crossing-optimal acyclic
HP-completion set. Let $V^l = \{ v^l_1,~\ldots, v^l_k\}$ and $V^r =
\{ v^r_1,~\ldots, v^r_m\}$, where the subscripts indicate the order
in which the vertices appear on the left (right) boundary of $o$.
Suppose that $I:V \times V \rightarrow \{0,1\}$ is an indicator
function such that $I(u,v)=1 \iff (u,v) \in E$.

The only two possible  HP-completion sets consisting of exactly one
edge are $\left\{(v^l_k, v^r_1)\right\}$ and $\left\{(v^r_m,
v^l_1)\right\}$.

Edge $(v^l_k, v^r_1)$ crosses  all edges connecting $t$ with
vertices in $V_l \setminus \{v^l_k\} $, the median (provided it
exists), and all edges connecting $s$ with vertices in $V_r\setminus
\{v^R_1\}$ (see Figure~\ref{fig:HPcompletionPolygon}.a). It follows
 that:
$$c^1(o,R)= I(s,t) + \sum_{i=2}^{k-1}{I(v^\ell_i,t)}
+\sum_{i=2}^{m-1}{I(s,v^r_i)}$$

Similarly, edge $(v^r_m, v^l_1)$ crosses all edges connecting $t$
with vertices in $V_r \setminus \{v^r_m\}$, the median (provided it
exists), and all  edges connecting $s$ with vertices in
$V_l\setminus \{v^l_1\}$ (see
Figure~\ref{fig:HPcompletionPolygon}.b). It follows that:
$$c^1(o,L)=I(s,t)+\sum_{i=2}^{m-1}{I(v^r_i,t)}
+\sum_{i=2}^{k-1}{I(s,v^\ell_i)}$$

\begin{figure}[htb]
    \centering
    \includegraphics[width=.5\textwidth]{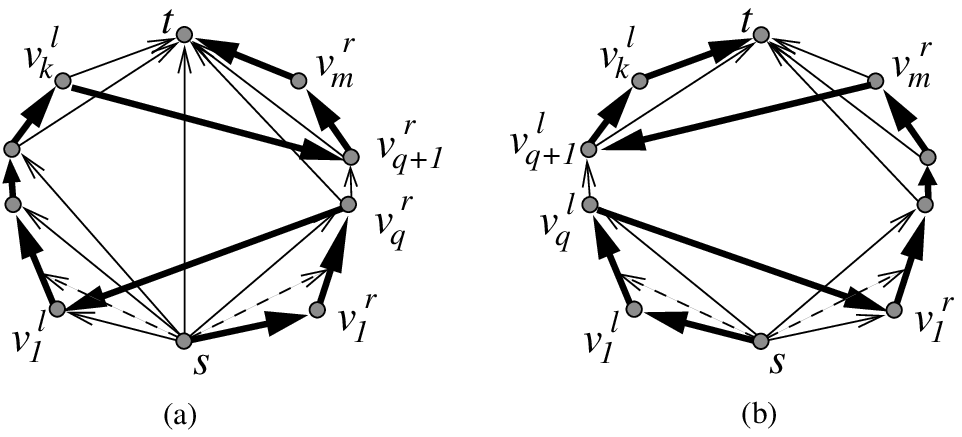}
    \caption{HP-completion set of an $st$-polygon with two edges.}
    \label{fig:HPcompletionPolygon_twoedges}
\end{figure}

Consider now an acyclic HP-completion set of size 2. Assume that the
lowermost of its edges leaves  node $v_q^r$ on the right side of $o$
(see Figure~\ref{fig:HPcompletionPolygon_twoedges}.a). Then,  it
must enter vertex $v_1^l$. Moreover, the second edge of the acyclic
HP-completion set must leave vertex $v_k^l$ and enter vertex
$v_{q+1}^r$. Thus, the HP-completion set is $\left\{ (v_q^r,
v_1^l),~(v_k^l,v_{q+1}^r) \right\}$ and, as we observe, it can be
put into correspondence with edge $(v_q^r,v_{q+1}^r)$ on the right
side of $o$. In addition, we observe that the hamiltonian path
enters $t$ from the right side. An analogous situation occurs when
the lowermost edge leaves the left side of $o$ (see
Figure~\ref{fig:HPcompletionPolygon_twoedges}.b).

We denote by $c_q^2(o,R)$ the  number of crossings caused by the
completion set associated with the edge originating at the $q^{th}$
lowermost vertex on the right side of $o$. Similarly we define
$c_q^2(o,L)$. $c_q^2(o,R)$  can be computed as follows:

\vspace*{-0.4cm} \small{
$$c_q^2(o,R)=2\cdot I(s,t)+\sum_{i=1}^{k-1}{I(v^\ell_i,t)}+\sum_{i=2}^{k}{I(s,v^\ell_i)}+2\cdot\sum_{i=1}^{q-1}{I(v^r_i,t)}
+2\cdot\sum_{i={q+2}}^{m}{I(s,v^r_i)} + I(v^r_{q},t) +
I(s,v^r_{q+1})$$}

\vspace*{-0.4cm}
 Then, the optimal solution where the hamiltonian path terminates on the
right side of $o$ can be taken as  the  minimum over all
$c_q^2(o,R), ~1\leq q \leq m-1$:
$$c^2(o,R) = \min_{1\leq q \leq m-1}\{c_q^2(o,R)\}$$

Similarly, $c_q^2(o,L)$ can be computed  as follows:

\vspace*{-0.4cm} \small{
$$c_q^2(o,L)=2\cdot I(s,t)+\sum_{i=1}^{m-1}{I(v^r_i,t)}+\sum_{i=2}^{m}{I(s,v^r_i)}+2\cdot\sum_{i=1}^{q-1}{I(v^\ell_i,t)}
+2\cdot\sum_{i={q+2}}^{k}{I(s,v^\ell_i)} + I(v^\ell_{q},t) +
I(s,v^\ell_{q+1})$$}

\vspace*{-0.4cm}
 Then, the optimal solution where the hamiltonian path terminates on the
left side of $o$ can be taken as  the  minimum over all
$c_q^2(o,R),~1\leq q \leq k-1$:
 $$c^2(o,L) = \min_{1\leq q \leq
k-1}\{c_q^2(o,L)\}$$

So, now, the number of crossings that corresponds to the optimal
solution can be computed as follows:
$$c(o)=\min\{c^1(o,L),c^1(o,R),c^2(o,L),c^2(o,R)\}$$

It is evident that $c^1(o,R)$ and $c^1(o,L)$ can be computer in time
$O(n)$. It is also easy to see that any $c_q^2(o,R)$ can be computed
from $c_{q-1}^2(o,R)$ in constant time, while $c_1^2(o,R)$ can be
computed in time $O(n)$. Therefore, $c^2(o,R)$, as well as
$c^2(o,L)$, can be computed in linear time. Thus, we conclude that a
crossing-optimal acyclic HP-completion set for any $n$ vertex
$st$-polygon $o$ and its corresponding number of crossings can be
computed in $O(n)$ time. \qed
\end{proof}

Let $\mathcal{D}(G)= \{ o_1,~ \ldots, o_\lambda \}$ be the
$st$-polygon decomposition of $G$, where element $o_i,~ 1\leq i \leq
\lambda$ is either an $st$-polygon or a free vertex. Recall that, we
denote by $G_i,~1\leq i \leq \lambda $  the graph induced by the
vertices of elements $o_1, \ldots, o_i$. Graph $G_i$ is also an
outerplanar $st$-digraph. The same holds for the subgraph of $G$
that is induced by any number of consecutive elements of
$\mathcal{D}(G)$.

\begin{lemma}
\label{lem:optimalSubsolution} Assume an outerplanar $st$-digraph
$G$ and let $\mathcal{D}(G)= \{ o_1,~ \ldots, o_\lambda \}$ be its
$st$-polygon decomposition. Consider any two consecutive elements
$o_i$ and $o_{i+1}$ of $\mathcal{D}(G)$  that   share at most one
vertex. Then, the following statements hold:\vspace*{-.5cm}
\begin{enumerate}[(i)]
\item $S(G_{i+1})=S(G_i) \oplus S(o_{i+1})$,  and
\item $c(G_{i+1})= c(G_i) + c(o_{i+1})$.
\end{enumerate}
\end{lemma}

\begin{figure}[htb]
    \centering
    \includegraphics[width=1\textwidth]{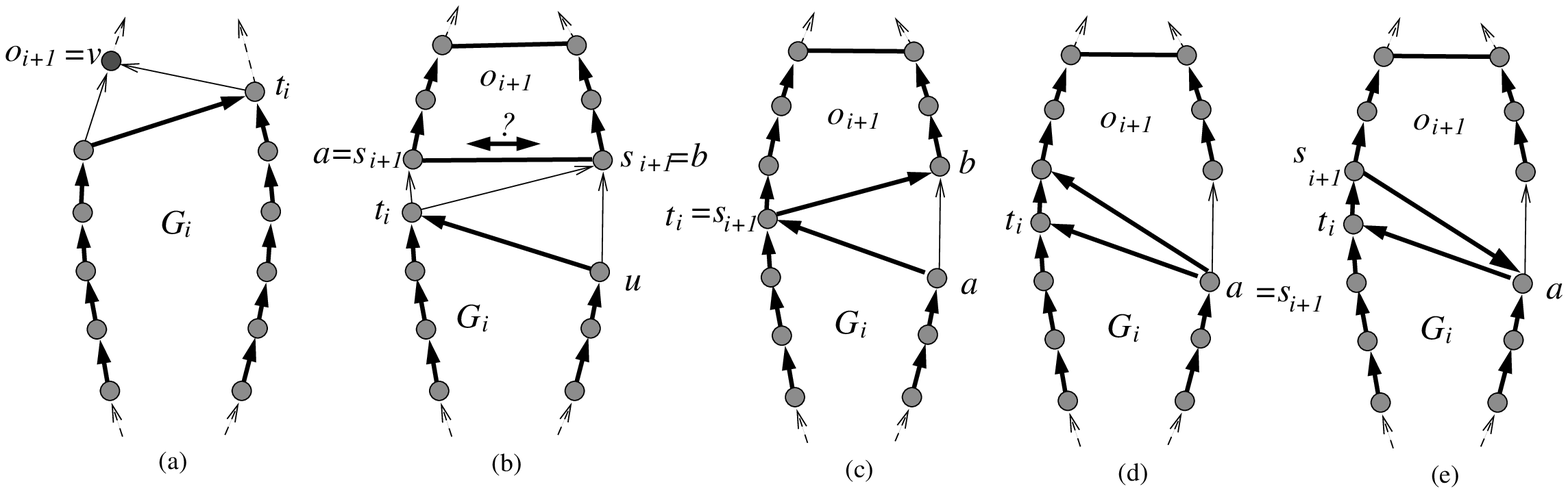}
    \caption{Configurations used in the proof of Lemma~\ref{lem:optimalSubsolution}.}
    \label{fig:optimalSubsolution}
\end{figure}

\begin{proof}
We proceed to prove first statement (i). There are three cases to
consider in which 2 consecutive elements of
$\mathcal{D}(G)$ share at most 1 vertex.\\
\textbf{Case-1}: Element $o_{i+1}=v$ is a free vertex (see
Figure~\ref{fig:optimalSubsolution}.a). By
Lemma~\ref{lem:pahtProperties}, if $o_i$ is either a free vertex or
an $st$-polygon, there is an edge connecting the sink of $o_i$ to
$v$. Also observe that if $v$ was not the last vertex of
$S(G_{i+1})$ then the crossing-optimal HP-completion set had to
include an edge from $v$ to some vertex of $G_i$. This is impossible
since it would create a cycle in the HP-extended digraph of
$S(G_{i+1})$.

\textbf{Case-2}: Element $o_{i+1}$ is an $st$-polygon that shares no
common vertex with $G_i$ (see
Figure~\ref{fig:optimalSubsolution}.b). Without loss of generality,
assume that the sink of $G_i$ is located on its left side. We first
observe that edge $(t_i, s_{i+1})$ exists in $G$. If $s_{i+1}$ is on
the left side of $G$, we are done. Note that there can be no other
vertex between $t_i$ and $s_{i+1}$ in this case, because then $o_i$
and $o_{i+1}$ would not be consecutive. If $s_{i+1}$ is on the right
side of $G$, realize that the area between two $st$-polygons $o_i$
and $o_{i+1}$ can not be free of edges, as it is a weak
$st$-polygon. Note also that the edge $(u,a)$ cannot exist in $G$,
since, if it existed, the area between the two polygons would be a
strong $st$-polygon with $(u,a)$ as its median. Thus, that area can
only contain the edge $(t_i, s_{i+1})$. Thus, as indicated in
Figure~\ref{fig:optimalSubsolution}.b, each of the end-vertices of
the lower limiting edge of $o_{i+1}$ can be its source. Since edge
$(t_i, s_{i+1})$ exists, the solution $S(o_{i+1})$ can be
concatenated to $S(G_i)$ and yield a valid hamiltonian path for
$G_{i+1}$. Now notice that in $S(G_{i+1})$ all vertices of $G_i$
have to be placed before the vertices of $o_{i+1}$. If this was not
the case, then the crossing-optimal HP-completion set had to include
an edge from a vertex $v$ of $o_{i+1}$ to some vertex $u$ of $G_i$.
This is impossible since it would create a cycle in the HP-extended
digraph of $S(G_{i+1})$.

\textbf{Case-3}: Element $o_{i+1}$ is an $st$-polygon that shares
one common vertex with $G_i$ (see
Figure~\ref{fig:optimalSubsolution}.c). Without loss of generality,
assume that the sink $t_i$ of $G_i$ is located on its left side.
Firstly, notice that the  vertex shared by $G_i$ and $o_{i+1}$ has
to be vertex $t_i$. To see that let $a$ be upper vertex at the right
side of $G_i$. Then, edge $(a,t_i)$ exists since $t_i$ is the sink
of $G_i$. For the sake of contradiction assume that $a$ was the
vertex shared between $G_i$ and $o_{i+1}$. If $a$ was also the
source of $o_{i+1}$ (see Figure~\ref{fig:optimalSubsolution}.d) then
$o_{i+1}$ wouldn't be maximal (edge $(a,t_i)$ should also belong to
$o_{i+1}$). If  $s_{i+1}$ was on the left side (see
Figure~\ref{fig:optimalSubsolution}.e), then a cycle would be formed
involving edges $(t_i),~(t_1, s_{i+1})$ and $(s_{i+1},a)$, which is
impossible since $G$ is acyclic. Thus, the vertex shared by $G_i$
and $o_{i+1}$ has to be vertex $t_i$. Secondly, {observe that $t_i$
must coincide with vertex $s_{i+1}$ (see
Figure~\ref{fig:optimalSubsolution}.c). If $s_{i+1}$ coincided with
vertex $b$, then the $st$-polygon $o_i$ wouldn't be maximal since
edge $(b,t_i)$ should also belong to $o_{i}$.}

We conclude that $t_i$ coincides with $s_{i+1}$ and, thus, the
solution $S(o_{i+1})$ can be concatenated to $S(G_i)$ and yield a
valid hamiltonian path for $G_{i+1}$. To complete the proof for this
case, we can show by contradiction (on the acyclicity of $G$; as in
Case-2)  that in $S(G_{i+1})$ all vertices of $G_i$ have to be
placed before the vertices of $o_{i+1}$.

Now observe that statement (ii) is trivially true since, in all
three cases, the hamiltonian paths $S(G_i)$ and $S(o_{i+1})$ were
concatenated by using at most one additional edge of graph $G$.
Since $G$ is planar, no new crossings are created.\qed
\end{proof}

\begin{lemma}
\label{lem:dynProgSolution} Assume an outerplanar $st$-digraph $G$
and let $\mathcal{D}(G)= \{ o_1,~ \ldots, o_\lambda \}$ be its
$st$-polygon decomposition. Consider any two consecutive elements
$o_i$ and $o_{i+1}$ of $\mathcal{D}(G)$  that   share an edge. Then,
the following statements hold:\vspace*{-.3cm}
\begin{enumerate}[(1)]
\item $~t_{i} \in V^l ~\Rightarrow~ c(G_{i+1},L)=\min \{
    c(G_{i},L)+c^1(o_{i+1},L)+1,~
    c(G_{i},R)+c^1(o_{i+1},L),~
    c(G_{i},L)+c^2(o_{i+1},L),~
    c(G_{i},R)+c^2(o_{i+1},L)
\} $
\\
\item $~t_{i} \in V^l ~\Rightarrow~  c(G_{i+1},R)=\min \{
    c(G_{i},L)+c^1(o_{i+1},R),~
    c(G_{i},R)+c^1(o_{i+1},R),~
    c(G_{i},L)+c^2(o_{i+1},R)+1,~
    c(G_{i},R)+c^2(o_{i+1},R)
\} $
\\
\item $~t_{i} \in V^r ~\Rightarrow~  c(G_{i+1},L)=\min \{
    c(G_{i},L)+c^1(o_{i+1},L),~
    c(G_{i},R)+c^1(o_{i+1},L),~
    c(G_{i},L)+c^2(o_{i+1},L),~
    c(G_{i},R)+c^2(o_{i+1},L)+1
\} $
\\
\item $~t_{i} \in V^r ~\Rightarrow~  c(G_{i+1},R)=\min \{
    c(G_{i},L)+c^1(o_{i+1},R),~
    c(G_{i},R)+c^1(o_{i+1},R)+1,~
    c(G_{i},L)+c^2(o_{i+1},R),~
    c(G_{i},R)+c^2(o_{i+1},R)
\} $
\end{enumerate}
\end{lemma}

\begin{figure}[htb]
  \begin{minipage}{\textwidth}
    \centering
    \includegraphics[width=0.8\textwidth]{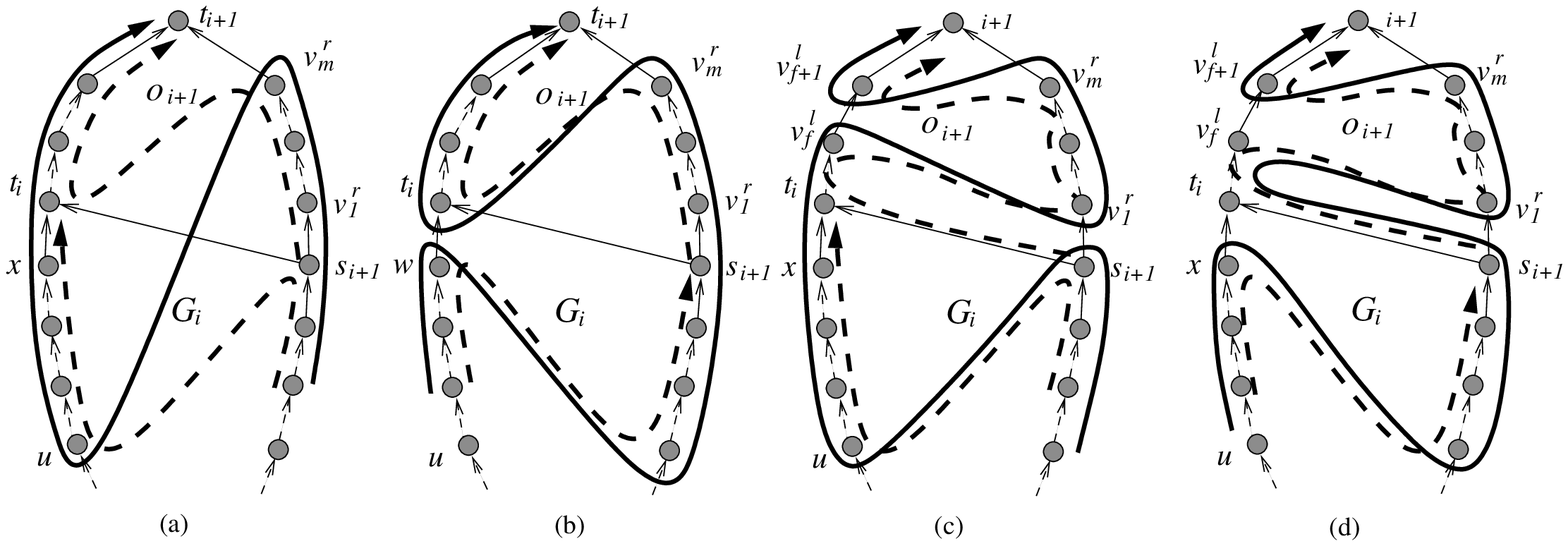}
    \caption{The hamiltonian paths for statement~(1) of Lemma~\ref{lem:dynProgSolution}.}
    \label{fig:dynProgSolutionLEFT-L}
  \end{minipage}
\end{figure}

\begin{figure}[htb]
    \begin{minipage}{\textwidth}
    \centering
    \includegraphics[width=0.8\textwidth]{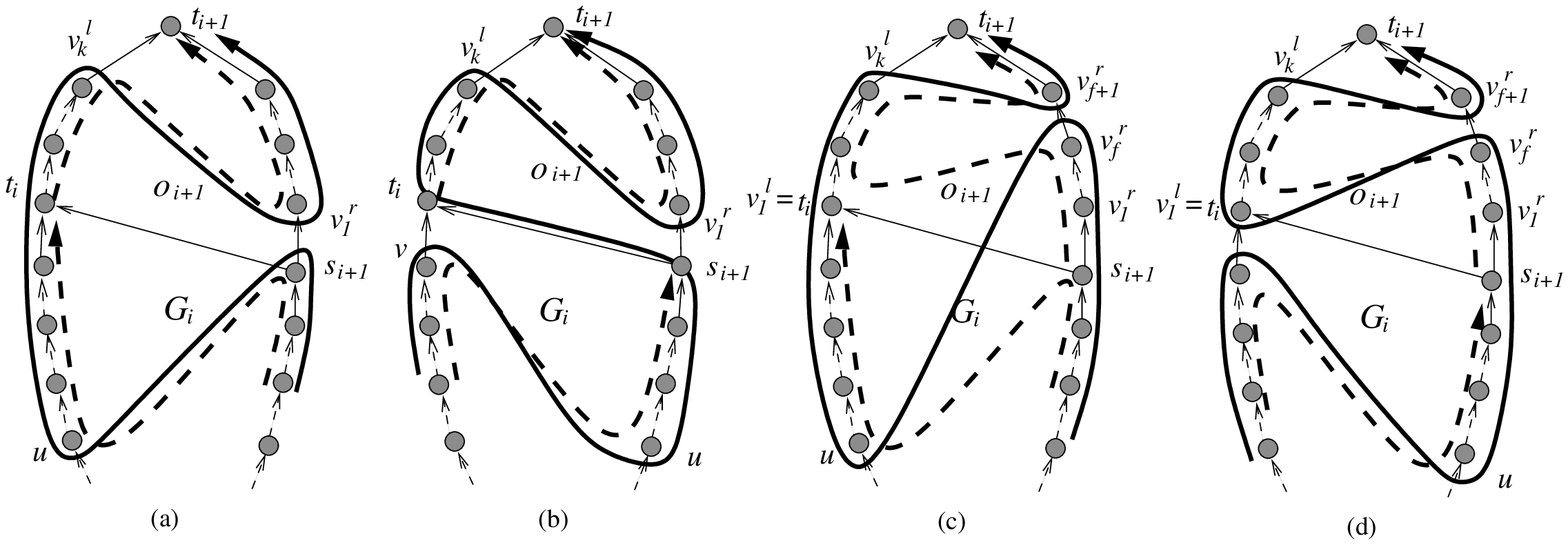}
    \caption{The hamiltonian paths for statement~(2) of Lemma~\ref{lem:dynProgSolution}.}
    \label{fig:dynProgSolutionLEFT-R}
  \end{minipage}
\end{figure}

\begin{proof}
We first show how to build hamiltonian paths that infer HP-completions
sets of the specified size. \eat{Then, we prove that these HP-completion sets are crossing-optimal.}
For each of the statements, there are
four cases to consider. The minimum number of crossings, is then
determined by taking the minimum over the four sub-cases.

 (1) $t_{i}
\in V^l \Rightarrow c(G_{i+1},L)=\min \{
    c(G_{i},L)+c^1(o_{i+1},L)+1,~
    c(G_{i},R)+c^1(o_{i+1},L),~
    c(G_{i},L)+c^2(o_{i+1},L),~
    c(G_{i},R)+c^2(o_{i+1},L)
\}$.
\begin{quote}
\hspace*{-.4cm} \emph{Case 1a}. The hamiltonian path
  enters $t_{i}$ from a vertex on the left side of $G_i$ and the size of the HP-completion of $G_{i+1}$
  is one.
 Figure~\ref{fig:dynProgSolutionLEFT-L}.a shows the hamiltonian paths for $G_i$ (lower dashed
 path) and
  $o_{i+1}$ (upper dashed path) as well as the   resulting hamiltonian path for $G_{i+1}$ (shown in bold).
 From the figure, it follows that $c(G_{i+1},L)=c(G_{i},L)+c^1(o_{i+1},L)+1$. To see that, just
 follow the
 edge $(v_m^r,u)$ that becomes part of the completion set of $G_{i+1}$.
 Edge $(v_m^r,u)$ is involved in as many edge crossings as edge $(v_m^r,t_i)$
 (the only edge in the HP-completion set of $o_{i+1}$),
 plus as many edge crossings as edge $(s_{i+1},u)$
 (an edge in the HP-completion set of $G_{i}$), plus one (1) edge
 crossing of the lower
 limiting edge of $o_{i+1}$.  \\

\hspace*{-.4cm}\emph{Case 1b}. The hamiltonian path  reaches
$t_{i}$
 from a vertex on the right side of $G_i$ and the size of the HP-completion of $G_{i+1}$ is one.
 Figures~\ref{fig:dynProgSolutionLEFT-L}.b shows the resulting path.
 From the figure, it follows that
 $c(G_{i+1},L)=c(G_{i},R)+c^1(o_{i+1},L)$, that is the simple concatenation of two solutions.\\

\hspace*{-.4cm} \emph{Case 1c}. The hamiltonian path
  enters $t_{i}$ from a vertex on the left side of $G_i$ and the size of the HP-completion of $G_{i+1}$ is two.
 Figure~\ref{fig:dynProgSolutionLEFT-L}.c shows the resulting path.
 From the figure, it follows that $c(G_{i+1},L)=c(G_{i},L)+c^2(o_{i+1},L)$, which is just concatenation
of two solutions. \\

\hspace*{-.4cm}\emph{Case 1d}. The hamiltonian path  reaches $t_{i}$
 from a vertex on the right side of $G_i$ and the size of the HP-completion of $G_{i+1}$ is two.
 Figure~\ref{fig:dynProgSolutionLEFT-L}.d  shows the resulting pathes.
 From the figure, it follows that
 $c(G_{i+1},L)=c(G_{i},R)+c^2(o_{i+1},L)$, which is again a simple
 concatenation of two solutions.

 \end{quote}

 (2) $t_{i} \in V^l \Rightarrow  c(G_{i+1},R)=\min \{
    c(G_{i},L)+c^1(o_{i+1},R),~
    c(G_{i},R)+c^1(o_{i+1},R),~
    c(G_{i},L)+c^2(o_{i+1},R)+1,~
    c(G_{i},R)+c^2(o_{i+1},R)
\} $.
 \begin{quote}
\hspace*{-.4cm}\emph{Case 2a}. The hamiltonian path
  enters $t_{i}$ from a vertex on the left side of $G_i$ and the size of the HP-completion of $G_{i+1}$ is one.
 Figure~\ref{fig:dynProgSolutionLEFT-R}.a shows the resulting path.
 From the figure, it follows that $c(G_{i+1},R)=c(G_{i},L)+c^1(o_{i+1},R)$, that is, a simple
  concatenation of the two
 solutions.\\

\hspace*{-.4cm}\emph{Case 2b}. The hamiltonian path  reaches $t_{i}$
 from a vertex on the right side of $G_i$ and the size of the HP-completion of $G_{i+1}$ is one.
 Figure~\ref{fig:dynProgSolutionLEFT-R}.b shows the resulting path.
 From the figure, it follows that $c(G_{i+1},R)=
    c(G_{i},R)+c^1(o_{i+1},R)$,  that is, a simple
  concatenation of the two
 solutions.\\

\hspace*{-.4cm}\emph{Case 2c}. The hamiltonian path
  enters $t_{i}$ from a vertex on the left side of $G_i$ and the
  size of HP-completion set of $G_{i+1}$ is two.
 Figure \ref{fig:dynProgSolutionLEFT-R}.c shows the resulting path.
 From the figure, it follows that $c(G_{i+1},R)=c(G_{i},L)+c^2(o_{i+1},R)+1$.\\
 Note that the added edge
 $(v_i^r,u)$ creates one more crossing  than the number of crossings caused by  edges $(v_i^r,v_1^l)$,
 $(s_{i+1},u)$ taken together. The additional crossing is due to the crossing of the lower limiting edge
 of $o_{i+1}$.\\

\hspace*{-.4cm}\emph{Case 2d}. The hamiltonian path  reaches $t_{i}$
 from a vertex on the right side of $G_i$ and the size of HP-completion set of $G_{i+1}$ is two.
 Figure~\ref{fig:dynProgSolutionLEFT-R}.d shows the resulting path.
 From the figure, it follows that $c(G_{i+1},R)=
    c(G_{i},R)+c^2(o_{i+1},R)$,  that is, a simple
  concatenation of the two
 solutions.

 \end{quote}

\begin{figure}[htb]
  \begin{minipage}{\textwidth}
    \centering
    \includegraphics[width=0.8\textwidth]{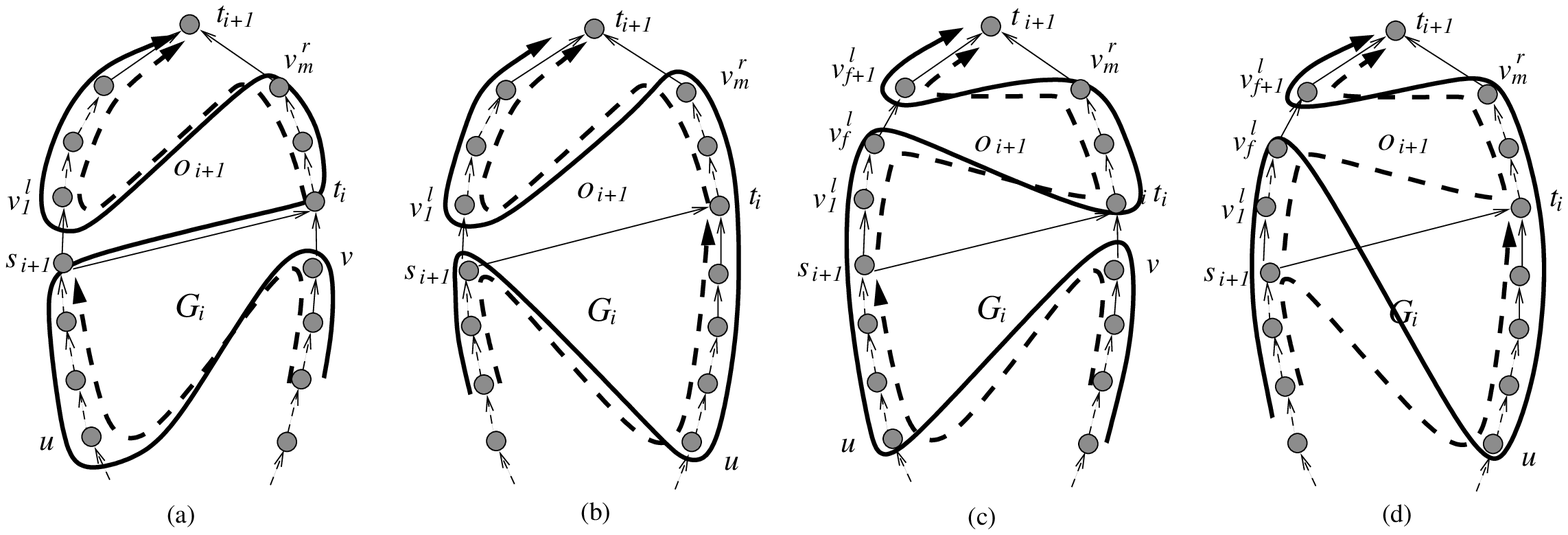}
    \caption{The hamiltonian paths for statement~(3) of Lemma~\ref{lem:dynProgSolution}.}
    \label{fig:dynProgSolutionRIGHT-L}
  \end{minipage}
\end{figure}

\begin{figure}[htb]
    \begin{minipage}{\textwidth}
    \centering
    \includegraphics[width=0.8\textwidth]{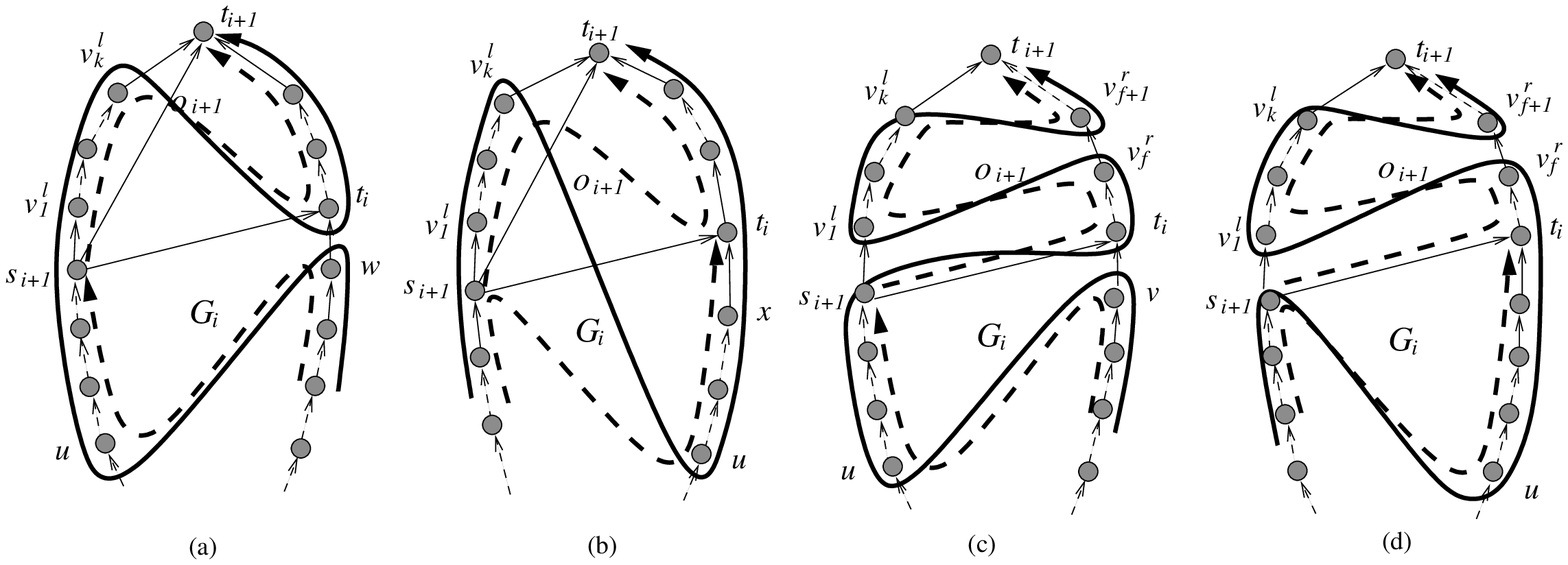}
    \caption{The hamiltonian paths for statement~(4) of Lemma~\ref{lem:dynProgSolution}.}
    \label{fig:dynProgSolutionRIGHT-R}
  \end{minipage}
\end{figure}

The proofs for statements (3) and (4) are symmetric to those of
statements (2) and (1), respectively. Figures
~\ref{fig:dynProgSolutionRIGHT-L}
and~\ref{fig:dynProgSolutionRIGHT-R} show how to construct the
corresponding hamiltonian paths in each  case.

In order to complete the proof, we need to also show that the
constructed hamiltonian paths which cause the stated number of
crossings are optimal. The basic idea of the proof is the following:
we assume a crossing optimal solution $P^{\mathrm{opt}}_{G_{i+1}}$
for $G_{i+1}$ and, based on it,  we identify   two solutions
$P_{G_{i}}$ and $P{o_{i+1}}$ for $G_i$ and $o_{i+1}$, respectively,
and we prove that they are crossing optimal. In addition, we observe
that $P^{\mathrm{opt}}_{G_{i+1}}$ can be obtain  from $P_{G_{i}}$
and $P{o_{i+1}}$ as one of the four cases in the statement of the
Lemma. The proof of optimality traces backwards the construction of
the hamiltonian paths given for each of the four cases in the lemma.
For this reason, a detailed proof is omitted. \qed
\end{proof}

\eat{ 
Next we will show two lemmata that prove that the solutions found by
Lemma~\ref{lem:dynProgSolution} and
Lemma~\ref{lem:optimalSubsolution} give the optimal solution.

\begin{lemma}
\label{lem:correctness_commonedge}  Assume an outerplanar
$st$-digraph $G=(V^l \cup V^r \cup \{s,t\}, E)~$ and
 its  $st$-polygon decomposition
$\mathcal{D}(G)= \{ o_1,~ \ldots, o_\lambda \}$. Suppose $G_i$ to be
the graph induced by the vertices of elements $o_1, \ldots, o_i$ and
$o_{i+1}$ be the next component of the $st$-polygon decomposition,
which is an $st$-polygon. Suppose also that  $G_i$ and $o_{i+1}$
have a common edge. Denote by $P_{\mathrm{opt}}$ the optimal
solution of $st$-digraph $G_i \cup o_{i+1}$. We will prove that
$P_{\mathrm{opt}}$ can be constructed from the optimal solutions for
$G_i$ and $o_{i+1}$ using Lemma~\ref{lem:dynProgSolution}.
\end{lemma}
\begin{proof}

The basic idea of the proof is as follows: we suppose an optimal
solution $P_{\mathrm{opt}}$ for the $G_{i+1}=G_i \cup o_{i+1}$ ,
construct from it two solutions $P_i$ and $P_{i+1}$ for $G_i$ and
$o_{i+1}$ respectively and then show that $P_i$ and $P_{i+1}$ are
optimal and that $P_{\mathrm{opt}}$ can be got from them by one of
the cases of Lemma~\ref{lem:dynProgSolution}. We proceed in the
proof by taking the cases based on the structure of
$P_{\mathrm{opt}}$.

We will call the common edge of $G_i$ and $o_{i+1}$ a limiting edge
and denote by $\tilde{e}$. Suppose, without lost of generality, that
the limiting edge $\tilde{e}$ has its direction from the right to
the left. From Property~\ref{prop:commonedge} it follows that
$\tilde{e}=(s_{i+1},t_i)$. So by assumption, $s_{i+1}$ is on the
right side, while $t_i$ on the left.

\emph{Case 1}. \emph{The solution $P_{\mathrm{opt}}$ contains an
 edge $e$ connecting a vertex of $o_{i+1}$ with a vertex of
$G_i$.}

It follows from Property~\ref{prop:2edgeat-most-1edge} that there
can be only one such edge in  $P_{\mathrm{opt}}$ and from
Property~\ref{prop:1edge-out} that it can not be directed from $G_i$
to $o_{i+1}$. Also from Property~\ref{prop:1edgedirection} it
follows that $e$ can not be directed from the left of $o_{i+1}$ to
the right of $G_i$.

\eat{ Note that $o_{i+1}$ is not a simple vertex, because otherwise
the edge $e$ could not exists. We will prove first that $G_i$ and
$o_{i+1}$ share an edge. Really, suppose they do not. We have three
cases:
\begin{itemize}
\item[$(i)$] $G_i$ and $o_{i+1}$ share a vertex. \\
As it was proved in Lemma~\ref{lem:optimalSubsolution} the only
vertex they can share is the sink of $G_i$, denote it $t_i$ and the
source of $o_{i+1}$, denote it $s_{i+1}$. As $G_i$ is an
$st$-polygon, it is true that $\forall u \in G_i $ there is a path
from $u$ to $t_i$. The subgraph $o_{i+1}$ is also an $st$-polygon,
so there is a path from $s_{i+1}$ to each $v \in o_{i+1}$. Hence
each edge $(v,u)$ connecting a vertex $v$ of $o_{i+1}$ to a vertex
$u$ of $G_i$ will create a cycle.
\item[$(ii)$] $G_i$ and $o_{i+1}$ do not share any vertex and the set of
free vertices between them $\mathcal{F}_i$ is the empty set:
$\mathcal{F}_i=\emptyset$. As it was proved in
Lemma~\ref{lem:optimalSubsolution} there is an edge from $s_i$ to
$t_{i+1}$. So the argument of the previous case can be applied.
\item[$(iii)$] $G_i$ and $o_{i+1}$ do not share any vertex but the set of
free vertices between them $\mathcal{F}_i$ is not empty:
$\mathcal{F}_i\neq\emptyset$. As we know from the
Lemma~\ref{lem:pahtProperties}, for each $w \in \mathcal{F}_i$ there
is path from $t_i$ to $w$ and from $w$ to $s_{i+1}$. So, by the
argument of case $(i)$ we have a contradiction again.
\end{itemize}
So we have proved that $G_i$ and $o_{i+1}$ share an edge. } 

\begin{figure}[htb]
  \begin{minipage}{\textwidth}
    \centering
    \includegraphics[width=0.8\textwidth]{images/correctness_proof.eps}
    \caption{The construction of Lemma~\ref{lem:correctness_commonedge}.}
    \label{fig:correctness_commonedge}
  \end{minipage}
\end{figure}

So we have that the edge of the HP-completion set $e$ is directed
from the right to the left and from $o_{i+1}$ to $G_i$ (see
Figure~\ref{fig:correctness_commonedge}.a). Suppose, as showed in
the figure, that $u=v_i^r$ and that $x$ is the vertex positioned
just below the vertex $v$. In the next we construct two solutions,
$P_i$ and $P_{i+1}$ for $G_i$ and $o_{i+1}$ respectively from the
given optimal solution $P_{\mathrm{opt}}$  and show that they are
optimal.

Observe first that the solution $P_{\mathrm{opt}}$ has a single
edge, $e$, of the HP-completions set that connects $o_{i+1}$ to
$G_i$. Hence the $P_{\mathrm{opt}}$ approaches the vertex $u$ by the
right side from a vertex $w\in G_i$ that is placed on the right
side. Suppose $w$ is the last vertex of this path - looking
backward. Then the previous vertex of the hamiltonian path is on the
other side. And it have to be vertex placed just below $v$, that is
$x$. Otherwise $P_{\mathrm{opt}}$ is not acyclic. Now, let us see
how is the $P_{\mathrm{opt}}$ after the edge $e$. It is surely
continue on the left side and leave it above $t_i$, because as we
supposed $e$ is the single edge connecting $G_i$ and $o_{i+1}$. On
Figure~\ref{fig:correctness_commonedge}.a the solution
$P_{\mathrm{opt}}$ is shown by a dashed bold line. Now we set $P_i$
to be the following path: it is as $P_{\mathrm{opt}}$ till the
vertex $x$ on the left side of $G_i$ (see
Figure~\ref{fig:correctness_commonedge}.b), then it goes right, as
$P_{\mathrm{opt}}$ does, to the $w$, continues on the right side
till the vertex $s_{i+1}$, goes to the left to the vertex $v$ and
continues on the left side till the vertex $t_i$. The $P_{i+1}$ is
as follows: the path starts on the right side on $s_{i+1}$,
continues till the vertex $u$, change the side going to the vertex
$t_i$ and after continues following to $P_{\mathrm{opt}}$. Note that
if the vertex $u$ coincide with the last vertex of the right side
then the solution $P_{i+1}$ terminates on the left side - see
Figure~\ref{fig:correctness_commonedge}.c). Note also that
$P_{\mathrm{opt}}$ can be got from the solutions $P_i$ and $P_{i+1}$
by cases 1.a and 2.c of Lemma~\ref{lem:dynProgSolution}(see
Figures:~\ref{fig:dynProgSolutionLEFT-L}.a
and~\ref{fig:dynProgSolutionLEFT-R}.c).

Now suppose that $P_{i+1}$ is not the optimal solution. Then there
is an $P^\prime_{i+1}$ which has less number of crossings than
$P_{i+1}$. We now compose $P^\prime_{i+1}$ with $P_i$ getting a
solution $P^\prime$. If $P^\prime_{i+1}$ starts on the left side of
the $o_{i+1}$ then it can be trivially composed with $P_i$ by simple
concatenation of this pathes. Suppose now that $P^\prime_{i+1}$
starts on the right. We now have two cases: HP-completions set of
$P^\prime_{i+1}$ contains one or two edges. This is because as we
have  proved in Lemmata~\ref{lemma:3to1} and~\ref{lemma:4to2} for
each optimal solution of an $st$-polygon, there is another one, also
optimal, with at most two edges in the HP-completion set.

$(i)$ Suppose now that $P^\prime_{i+1}$ has two edges.  We suppose
that $v_k^r,~v_{k+1}^r$ are the vertices of the right side where the
path $P^\prime_{i+1}$ leaves and returns to the right side
respectively. Figure~\ref{fig:correctness_optimal}.a shows $P_{i+1}$
and $P^\prime_{i+1}$ by dashed and solid bold lines respectively.
Note that both of this solutions cross all the edges that ``cover''
the vertices $t_i$ and $v_m^l$. Therefore, as we supposed that
$P^\prime_{i+1}$ has less number of crossings, we infer that the
edges $(v_k^r,v)$, $(v_m^l,v_{k+1}^r)$ cross less edges than
$(v_i^r,v)$, $(v_m^l,v_{i+1}^r)$ do.

We now compose the $P^\prime_{i+1}$ with $P_i$ getting a solution
$P^\prime$ using the case of
Figure~\ref{fig:dynProgSolutionLEFT-R}.c. The solution $P^\prime$ do
the same number of crossing that $P_{i+1}^\prime$ and $P_i$ do plus
one, while $P_{\mathrm{opt}}$ has the number of crossing that
$P_{i+1}$ and $P_i$ plus one. As $P_{i+1}^\prime$ is better than
$P_{i+1}$ we conclude that $P^\prime$ is better than
$P_{\mathrm{opt}}$ - contradiction.

\eat{
 take $P_i$ as it is till the vertex
$s_{i+1}$, continue on the right side till the vertex $v^r_k$, go to
the left side - to the vertex $v$, continue on the left side till
the vertex $v_m^l$, go to the right side to the vertex $v_{k+1}$ and
go up to $t_{i+1}$ (see Figure~\ref{fig:correctness_optimal}.b solid
bold curve). As the solution $P^\prime$ contains the edges
$(v_k^r,v)$ and $(v_m^l,v_{k+1}^r)$, while $S$ contains the edges
$(v_i^r,v)$ and $(v_m^l,v_{i+1}^r)$  we have that $S^\prime$ has
less number of crossings than $S$ do. }

\begin{figure}[htb]
  \begin{minipage}{\textwidth}
    \centering
    \includegraphics[width=\textwidth]{images/correctness_proof_optimal.eps}
    \caption{The construction of Lemma~\ref{lem:correctness_commonedge}. Case 1.}
    \label{fig:correctness_optimal}
  \end{minipage}
\end{figure}

$(ii)$ The case when $P^\prime_{i+1}$ has one HP-completion edge can
be treated as a special case, when the edge $(v_{k},v_{k+1})$ is the
last edge of the right side and for composition of the solutions
$P^\prime_{i+1}$ and $P_i$ the case of
Figure~\ref{fig:dynProgSolutionLEFT-L}.a is used.

Now suppose that $P_{i}$ is not the optimal solution. So there is a
$P^\prime_{i}$ which has less number of crossings. As we have in
this case, the edge $e$ is from the right to the left, so the path
$P_{\mathrm{opt}}$ approaches $t_{i}$ by the left side, therefore we
are interested only in $P^\prime_{i}$ that terminates on the left
side, so that the composed solution $P^\prime$ also approaches
$t_{i}$ by the left side.

So let $P^\prime_{i}$ to terminate on the left side. Suppose that
$v^\prime$ is the last vertex (looking to the $P^\prime_{i}$
backwards) of $P^\prime_{i}$ on the left side. It means that before
$v^\prime$, $P^\prime_{i}$ passes through the right side, suppose
that $w^\prime$ is the last vertex of $P^\prime_{i}$ on the right
side.  As the $P^\prime_{i}$ is acyclic, the vertex that precede
$w^\prime$ is the vertex $x^\prime$ on the left side. Note that
vertex $x^\prime$ is the vertex that precedes $v^\prime$ on the left
side. Figure~\ref{fig:correctness_optimal}.c shows $P^\prime_{i}$
drawn by a dashed line, and the solution $P_{i}$ of $G_i$ by a bold
solid line. As we supposed $P^\prime_{i}$ is better than $P_{i}$. It
means that the HP-completion set of $P^\prime_{i}$ makes less
crossing with the edges of $G_i$ than the HP-completion set of
$P_{i}$. Let denote by the values $p_0$, $p_1$, $p_2$, $p_3$ the
number of crossing made by edges $(v_m^l,v_{i+1}^r)$, $(v_i^r,
t_i)$, $(s_{i+1},v)$, $(x,w)$ of $P_i$ and $P_{i+1}$, also let $p$
be the rest number of crossing made by $P_i$. Denote by
$p_2^\prime$, $p_3^\prime$ the number of crossing made by  the edges
$(s_{i+1},v^\prime)$, $(x^\prime,w^\prime)$ of $P_i^\prime$. Now let
us look to the total number of crossing made by $P_{\mathrm{opt}}$
and
$P^\prime$:\\
For $P_{\mathrm{opt}}$: $p_0 +p_1+ p_2 + p_3 + 1+ p \equiv
c_{P_{\mathrm{opt}}}$. While for $P^\prime$: $p_0 +p_1+ p_2^\prime +
p_3^\prime +1+ p \equiv c_{P^\prime}$. We have that: $p_2 + p_3 + p
> p_2^\prime + p_3^\prime + p$ therefore $c_{P_{\mathrm{opt}}} > c_{P^\prime}$ which
is a contradiction, as $P_{\mathrm{opt}}$ is supposed to be the
optimal solution. Figure~\ref{fig:correctness_optimal}.d
 shows $P_{\mathrm{opt}}$ and $P^\prime$ by solid and dashed bold
 lines respectively.

\emph{Case 2}. \emph{The solution $P_{\mathrm{opt}}$ does not
contain any edge connecting vertices of $o_{i+1}$ with a vertices of
$G_i$.}
\\\emph{Case 2a}. \emph{The last vertex of $P_{\mathrm{opt}}$ before
$t_i$ is on the left side.} The only possible such vertex is the
vertex that is placed just below the vertex $t_i$ on the left side.
Let denote it by $x$. (see
Figures:~\ref{fig:dynProgSolutionLEFT-L}.c
and~\ref{fig:dynProgSolutionLEFT-R}.a). Let see now how the path
$P_{\mathrm{opt}}$ is before the vertex $x$. Suppose $u$ is the last
vertex of the $P_{\mathrm{opt}}$ on the left side before $x$, so
before it $P_{\mathrm{opt}}$ is on the right side, the first vertex
on the right side is $s_{i+1}$. If it was below the vertex $s_{i+1}$
then $P_{\mathrm{opt}}$ would contain a cycle, otherwise, if it was
above the vertex $s_{i+1}$ then $P_{\mathrm{opt}}$ would contain an
edge that connects a vertex of $o_{i+1}$ and $G_i$.

Now note that the path $P_{\mathrm{opt}}$ after $t_i$ is on the left
side. Set $P_i$ and $P_{i+1}$ to be as $P_{\mathrm{opt}}$ before and
after $t_i$ respectively. Note that $P_{\mathrm{opt}}$ can be got
from $P_i$ and $P_{i+1}$ by the cases 1.c and 2.a of
Lemma~\ref{lem:dynProgSolution}(see
Figures:~\ref{fig:dynProgSolutionLEFT-L}.c
and~\ref{fig:dynProgSolutionLEFT-R}.a). It is easy to see that $P_i$
and $P_{i+1}$ are optimal: the only solutions $P_i^\prime$
$P_{i+1}^\prime$ for $G_i$ and $o_{i+1}$ that gives $P^\prime$ with
its last vertex  before $t_i$ on the left side are those that
finishes on the left and starts on the left respectively. If we
suppose that one of  $P_i^\prime$, $P_{i+1}^\prime$ is better than
$P_i$, $P_{i+1}$ respectively then by the contraction of cases of
Figures~\ref{fig:dynProgSolutionLEFT-L}.c
and~\ref{fig:dynProgSolutionLEFT-R}.a gives us a better solution -
contradiction.

\emph{Case 2b}. \emph{The last vertex of $P_{\mathrm{opt}}$ before
$t_i$ is on the right side.} $(i)$ \emph{The vertex before $t_i$ is
$s_{i+1}$.} This case corresponds to the cases  1.d and 2.b of
Lemma~\ref{lem:dynProgSolution}(see
Figures:~\ref{fig:dynProgSolutionLEFT-L}.d
and~\ref{fig:dynProgSolutionLEFT-R}.b). So, by the argument of the
previous case we have that $P_{\mathrm{opt}}$ can be got from the
optimal solutions for $G_i$ and $o_{i+1}$ by
Lemma~\ref{lem:dynProgSolution}.  $(ii)$ \emph{The vertex before
$t_i$ is a vertex $v_i^r$ on the right side which is above
$s_{i+1}$.} This case corresponds to the cases  1.b and 2.d of
Lemma~\ref{lem:dynProgSolution}(see
Figures:~\ref{fig:dynProgSolutionLEFT-L}.b
and~\ref{fig:dynProgSolutionLEFT-R}.d), and can be continued as all
previous cases.

Note that, there is not necessarily only one hamiltonian path that
yields an optimal solution. Since in our construction, we apply a
``minimum'' operator,  if both of the involved hamiltonian paths
yield the same number  of crossings, then we should have at least
two different but equivalent (wrt edge crossings) hamiltonian paths.
\qed
\end{proof}

}

Algorihtm~\ref{alg:AHPCCM} is a dynamic programming algorithm, based
on Lemmata~\ref{lem:optimalSubsolution}
and~\ref{lem:dynProgSolution}, which computes the minimum number of
edge crossings $c(G)$ resulting from the addition of a
crossing-optimal  HP-completion set  to an outerplanar $st$-digraph
$G$. The algorithm can be easily extended to also compute the
corresponding hamiltonian path $S(G)$.

\begin{algorithm2e}[tb]

\Input{An Outerplanar $st$-digraph $G(V^l \cup V^r
\cup \{s,t\}, E)$.}

\Output{The minimum number of edge crossing $c(G)$ resulting from
the addition of a crossing-optimal  acyclic HP-completion set to
graph $G$.}

\caption{\textsc{Acyclic-HPC-CM($G$)}}

\label{alg:AHPCCM}

\begin{enumerate}
\item
Compute the $st$-polygon decomposition $\mathcal{D}(G)= \{ o_1,~
\ldots, o_\lambda \}$ of $G$;\;
\item For each element $o_i \in \mathcal{D}(G),~ 1 \leq i \leq
\lambda$, \\
compute $c^1(o_i,L)$, $c^1(o_i,R)$ and $c^2(o_i,L)$, $c^2(o_i,R)$:\\
\hspace*{.7cm}\textbf{if} $o_i$ is a free vertex, \textbf{then}
$c^1(o_i,L)=c^1(o_i,R)=c^2(o_i,L)=c^2(o_i,R)=0$. \\
\hspace*{.7cm}\textbf{if} $o_i$ is an $st$-polygon, \textbf{then}
$c^1(o_i,L)$, $c^1(o_i,R)$, $c^2(o_i,L)$, $c^2(o_i,R)$ are computed
\hspace*{0.4cm} based on Lemma~\ref{lem:spineCrossingSTpolygon}.\\

\item \textbf{if} $o_1$ is a free vertex, \textbf{then} $c(G_1,L)=c(G_1,R)=0$;\\
\textbf{else}  $c(G_1,L)= \min\{c^1(o_1,L),c^2(o_1,L)\}$ and $
c(G_1,R)=\min\{c^1(o_1,R),c^2(o_1,R)\}$;
\item
 For $i=1\ldots \lambda -1,$ compute $c(G_{i+1},L)$ and $c(G_{i+1},R)$ as
 follows:\\
\hspace*{.7cm}  \textbf{if} $o_{i+1}$ is a free vertex,
\textbf{then}
\\ \hspace*{1.4cm} $c(G_{i+1},L)=c(G_{i+1},R)=\min\{ c(G_{i},L), c(G_{i},R)\}$;

\hspace*{.7cm}  \textbf{else-if} $o_{i+1}$ is an $st$-polygon sharing \textbf{at most} one vertex with $G_{i}$, \textbf{then}\\
     \hspace*{1.4cm} $c(G_{i+1},L)=\min\{ c(G_{i},L), c(G_{i},R)\} + \min\{c^1(o_{i+1},L),c^2(o_{i+1},L)\} $;
     \hspace*{1.4cm} $c(G_{i+1},R)=\min\{ c(G_{i},L), c(G_{i},R)\}  +
     \min\{c^1(o_{i+1},R),c^2(o_{i+1},R)\}
     $;\\
\hspace*{.7cm}  \textbf{else} \{ $o_{i+1}$ is an $st$-polygon sharing \textbf{exactly} two vertices with $G_{i}$\}, \\

    \hspace*{1.4cm} \textbf{if} $t_{i} \in V^l$, \textbf{then} \\
        \hspace*{2.1cm}  $c(G_{i+1},L)=\min \{ c(G_{i},L)+c^1(o_{i+1},L)+1,~c(G_{i},R)+c^1(o_{i+1},L),$\\
        \hspace*{4.7cm} $c(G_{i},L)+c^2(o_{i+1},L),~c(G_{i},R)+c^2(o_{i+1},L)\}
        $\\
        \hspace*{2.1cm}  $c(G_{i+1},R)=\min \{
        c(G_{i},L)+c^1(o_{i+1},R),~c(G_{i},R)+c^1(o_{i+1},R)$,
        \\\hspace*{4.8cm}$c(G_{i},L)+c^2(o_{i+1},R)+1,~c(G_{i},R)+c^2(o_{i+1},R)\}$\\

    \hspace*{1.4cm} \textbf{else} \{ $t_{i} \in V^r$ \} \\
        \hspace*{2.1cm} $c(G_{i+1},L)=\min \{
        c(G_{i},L)+c^1(o_{i+1},L),~c(G_{i},R)+c^1(o_{i+1},L),$
        \\\hspace*{4.8cm}$c(G_{i},L)+c^2(o_{i+1},L),~c(G_{i},R)+c^2(o_{i+1},L)+1
\} $\\
        \hspace*{2.1cm}  $c(G_{i+1},R)=\min
        \{c(G_{i},L)+c^1(o_{i+1},R),~c(G_{i},R)+c^1(o_{i+1},R)+1$,
        \\\hspace*{4.8cm}$c(G_{i},L)+c^2(o_{i+1},R),~c(G_{i},R)+c^2(o_{i+1},R)
\} $\\
\item \textbf{return} $c(G)= \min\{ c(G_\lambda, L), c(G_\lambda,R) \}$
\end{enumerate}
\end{algorithm2e}

\begin{theorem}
\label{thm:optimalAcyclicHPCCM} Given an  $n$ node outerplanar
$st$-digraph $G$, a  crossing-optimal HP-completion set for $G$ and
the corresponding number of edge-crossings can be computed in $O(n)$
time.
\end{theorem}

\begin{proof}
Algorithm~\ref{alg:AHPCCM} computes the number of crossings in an
acyclic HP-completion set. Note that it is easy to be extended so
that it computes the actual hamiltonian path (and, as a result, the
acyclic HP-completion set). To achieve this, we only need to store
in an auxiliary array the term that resulted to the minimum values
in Step~4 of the algorithm, together with the endpoints of the edge
that is added to the HP-completion set for each $st$-polygon in the
$st$-polygon decomposition $\mathcal{D}(G)= \{ o_1,~ \ldots,
o_\lambda \}$ of $G$. The correctness of the algorithm follows
immediately from Lemmata~\ref{lem:optimalSubsolution}
and~\ref{lem:dynProgSolution}.

From Lemma~\ref{lem:medianComputation} and
Theorem~\ref{thm:STpolygonDecomposition}, it follows that Step~1 of
the algorithm needs $O(n)$ time. The same hold for Step~2 (due to
Lemma~\ref{lem:spineCrossingSTpolygon}). Step~3 is an initialization
step that needs $O(1)$ time. Finally, Step~4 takes $O(\lambda)$
time. In total, the running time of  Algorithm~\ref{alg:AHPCCM} is
$O(n)$. Observe that $O(n)$ time is enough to also recover the
acyclic HP-completion set.\qed
\end{proof}

\section{Spine Crossing Minimization for Upward Topological 2-Page Book Embeddings of OT-$st$
Digraphs}

In this section,  we establish for the class of $st$-digrpahs an
equivalence (through a linear time transformation) between the
Acyclic-HPCCM problem and the problem of obtaining an upward
topological 2-page book embeddings with minimum number of spine
crossings and at most one spine crossing per edge. We exploit this
equivalence to develop an optimal (wrt spine crossings) book
embedding for  OT-$st$ digraphs.

\begin{theorem}
\label{thm:BEequivHPcompletionSet} Let $G=(V,E)$ be an $n$ node
$st$-digraph. $G$ has a crossing-optimal   HP-completion set $E_c$
with Hamiltonian path $P=(s=v_1, v_2, \ldots, v_n=t)$ such that the
corresponding optimal drawing $\Gamma(G^\prime)$ of $G^\prime=(V,E
\cup E_c)$  has $c$ crossings \textbf{if and only if} $~G$ has an
optimal (wrt the number of spine crossings) upward topological
2-page book embedding with $c$ spine crossings where the vertices
appear on the spine in the order $\Pi=(s=v_1, v_2, \ldots, v_n=t)$.
\end{theorem}

\begin{proof}

\begin{figure}[htb]
    \begin{minipage}{\textwidth}
    \centering
    \includegraphics[width=.75\textwidth]{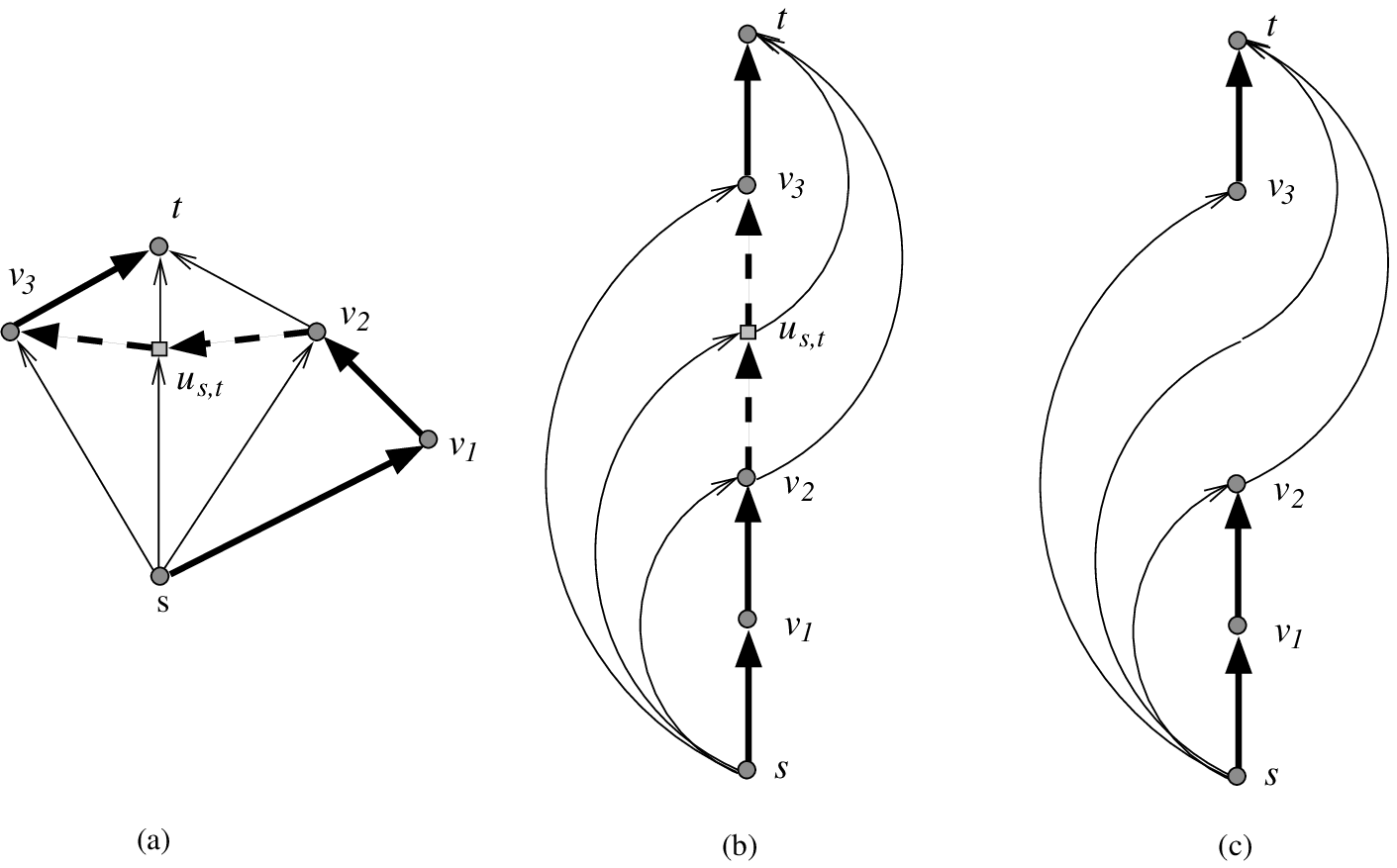}
    \caption{(a) A drawing of an HP-extended digraph for an  $st$-digraph $G$.
     The dotted segments correspond to the single edge $(v_2, v_3)$ of the HP-completion set for $G$.
     (b)An upward topological 2-page book embedding of $G_c$ with
     its
     vertices   placed on the spine in the order they appear
     on a hamiltonian path of $G_c$.
     (c)An upward topological 2-page book embedding of $G$. }
    \label{fig:HPcompletionExample}
  \end{minipage}
\end{figure}

We  show how to obtain from an HP-completion set with $c$ edge
crossings  an upward topological 2-page book embedding with $c$
spine crossings and vice versa. From this is follows that   a
crossing-optimal  HP-completion set for $G$ with $c$ edge crossing
corresponds to an optimal upward
topological 2-page book embedding  with the same number of spine crossings.\\
``$\Rightarrow$''$~~~$ We assume that we have an HP-completion set
$E_c$  that satisfies the conditions stated in the theorem. Let
$\Gamma(G^\prime)$ of $G^\prime=(V,E \cup E_c)$ be the corresponding
 drawing that has $c$ crossings and let $G_c=(V\cup V_c,
E^\prime \cup E_c^\prime)$ be the acyclic HP-extended digraph of $G$
wrt $\Gamma(G^\prime)$.   $V_c$ is the set of new vertices placed at
each edge crossing. $E^\prime$  and $E_c^\prime$ are the edge sets
resulting from $E$ and $E_c$,
 respectively, after splitting their edges
 involved in crossings and maintaining their orientation
 (see Figure~\ref{fig:HPcompletionExample}(a)). Note that $G_c$ is also an $st$-planar digraph.

 Observe that in $\Gamma(G^\prime)$ we have no crossing involving two edges of $G$.
 If this was the
 case, then $\Gamma(G^\prime)$ would not preserve $G$.
 Similarly, in $\Gamma(G^\prime)$ we have no crossing involving two edges of the HP-completion set $E_c$.
 If this was the
 case, then $G_c$ would contain a cycle.

 The hamiltonian path $P$ on $G^\prime$ induces a hamiltonian path
 $P_c$ on the HP-extended digraph $G_c$. This is due to the facts that all
 edges of $E_c$ are used in hamiltonian path $P$ and all vertices of
 $V_c$ correspond to crossings involving edges of $E_c$. We  use
 the hamiltonian path $P_c$ to construct an upward topological
 2-page book embedding for graph $G$ with exactly $c$ spine
 crossings. We place the vertices of $G_c$ on the spine in the order
 of hamiltonian path $P_c$, with vertex $s=v_1$ being the lowest.
 Since the HP-extended digraph $G_c$ is a planar $st$-digraph with vertices $s$ and $t$ on the external face,
 each edge of $G_c$
 appears either to the left or to the right of the hamiltonian path
 $P_c$.
 We place the  edges of $G_c$  on the left (resp. right) page of the book embedding if
 they appear to the left (resp. right) of path $P_c$. The edges of
 $P_c$ are drawn on the spine (see Figure~\ref{fig:HPcompletionExample}(b)).
 Later on they can be moved to any of
 the two book pages.

 Note that all edges of $E_c$ appear on the spine. Consider any vertex  $v_c \in V_c$.
 Since $v_c$ corresponds to a crossing between an edge of $E$ and an edge of
 $E_c$, and the edges of $E_c^\prime$ incident to it have been drawn
 on the spine, the two remaining edges of $E^\prime$ correspond to (better, they are parts of) an edge $e \in E$
 and  drawn on
 different pages of the book.   By removing vertex $v_c$ and merging
 its two incident edges of $E^\prime$ we create a crossing of edge
 $e$ with the spine. Thus, the constructed book embedding has as
 many spine crossings as the number of edge crossings of
 HP-completed graph $G^\prime$ (see Figure~\ref{fig:HPcompletionExample}(c)).

 It remains to show that the
 constructed book embedding is upward. It is sufficient to show that the constructed book
 embedding of $G_c$ is upward. For the sake of contradiction, assume
 that there exists a downward edge $(u,w) \in E_c^\prime$. By the construction, the fact
 that $w$ is drawn below $u$ on the spine implies that there is a
 path in $G_c$ from $w$ to $u$. This path, together with edge
 $(u,w)$ forms a cycle in $G_c$, a clear contradiction since $G_c$
 is acyclic.

``$\Leftarrow$''$~~~$ Assume that we have an upward 2-page
topological book embedding of $st$-digraph $G$ with $c$ spine
crossings where the vertices appear on the spine in the order
$\Pi=(s=v_1, v_2, \ldots, v_n=t)$. Then, we construct an
HP-completion set $E_c$ for $G$ that satisfies the condition of the
theorem as follows: $E_c = \{ (v_i, v_{i+1}) ~|~ 1 \leq i <n
\mbox{~and~} (v_i, v_{i+1}) \not\in E \}$, that is, $E_c$ contains
an edge for each consecutive pair of vertices of the spine that (the
edge) was not present in $G$. By adding/drawing these edges on the
spine of the book embedding we get a drawing  $\Gamma(G^\prime)$ of
$G^\prime=(V,E \cup E_c)$ that has $c$ edge crossings. This is due
to the fact that all spine crossing of the book embedding are
located, (i) at points of the spine above vertex $s$ and below
vertex $t$, and (ii) at points of the spine between consecutive
vertices that are not connected by an edge. By inserting at  each
crossing of $\Gamma(G^\prime)$  a new vertex and by splitting the
edges involved in the crossing while maintaining their orientation,
we get an HP-extended digraph $G_c$. It remains to show that $G_c$
is acyclic. For the sake of contradiction, assume that $G_c$
contains a cycle. Then, since graph $G$ is acyclic, each cycle of
$G_c$ must contain a segment resulting from the splitting of an edge
in $E_c$. Given that in $\Gamma(G^\prime)$ all vertices appear on
the spine and all edges of $E_c$ are drawn upward, there must be a
segment of an edge of $G$ that is downward in order to close the
cycle. Since, by construction, the book embedding of $G$ is a
sub-drawing of $\Gamma(G^\prime)$,  one of its edges (or just a
segment of it) is downward. This is a clear contradiction since we
assume that the topological 2-page book embedding of $G$ is upward.
\qed
\end{proof}

\begin{theorem}
Given an  $n$ node outerplanar $st$-digraph $G$, an upward 2-page
topological book embedding for $G$ with minimum number of spine
crossings and the corresponding number of edge-crossings can be
computed in $O(n)$ time.
\end{theorem}

\begin{proof}
By Theorem~\ref{thm:BEequivHPcompletionSet} we know that by solving
the Acyclic-HPCCM problem on $G$, we can deduce the wanted upward
book embedding. By Theorem~\ref{thm:optimalAcyclicHPCCM}, the
Acyclic-HPCCM problem can be solved in $O(n)$ time.\qed
\end{proof}


\newpage
\section{Conclusions - Open Problems}
We have studied the problem of Acyclic-HPCCM and we have presented a
linear time algorithm that computes a crossing-optimal acyclic
HP-completion set for outerplanar  $st$-digraphs. Future research
topics include the study of the Acyclic-HPCCM on the larger class of
$st$-digraphs.


\bibliographystyle{abbrv}

\end{document}